\documentclass[12pt]{article}
\usepackage{natbib, bm, amssymb, amsbsy,amsmath,graphicx}
\usepackage{algorithm}
\usepackage{algpseudocode}
\usepackage{colortbl,xcolor}
\usepackage{url}
\usepackage{amsfonts}
\usepackage{multirow}
\usepackage[margin=1in]{geometry}
\usepackage[colorlinks=true,allcolors=blue]{hyperref}%
\usepackage{setspace}

\newtheorem{theorem}{Theorem}
\newtheorem{lemma}{Lemma}%
\newtheorem{corollary}{Corollary}%
\newtheorem{proposition}{Proposition}

\newcommand{\cA}{\mathcal{A}}

\newcommand{\bcB}{\bm{\mathcal{B}}}

\newcommand{\balpha}{{\bm\alpha}}

\newcommand{\bfeta}{{\bm\eta}}
\newcommand{\bmu}{{\bm\mu}}

\newcommand{\bDelta}{{\bm\Delta}}

\newcommand{\bLambda}{{\bm\Lambda}}
\newcommand{\bSigma}{{\bm\Sigma}}
\newcommand{\bTheta}{{\bm\Theta}}

\def\b{{\bm b}}
\def\c{{\bm c}}
\def\d{{\bm d}}

\def\u{{\bm u}}

\def\x{{\bm x}}

\def\A{{\bm A}}
\def\B{{\bm B}}
\def\C{{\bm C}}
\def\D{{\bm D}}

\def\H{{\bm H}}
\def\I{{\bm I}}
\def\M{{\bm M}}

\def\X{{\bm X}}
\def\U{{\bm U}}
\def\V{{\bm V}}
\def\W{{\bm W}}
\def\X{{\bm X}}
\def\Y{{\bm Y}}

\begin{document}
\title{\Large \bf Generalized Connectivity Matrix Response Regression with Applications in Brain Connectivity Studies}

\author{
Jingfei Zhang$^{1}$, Will Wei Sun$^{2}$ and Lexin Li$^{3}$ \bigskip \\
\fontsize{10}{10}\selectfont\itshape
$^{1}$\,Department of Management Science, Miami Herbert Business School, \\ 
\fontsize{10}{10}\selectfont\itshape
University of Miami, Miami, FL, 33146. \medskip \\
\fontsize{10}{10}\selectfont\itshape
$^{2}$\,Krannert School of Management, \\ 
\fontsize{10}{10}\selectfont\itshape
Purdue University, West Lafayette, IN, 47906. \medskip \\
\fontsize{10}{10}\selectfont\itshape
$^{3}$\,Department of Biostatistics and Epidemiology, School of Public Health, \\ 
\fontsize{10}{10}\selectfont\itshape
University of California at Berkeley, Berkeley, CA, 94720. \\
}
\date{}
\maketitle

\begin{abstract}
Multiple-subject network data are fast emerging in recent years, where a separate connectivity matrix is measured over a common set of nodes for each individual subject, along with subject covariates information. In this article, we propose a new generalized matrix response regression model, where the observed networks are treated as matrix-valued responses and the subject covariates as predictors. The new model characterizes the population-level connectivity pattern through a low-rank intercept matrix, and the effect of subject covariates through a sparse slope tensor. We develop an efficient alternating gradient descent algorithm for parameter estimation, and establish the non-asymptotic error bound for the actual estimator from the algorithm, which quantifies the interplay between the computational and statistical errors. We further show the strong consistency for graph community recovery, as well as the edge selection consistency. We demonstrate the efficacy of our method through simulations and two brain connectivity studies.
\end{abstract}


\section{Introduction}
\label{sec::intro}

\noindent
Network data are now ubiquitous in a wide range of scientific and business applications. More recently, multiple-subject network data are fast emerging, in which a separate connectivity network is measured over a common set of nodes for each individual subject. Examples include social cognitive science \citep{brands2013cognitive}, genetics \citep{dai2019cell}, and our motivating brain connectivity analysis. Brain connectivity concerns functional and structural architectures of the brain \citep{varoquaux2013learning}. A typical connectivity study collects imaging scans, e.g., functional magnetic resonance imaging (fMRI), or diffusion tensor imaging (DTI), from multiple subjects. Based on the scan, a connectivity network is constructed for each subject, with the nodes corresponding to a common set of brain regions, and the edges encoding functional or structural associations between the regions. In addition, the study collects subject features such as age, sex and other traits. A fundamental scientific question of interest is to characterize the brain connectivity at both the population-level and subject-level, and to {ascertain how subject features modulate the subject-level connectivity changes. Characterizing such individualized brain connectivity networks is central in developing personalized treatment for neurological disorders \citep{sylvester2020individual}.} 

{There is some recent literature on modeling a collection of networks, such as \citet{ChenKang2015,Kang2016,WangKang2016,ZhangCao2017,kundu2018estimating,wang2019hierarchical}.} 
However, these methods may not flexibly associate network connectivity with external covariates.
\citet{wang2017bayesian, durante2017nonparametric} considered Bayesian network models with covariates, which are flexible, but may be computationally intensive, especially for large networks and/or a large number of covariates. 
There is another line of related work in matrix and tensor data analysis. \citet{sun2017store} developed a tensor response regression models, and \citet{kong2019l2rm} proposed a matrix response linear regression model. Both models could only handle a continuous-valued response (see discussions in Section \ref{sec:bene}), and they imposed different structures on the coefficients as our model. \citet{zhang2017tensor, li2017parsimonious, tang2019individualized} considered tensor models, where the tensor was treated as a predictor and the response variable was a scalar.

In this article, we propose a new connectivity matrix response generalized linear regression model for a collection of network samples with network-level covariates. We represent the observed networks as matrix-valued response variables, and the subject covariates as predictors. We then adopt the form of generalized linear model (GLM), and formulate the population-level connectivity, after a proper transformation, as the sum of two high-dimensional components. 
The first component is the intercept matrix and is assumed to possess a low-rank structure. The second component involves the slope coefficient tensor, which models the effects of covariates on the connectivity and is assumed to be sparse. These structural assumptions substantially reduce the number of free parameters, and the subsequent modeling and computation complexity. Moreover, they are scientifically plausible, and are frequently employed in scientific applications \citep{bi2018multilayer}. 

Our proposal makes some useful contributions to both methodology and theory. As to methodology, we develop a systematic approach to model independent connectivity matrices with covariates. The proposed model framework preserves the intrinsic characteristics of networks, facilitates a scalable computation, and allows an explicit quantification of the computational and statistical errors. 
As to theory, we establish several useful statistical properties. We obtain an explicit non-asymptotic error bound for the actual estimator of our algorithm.
This error bound reveals an interesting interplay between the computational efficiency and the statistical rate of convergence. It shows that the computational error decays geometrically with the number of iterations, while the statistical error matches with existing rates for sparse regressions and and low-rank regressions. Built on this error bound, we further establish the consistency of a community detection procedure and the selection consistency, in that we can consistently identify the edges that are affected by the covariates, and exclude those that are not affected. These theoretical analyses are highly nontrivial, involving the alternating gradient descent, the factorization of the low-rank component, the hard-thresholding operator for sparsity, and the non-quadratic form of the loss function. 
{Finally, although our motivating applications are brain connectivity studies, our method is applicable to other problems such as the genetic study that investigates the gene regulatory relationships among gene-gene networks based on single-cell samples \citep{dai2019cell}. }

The rest of the article is organized as follows. Section~\ref{sec::model} introduces the generalized matrix response model. 
Section~\ref{sec::est} develops the estimation algorithm, and Section~\ref{sec::theory} investigates the statistical properties. Section~\ref{sec::sim} presents the simulations and Section~\ref{sec::brain} illustrates with two studies of brain functional and structural connectivity. 
Section~\ref{sec::discuss} concludes the paper with a short discussion. 
All technical proofs are relegated to the supplement.

\section{Generalized Connectivity Matrix Response Model}
\label{sec::model}

\subsection{Notation}
Let $\I_{n\times n}$ denote the $n\times n$ identity matrix. For a vector $\b\in\mathbb{R}^{d_1}$, let $\|\b\|_2$ denote its $\ell_2$ norm. For a matrix $\B\in\mathbb{R}^{d_1\times d_2}$, let $\B_{i\cdot}$ and $\B_{\cdot j}$ denote its $i$th row and $j$th column, and let $\|\B\|_2$, $\|\B\|_*$, $\|\B\|_F$, and $\|\B\|_{\infty}$ denote its spectral norm, nuclear norm, Frobenius norm, and entry-wise infinity norm, respectively. Let $\text{SVD}_r(\B)$ denote the rank-$r$ singular value decomposition of $\B$ such that $\text{SVD}_r(\B)=[\U,\bSigma,\V]$, where $\bSigma_{r\times r}$ is a diagonal matrix with the largest $r$ singular values and $\U_{d_1\times r},\V_{d_2\times r}$ collect the corresponding left and right singular vectors, respectively. For a tensor $\bcB \in \mathbb{R}^{d_1\times d_2\times d_3}$, let $\bcB_{ijk}$, $\bcB_{ij\cdot}$ and $\bcB_{\cdot\cdot k}$ denote its $(i,j,k)$th entry, $(i,j)$th tube fiber, and $k$th frontal slice, respectively. Let $\|\bcB\|_F=\sqrt{\sum_{ijk}\bcB^2_{ijk}}$ and $\|\bcB\|_0$ denote the number of nonzero entries. Lastly, define the tensor matrix product $\langle\B,\bcB\rangle=\sum_{ijk}\bcB_{ijk}\B_{ij}$ for $\B\in\mathbb{R}^{d_1\times d_2}$ and $\bcB \in \mathbb{R}^{d_1\times d_2\times d_3}$.

\subsection{Model Formulation}
\label{sec::net}

\noindent
Consider a network with $n$ nodes, and the $n \times n$ adjacency matrix $\A$, where $\A_{jj'}$ denotes the edge from node $j$ to $j'$, $1 \le j, j' \le n$. If the edge is undirected, then $\A_{jj'} = \A_{j'j}$. The edge can be binary, i.e., $\A_{jj'} \in \{0, 1\}$, or count, i.e., $\A_{jj'}$ is a nonnegative integer. We consider independent network samples observed from $N$ individuals, with corresponding $n\times n$ adjacency matrices $\A^{(1)}, \ldots, \A^{(N)}$. Here we assume all $N$ networks share a common set of $n$ nodes. Additionally, for each subject, we observe a vector of $p$ covariates, denoted by $\x_i=(x_{i1},\ldots,x_{ip})^\top$. 

Denote $\bmu^{(i)} = \mathbb{E}\{ \A^{(i)} | \x_i \}$, where the expectation $\mathbb{E}(\cdot)$ is applied element-wise to the entries in $\A^{(i)}$. We assume that $\A^{(i)}$ conditional on $\x_i$ follows an exponential family distribution with a canonical link function, i.e., 
\begin{equation}\label{eqn::exp}
f(\A^{(i)}|\bmu^{(i)}) = \prod^n_{j\ne j'}h(\A_{jj'}^{(i)})\left[ \A^{(i)}_{jj'}\bfeta^{(i)}_{jj'} - \psi\left\{ \bfeta^{(i)}_{jj'} \right\} \right],
\end{equation}
where $\bfeta^{(i)}=g(\bmu^{(i)})$, $g(\cdot)$ is a known invertible link function in usual GLM and is applied element-wise to the entries of $\bmu^{(i)}$, and $\psi(\cdot)$ is the cumulant function with its first derivative $\psi'(\cdot) = g(\cdot)^{-1}$. Furthermore, we postulate that, 
\begin{equation}\label{eqn::model}
g\left\{ \bmu^{(i)} \right\} = \bTheta + \bcB \times_3 \x_i, \quad i=1,\ldots,N,
\end{equation}
where $\bTheta \in \mathbb{R}^{n\times n}$ is the intercept matrix that characterizes the population level connectivity, $\bcB \in \mathbb{R}^{n\times n\times p}$ is the slope tensor that encodes the effects of subject covariates on the connectivity matrix, and $\bcB\times_{3}\x_i=\sum_{l=1}^{p} x_{ip} \, \bcB_{\cdot\cdot l}$.

We assume the population level connectivity $\bTheta$ to be low-rank, which reduces the number of free parameters {and is plausible in neuroscience applications \citep{bi2018multilayer,kong2019l2rm}.}
Next, we assume that $\bcB$ is sparse, i.e., the effects of covariates concentrate only on a subset of connections. This sparsity assumption again reduces the number of free parameters, greatly facilitates the model interpretation, and is also well supported by empirical neurological studies \citep{vounou2010discovering}. As every subject has a unique sparse deviation $\bcB\times_3\x_i$ from the low-rank $\bTheta$, model \eqref{eqn::model} is identifiable. 
It is possible to impose more complex structures on $\bTheta$ and $\bcB$; e.g., $\bcB$ is low-rank and sparse, or $\bcB$ is slice sparse. These changes require some straightforward modifications to the estimation procedure. We choose to focus on the current setup as it offers a good balance between model complexity and model flexibility. 

To ensure the low-rank structure of $\bTheta$, we adopt the Burer-Monteiro factorization \citep{burer2003nonlinear}, in which the low-rank matrix is reparameterized as the product of two factor matrices, $\bTheta=\U\V^\top$, where $\U, \V\in\mathbb{R}^{n\times r}$, and $r$ is the rank of $\bTheta$. This reparameterization avoids repeatedly performing the computationally expensive SVD, which is often required in optimization with the low-rank constraint. If the adjacency matrix is symmetric, we reparameterize $\bTheta$ as $\bTheta=\U\bLambda\U^\top$, where $\U\in\mathbb{R}^{n\times r}$ and $\bLambda$ is a $r\times r$ diagonal matrix with diagonal entries $\{-1,1\}$. If $\bTheta$ is positive semi-definite (PSD), then $\bLambda$ becomes the identity matrix. We note that the intercept matrix $\bTheta$ may not be PSD.
To enforce the sparsity of $\bcB$, we adopt the hard-thresholding sparsity constraint, by setting $\|\bcB\|_0\le s$ for some positive integer $s$. Compared to the lasso type soft-thresholding constraint, the hard-thresholding constraint reduces bias and has been shown to enjoy superior performance in many high-dimensional problems \citep{zhang2018unified}.

\subsection{Benefits of Imposing Separate Structures}
\label{sec:bene}
{We discuss the benefits and necessity of imposing separate structures on $\bTheta$ and $\bcB$. At first glance, it seems that one could stack $\bTheta$ and $\bcB$ into one larger coefficient tensor of size $n\times n\times (p+1)$ and assume it to be both low-rank and sparse, as has been considered in \citet{sun2017store}. However, assuming $\bTheta$ to be sparse may not be plausible in a generalized linear model with a nonlinear $g(\cdot)$. 
For example, when the edges are binary and $g(\cdot)$ taken to be the logit link, $g(0)$ yields a connecting probability of 0.5; when the edges are counts and $g(\cdot)$ taken to be the log link, $g(0)$ is not well defined. 
Thus, in the nonlinear case, a sparse $\bTheta$ does not necessarily imply sparsity connectivity at the population-level and may not even be well defined. This is a unique challenge in using generalized linear models to model discrete-valued connectivity matrices.
In our proposal, we assume $\bTheta$ to be low rank, which effectively reduces the number of free parameters, and it connects \eqref{eqn::model} with several commonly used network models, as we further discuss in Section \ref{sec::con}; furthermore, we assume $\bcB$ to be sparse, which is scientifically plausible and also enables edge selection in which we identify the edges that are modulated by covariates. }

{Moreover, based on the Burer-Monteiro reparameterization of the low-rank $\bTheta$, we can detect clusters, or communities, of nodes, so that the nodes are more densely connected within the clusters and less so between the clusters (see Section \ref{sec::cluster}). 
This contributes to the network community detection literature as existing spectral clustering methods cannot handle network heterogeneity due to the network-level covariates.}

\subsection{Connections with Existing Models for a Single Network}
\label{sec::con}

\noindent
Our proposed model \eqref{eqn::model}, when applied to a single network sample, is connected to several prevalent network models, including the stochastic blockmodel \citep{holland1983stochastic}, the latent space model \citep{hoff2002latent}, and the latent factor model \citep{minhas2016inferential}. Similar to those models, our model \eqref{eqn::model} also assumes the low-rank structure, but is more general in that it imposes no additional structural constraint, e.g., the block structure. 

Consider a single observed adjacency matrix $\A$, and $\bmu = \mathbb{E}(\A)$. The stochastic blockmodel is one of the most popular network models. It imposes that the nodes form $K$ communities, and the edges are determined by the community memberships of the two end nodes and are independent given the community assignment. Accordingly, the model can be written as 
\begin{equation*}
g(\bmu) = \C \M \C^\top, 
\end{equation*}
where $\C$ is a $n\times K$ community assignment matrix, with $\C_{jk}=1$ if node $j$ belongs to the $k$th community,  and 0 otherwise, and $\M\in\mathbb{R}^{K\times K}$ characterizes the connecting probabilities within and between the $K$ communities. It is seen that the rank of the matrix $\C \M \C^\top$ is $K$, and may be viewed as a special case of model \eqref{eqn::model}.

The latent space model \citep{hoff2002latent} is another popular model thanks to its easy interpretation. It assumes the nodes are positioned in a $K$-dimensional latent space, and two nodes are likely to form a tie if their latent positions are close. The model can be written as 
\begin{equation*}
g(\bmu) = \alpha\bm{1}_n\bm{1}_n^\top + \C (\M \C)^\top,
\end{equation*}
where $\bm{1}_n$ is an $n$-dimensional vector of ones, $\C \in \mathbb{R}^{n \times K}$ has its $j$th row $\c_j^\top \in \mathbb{R}^{K \times 1}$ encoding the latent position of node $j$, and $\M \in\mathbb{R}^{n\times n}$ is a diagonal matrix with its $j$th diagonal entry equal to $1 / \| c_j\|$, $1 \leq j \leq n$ We see the rank of the matrix $\alpha\bm{1}\bm{1}^\top+\C(\M\C)^\top$ is $(K+1)$, and thus this model is again a special case of \eqref{eqn::model}. Relatedly, the latent factor model \citep{minhas2016inferential}, similar to the latent space model, imposes
\begin{equation*} 
g(\bmu) = \balpha \otimes \bm{1}_n^\top+\balpha^\top \otimes \bm{1}_n + \C\C^\top, 
\end{equation*}
where $\balpha \in \mathbb{R}^{n}$ encodes the additive effect, and $\C \in \mathbb{R}^{n \times K}$ encodes the multiplicative effect. In this case, the rank of the matrix $\balpha \otimes \bm{1}_n^\top+\balpha^\top \otimes \bm{1}_n + \C\C^\top$ is $(K+1)$.

\section{Estimation}
\label{sec::est}

\noindent
Denote the negative log-likelihood function of the connectivity matrix response model \eqref{eqn::model} by $\ell(\bTheta, \bcB)$, which, up to a constant, is of the form \citep{mccullagh1989generalized},
\begin{equation} \label{loglikelihood}
\ell(\bTheta, \bcB) = -\frac{1}{N} \sum^N_{i=1}\sum^n_{j\ne j'}\left[ \A^{(i)}_{jj'}\bfeta^{(i)}_{jj'} - \psi\left\{ \bfeta^{(i)}_{jj'} \right\} \right],
\end{equation}
where $\bfeta^{(i)}=\bTheta + \bcB \times_3 \x_i$. We propose to estimate the parameters $\bTheta$ and $\bcB$ through a non-convex regularized optimization. 
We first develop the optimization algorithm for the general case without the symmetry constraint,  {which is an easier scenario. Building upon this procedure, we further develop the algorithm} for the symmetric case. 

\begin{algorithm}[t!]
\caption{Optimization algorithm for \eqref{obj3}}
\begin{algorithmic}
\State Step 1: compute $\bar\A=\frac{1}{N}\sum_{i=1}^N \A^{(i)}$ and let $\text{SVD}_r\{g(\bar\A)\}=[\bar\U_0,\bar\bSigma_0,\bar\V_0]$. Set $\U^{(0)}=\bar\U_0\bar\bSigma_0^{1/2}$, $\V^{(0)}=\bar\V_0\bar\bSigma_0^{1/2}$, and $\bcB^{(0)}=\textbf{0}$. 
\Repeat
\State Step 2: update $\U^{(t+1)}=\U^{(t)} - \delta \nabla_{\U}\tilde\ell \left\{ \U{\V^{(t)}}^\top,\bcB^{(t)} \right\} \Big |_{\U=\U^{(t)}}$;
\State Step 3: update $\V^{(t+1)}=\V^{(t)} - \delta \nabla_{\V}\tilde\ell \left\{ \U^{(t+1)}\V^\top,\bcB^{(t)} \right\} \Big |_{\V=\V^{(t)}}$;
\State Step 4: update $\bcB^{(t+1)}=\text{Truncate}\left[\bcB^{(t)} - \tau\nabla_{\bcB}\tilde\ell \left\{ \U^{(t+1)}{\V^{(t+1)}}^\top,\bcB \right\} \Big |_{\bcB=\bcB^{(t)}},s\right]$. 
\Until{the objective function converges.}
\end{algorithmic}
\label{algo_asym}
\end{algorithm}

For the general case that $\bTheta$ is low-rank but not necessarily symmetric, we consider the factorization $\bTheta = \U\V^\top$ and the corresponding optimization problem, 
\begin{equation} \label{obj3}
\min_{\substack{{\U,\V\in\mathbb{R}^{n\times r}}\\{\bcB\in\mathbb{R}^{n\times n\times p}}}} \tilde\ell \left( \U\V^\top,\bcB \right),\quad\text{subject to}\quad\|\bcB\|_0\le s,
\end{equation}
where we augment the loss function $\ell(\bTheta, \bcB)$ with an additional regularizer, $\tilde\ell\left( \U\V^\top,\bcB \right)= \ell \left( \U\V^\top,\bcB \right)+\frac{1}{8}\|\U^\top\U-\V^\top\V\|_F^2$.
The regularizer $\|\U^\top\U-\V^\top\V\|_F^2 / 8$ is added to guarantee the uniqueness of the factorization of $\bTheta=\U\V^\top$ \citep{zheng2016convergence}; it does not change the optimization problem, but merely reduces the set of solutions from all possible factorizations to the ones that are balanced in that $\U$ and $\V$ have the same singular values.
To enforce sparsity along the solution path, we employ a truncation operator $\text{Truncate}(\bcB,s)$, 
\begin{equation*}
\left[ \text{Truncate}(\bcB,s) \right]_{jj'l}=
\begin{cases}
\bcB_{jj'l}       & \quad \text{if } (j,j',l)\in \text{supp}(\bcB,s),\\
0  & \quad \text{otherwise},
\end{cases}
\end{equation*}
for $\bcB\in R^{d_1\times d_2\times d_3}$ and $s\le d_1d_2d_3$. Here $\text{supp}(\bcB,s)$ is the set of indices of $\bcB$ corresponding to its largest $s$ absolute values. We then develop an alternating gradient descent algorithm for \eqref{obj3} to iteratively update $\U$, $\V$ and $\bcB$. We summarize the optimization procedure in Algorithm \ref{algo_asym}. In this algorithm, $\nabla_{\U} \tilde\ell(\U\V^\top, \bcB)$ denotes the gradient of the objective function $\tilde\ell(\U\V^\top, \bcB)$ with respect to $\U$, and $\nabla_{\V} \tilde\ell(\U\V^\top, \bcB)$, $\nabla_{\bcB} \tilde\ell(\U\V^\top, \bcB)$ are defined similarly. Explicit forms of these gradients are given in the supplement. 
In Section \ref{sec::theory}, some theoretical conditions are placed on $\delta$ and $\tau$ to ensure the linear convergence rate of the algorithm, based on which we discuss their empirical choices.

Next, for the case that $\bTheta$ is low-rank and symmetric, we consider the factorization $\bTheta = \U\bLambda\U^\top$ and the corresponding optimization problem, 
\begin{equation}
\min_{\substack{{\U\in\mathbb{R}^{n\times r}, \bLambda\in\mathcal{D}_r}\\{\bcB\in\mathbb{R}^{n\times n\times p}}}} \ell(\U\bm\Lambda\U^\top,\bcB),\quad\text{subject to}\quad\|\bcB\|_0\le s,
\label{obj2}
\end{equation}
where $\mathcal{D}_r$ denotes the set of all $r\times r$ diagonal matrices with diagonal entry values $\{-1,1\}$. The alternating gradient descent algorithm for \eqref{obj2} is summarized in Algorithm \ref{algo_symm}. In this algorithm, we have chosen not to update the estimate of $\bLambda$. This is because we initialize by first solving the optimization problem \eqref{obj3}, treating $\bTheta$ as a general matrix without the symmetry constraint. From solving \eqref{obj3}, the obtained $[\widetilde\U; \widetilde\V]$ consistently estimates $[\U^*;\bLambda{\U^*}^\top]$, as we show in Proposition~\ref{thm1} in the supplement, where $\bTheta^*=\U^*\bLambda{\U^*}^\top$ is the true coefficient. As such, the diagonal entries of $\bLambda$ can be accurately estimated using $\bLambda_{ii}=\text{sign}(\widetilde\U_{.i}^\top\widetilde\V_{.i})$.

\begin{algorithm}[t!]
\caption{Optimization algorithm for \eqref{obj2}}
\begin{algorithmic}
\State Step 1: first solve \eqref{obj3} using Algorithm \ref{algo_asym} and denote the output as $\widetilde\U, \widetilde\V, \widetilde\bcB$. Set $\bLambda_{ii}=\text{sign}(\widetilde\U_{.i}^\top\widetilde\V_{.i})$, $i=1,\ldots,r$,  $\U^{(0)}=(\widetilde\U + \bLambda{\widetilde\V}^\top) / 2$ and $\bcB^{(0)}=\widetilde\bcB$.
\Repeat
\State Step 2: update $\U^{(t+1)}=\U^{(t)} - \delta \nabla_{\U}\ell \left\{ \U\bLambda{\U}^\top,\bcB^{(t)} \right\} \Big |_{\U=\U^{(t)}}$; 
\State Step 3: update $\bcB^{(t+1)}=\text{Truncate}\left[\bcB^{(t)} - \tau\nabla_{\bcB}\ell \left\{ \U^{(t+1)}\bLambda{\U^{(t+1)}}^\top,\bcB \right\} \Big|_{\bcB=\bcB^{(t)}},s\right]$.
\Until{the objective function converges.}
\end{algorithmic}
\label{algo_symm}
\end{algorithm}

The rank $r$ and the sparsity $s$ in \eqref{obj3} and \eqref{obj2} are two tuning parameters. We select these parameters via the eBIC criterion proposed in \cite{chen2012extended}. Specifically, for the general case, among a set of working ranks and sparsity levels, we choose the combination of $(r, s)$ that minimizes
\begin{equation}\label{eqn::ebic}
\text{eBIC} = 2N\times\ell \bigl( \hat{\bTheta}, \hat\bcB \bigr)+\left[\log(n^2N) + \log\left \{ n^2(p+1)\right\} \right]\times\left(2nr+s\right),
\end{equation}
where $\ell$ is the loss function in \eqref{loglikelihood}, and $\hat{\bTheta}, \hat\bcB$ are the estimates of $\bTheta, \bcB$ under the working rank and sparsity level. The eBIC criterion for the symmetric case is computed similarly.

\section{Theory}
\label{sec::theory}

\noindent
We first derive the non-asymptotic error bound of the actual estimator from our algorithm, and then establish community detection consistency and edge selection consistency. We focus on the symmetric case and leave results for the asymmetric case to the supplement.

\subsection{Non-asymptotic error bound}
Assume the parameter space for $\{\bm{\Theta},\bcB\}$ is compact. Let $\bTheta^*$ denote the true coefficient matrix with rank $r^*$ and $\bcB^*$ the true coefficient tensor with $s^*$ nonzero entries. 
{Denote the nonzero singular values of $\bTheta^*$ as $\sigma^*_1\ge\ldots\ge\sigma^*_{r^*}>0$. Write $\bTheta^*=\U^*\bLambda{\U^*}^\top$, where $\U^*\in\mathbb{R}^{n\times r^*}$ and $\bLambda$ is a $r^*\times r^*$ diagonal matrix with diagonal entries in $\{-1,1\}$, collecting signs of the singular values of $\bTheta^*$.}
Let $\mathbb{B}_{\bTheta^*}(\kappa_1)\subset\mathbb{R}^{n\times n}$ and $\mathbb{B}_{\bcB^*}(\kappa_2)\subset\mathbb{R}^{n\times n\times p}$ denote the Frobenius-norm ball around $\bTheta^*$ with radius $\kappa_1>0$ and around $\bcB^*$ with radius $\kappa_2>0$, respectively. 
We next introduce several regularity conditions on covariate $\x_i$'s and the model. 

\begin{enumerate}
\item[(B1)] The samples $\x_i$'s are i.i.d.\ from a zero-mean distribution, and the covariance matrix $\bSigma_x$ satisfies that $b_l\le\lambda_{\min}({\bSigma_x})\le\lambda_{\max}({\bSigma_x})\le b_u$ for some positive constants $0<b_l\le b_u<\infty$, where $\lambda_{\min}(\bSigma_x)$ and $\lambda_{\max}(\bSigma_x)$ denote the smallest and largest eigenvalues of $\bSigma_x$, respectively. 
\item[(B2)] The covariates are bounded by some constant $M_x>0$, i.e., $|\x_{is}|\le M_x$.
\item[(B3)] Each element of $\A^{(i)}$ conditional on $\x_i$ follows an exponential family distribution with continuous $\psi''(\cdot)$. For $\bTheta\in\mathbb{B}_{\bTheta^*}(\sqrt{\sigma^*_{r^*}}/3)$ and $\bcB\in\mathbb{B}_{\bcB^*}(\sqrt{\sigma^*_{r^*}}/3)$, it holds that $\nu_0^{-1}\le\psi''(\bTheta_{jj'}+\x_i^\top\bcB_{jj'})\le \nu_0$, for any $j$ and some large constant $\nu_0>0$.
\item[(B4)] For $\bTheta\in\mathbb{B}_{\bTheta^*}(\sqrt{\sigma^*_{r^*}}/3)$, $\bcB\in\mathbb{B}_{\bcB^*}(\sqrt{\sigma^*_{r^*}}/3)$, we have $|\mathbb{E}\langle\psi''(\bfeta^{(i)})\circ\bTheta,\bcB\times_3\x_i\rangle|\le\kappa_0\|\bTheta\|_F\cdot\|\bcB\|_F$, where $\circ$ denotes Hadamard product and $\kappa_0=\sqrt{\lambda_{\min}({\bSigma_x})}/(18\nu_0)$.
\end{enumerate}

\noindent 
Condition (B1) places a regularity condition on the design matrix the and Condition (B2) is to bound the Hessian of the cumulant function in the neighborhood of $\bcB^*$. These two conditions are commonly assumed in high-dimensional generalized linear models \citep{negahban2012unified}. 
{
Condition (B3) is satisfied by most generalized linear models. In particular, the boundedness of $\psi''(\bTheta_{jj'}+\x_i^\top\bcB_{jj'})$ is directly implied by Condition (B2) as well as the compactness of the parameter space for $\{\bm{\Theta},\bcB\}$. Condition (B4) is to bound the Lipschitz gradient parameter. In the case of a linear model, (B4) is easily satisfied with $|\mathbb{E}\langle\psi''(\bfeta^{(i)})\circ\bTheta,\bcB\times_3\x_i\rangle|=0$, since $\psi''(\cdot)$ is a constant and $\x_i$ has mean zero. For a GLM, such as a logistic or multinomial model, $\psi''(\cdot)$ is not a constant, and (B4) requires the inner product of a sparse matrix and a low-rank matrix to be bounded. This is satisfied if the sparse entries are spread out so that $\bcB$ is not exactly low-rank, and the low-rank matrix is not spiky so that $\bm{\Theta}$ is not sparse. Such a conditions has been commonly assumed in the matrix factorization literature; see, e.g., \citet{zhang2018unified}. 
 
For any $\U$ and $\bcB$, we define the distance
\begin{eqnarray*}
D\left\{\U,\bcB\right\}=d^2(\U,\U^*)+\|\bcB-\bcB^*\|_F^2/\sigma^*_1, \;\; \text{ where } \; d(\U,\U^*)=\min_{\bm\Gamma\in\mathbb{Q}_{r^*}}\|\U-\U^*\bm\Gamma\|_F, 
\end{eqnarray*}
and $\mathbb{Q}_{r^*}$ denotes the set of $r^*\times r^*$ orthonormal matrices. The factor $1/\sigma^*_1$ in the distance metric comes from the difference between $\bTheta$ and $\U$, as it holds that $\|\bTheta-\bTheta^*\|_F^2\le c\sigma^*_1d^2(\U,\U^*)$ for a constant $c$ \citep{zhang2018unified}. 
The next theorem gives the non-asymptotic error bound of $\U^{(t)}$ and $\bcB^{(t)}$ from Algorithm \ref{algo_symm} at the $t$th iteration under the GLM loss function \eqref{loglikelihood}. 

\begin{theorem}
\label{thm3}
Assume (B1)-(B4) and define $\mu_1=\nu_0^{-1}$, $\mu_2=\lambda_{\min}(\bSigma_x)/(4\nu_0)$, $\alpha_1=\nu_0$ and $\alpha_2=7\lambda_{\max}(\Sigma_x)\nu_0/4$. 
Let $c_1$ and $c_2$ be constants such that $c_1\le\mu_1/(96\alpha_1^2)$, and $3c_1\alpha_2\le c_2\le\min\left\{1/3,\sqrt{\mu_1/(5\alpha_1)}\right\}$. 
Let the step sizes $\delta=c_1/\sigma^*_1$, $\tau=c_2/\alpha_2$, and $s=\gamma s^*$, where $\gamma \ge 1+\{ (3\alpha_2+\mu_2c_2)/(\mu_2c_2) \}^2$. 
When $N\ge c_3(r^*n\log n+s^*\log n)$ for some constants $c_3$ and $c_4$, for any initial estimator $\left\{ \U^{(0)},\bcB^{(0)} \right\}$ satisfying $D\{ \U^{(0)},\bcB^{(0)} \} \le c_2^2\sigma^*_{r^*}$, we have, with probability at least $1-c_4/n$, 
\begin{equation} \label{eqn::glm}
D\left\{ \U^{(t)},\bcB^{(t)} \right\} \le \rho^tD\left\{ \U^{(0)},\bcB^{(0)} \right\} + \phi_1 \frac{r^*n\log n}{N}+\phi_2\frac{s^*\log n}{N}, 
\end{equation}
where $\rho=\max\{1-\delta\mu_1\sigma^*_{r^*}/16,1-\tau\mu_2/18\}\in(0,1)$ is a contraction parameter, and $\phi_1$ and $\phi_2$ are constants that depend on $c_1$, $c_2$, $\nu_0$, $\lambda_{\min}(\bSigma_x)$ and $\lambda_{\max}(\bSigma_x)$.
\end{theorem}

\noindent  
Theorem~\ref{thm3} portrays the estimation error at each iteration. The defined $(\mu_1, \mu_2$) and $(\alpha_1,\alpha_2)$ are convexity and smoothness parameters that reflect the lower and upper bounds on the ``curvature" of the objective function around the true parameters, respectively.
The error bound consists of two terms that correspond to the computational error and the statistical error, respectively. 
It reveals an interesting interplay between the computational efficiency and the statistical rate of convergence. Note that the computational error decays geometrically with the iteration number $t$, whereas the statistical error remains the same. Therefore, as the iteration number increases, the computational error is to be dominated by the statistical error and the resulting estimator falls within the statistical precision of the true parameter.

We make a few remarks on the computational error, statistical error, initial condition and the choice of the step sizes $\delta$ and $\tau$. 
Firstly, the computational error $\rho^tD\{\U^{(0)},\bcB^{(0)}\}$ directly relies on the contraction parameter $\rho$, in that a smaller value of $\rho$ leads to a faster convergence. When the step sizes $\delta$ and $\tau$ increase, $\rho$ decreases. 
Secondly, the term $r^*n\log n/N$ is the statistical error from the low-rank matrix estimation, which, up to a logarithmic factor, matches with error rate for multi-response regression with a low-rank constraint \citep{raskutti2011minimax}, and the term $s^*\log n/N$ is the statistical error from the sparse tensor estimation, which matches with the error rate in sparse regressions \citep{negahban2012unified}.
Thirdly, in Theorem~\ref{thm3}, we require the initialization error to be bounded. Such an assumption is often needed in non-convex optimizations \citep{zhang2018tensorSVD}. An initialization method for low-rank models that satisfies this assumption is the spectral initialization \citep{zhang2018tensorSVD}. In Algorithm \ref{algo_asym}, we initialize with the truncated singular value decomposition, which we have found to enjoy a good empirical performance. 
Finally, Theorem \ref{thm3} offers useful guidance on the choice of the step sizes $\delta$ and $\tau$. Their bounds hinge on $\nu_0$, $\lambda_{\min}(\bSigma_x)$ and $\lambda_{\max}(\bSigma_x)$. These quantities can be estimated from the data. Specifically, we estimate $\bSigma_x$ by its usual sample covariance estimator, and approximate $\nu_0$ through the second derivative $\psi''$ that $\nu_0^{-1}\le\psi''(\bTheta_{jj'}+\x_i^\top\bcB_{jj'})\le \nu_0$.

\subsection{Consistency of Community Detection and Edge Selection}
\label{sec::cluster}

\noindent
One implication of our model is that we may recover the community structure of the nodes given the low-rank parameterization of $\bTheta$. We show that our solution can correctly recover the true community labels for all nodes with probability $1-O(K/n)$, while allowing the number of communities $K$ to grow sub-linearly with the number of nodes $n$.  

We first formally define the true underlying community structure. Based on $\U^*$ from the  decomposition $\bTheta^* = \U^*\bLambda\U^{*\top}$, the true community structure is determined by the rows of $\U^*$ in that there are $K$ distinct groups of rows, such that 
\begin{equation*} \label{eqn:beta3_cluster}
\U^* = \left( \U_{1\cdot}^{*}, \ldots, \U_{n\cdot}^{*} \right)^\top = \Bigl( \underbrace{ \u^*_{1},\ldots,  \u^*_{1}}_{l \textrm{~nodes}}, \; \underbrace{\u^*_{2}, \ldots, \u^*_{2}}_{l \textrm{~nodes}}, \; \ldots, \; \underbrace{\u^*_{K}, \ldots, \u^*_{K}}_{l \textrm{~nodes}} \Bigr)^\top  \in \mathbb R^{n\times r^*},
\end{equation*}
where $\u^*_{k}\in\mathbb{R}^{1\times r^*}$, $k=1,\ldots,K$. Here for notational simplicity, we assume there is an equal number of nodes, $l = n/K$, in each community. Accordingly, we define the true community assignments as $\cA_1^*:= \{1,\ldots, l\}, \ldots, \cA_K^*:= \{n-l, \ldots, n\}$. 

We propose to recover community labels by applying a distance-based clustering procedure, such as $K$-means to rows of the final estimate $\U^{(t)}$ obtained from Algorithm~\ref{algo_symm}. We show that the resulting clustering output achieves strong consistency, under the following regularity conditions.  

\begin{enumerate}
\item[(C1)] Assume that $\sigma^*_{r^*} > c_5$ for some constant $c_5 > 0$, where $\sigma^*_{r^*}$ is the smallest non-zero singular value of $\bTheta^*$.
\item[(C2)] Assume that $\min_{k\ne k^{'}} \|\u^*_{k} - \u^*_{k^{'}}\|^2_2 > c_6 e_0$ for some constant $c_6>0$, where $e_0=\phi_1 r^*n\log n / N + \phi_2 s^*\log n / N$, and $\phi_1$, $\phi_2$ are defined as in Theorem~\ref{thm3}.
\end{enumerate}

\noindent
Condition (C1) requires that the minimum non-zero singular value of $\bTheta^*$ is bounded below by a positive constant.
Condition (C2) ensures that the minimal gap between different cluster centers does not tend to zero too fast. 

\begin{theorem}
\label{thm:cluster}
Suppose the conditions in Theorem~\ref{thm3} and (C1)-(C2) hold. Then after $t$ iterations, with $t \ge \log_{\rho}\left(e_0 / D\{\U^{(0)},\bcB^{(0)}\}\right)$, we have, with probability at least $1-c_4K/n$, $\widehat{\cA}_k^{(t)} = \cA_k^*$, for all $k=1,\ldots,K$, where $c_4$ is the constant defined as in Theorem~\ref{thm3}. 
\end{theorem}

\noindent
Theorem~\ref{thm:cluster} shows that our community detection procedure achieves the strong consistency as long as $K=o(n)$. 
Note that $\hat\bTheta$ is estimated after the covariate effects have been removed from the connectivity matrix. Existing spectral clustering methods, either for a single network or for multiple networks, cannot handle heterogeneity due to the network-level covariates. 
Our result allows $K$ to grow at a sub-linear rate with $n$, which is achievable as we have $N$ network samples, which provides more information than a single sample.

Another property of our estimator is that we can select the edges that are affected by the covariates consistently. 
\begin{corollary}
\label{thm:selection}
Assume all conditions in Theorem~\ref{thm3} hold and $\min_{ijk}| \bcB_{ijk}^*| > 2 \sqrt{\sigma^*_1e_0}$. Then after $t$ iterations with $t \ge \log_{\rho}\left(e_0 / D\{\U^{(0)},\bcB^{(0)}\}\right)$, we have, with probability at least $1-c_4/n$, for any $\bcB_{ijk}^* \ne 0$, the estimate $\bcB^{(t)}_{ijk} \ne 0$, and for any $\bcB_{ijk}^* = 0$, the estimate $\bcB^{(t)}_{ijk} = 0$.
\end{corollary}

\noindent
Corollary \ref{thm:selection} is a direct consequence of Theorem \ref{thm3}, and thus we omit its proof. The condition on $\min_{ijk}| \bcB_{ijk}^*|$ is a minimal signal condition, which is commonly employed to establish selection consistency \citep{kong2019l2rm}. 
This result has an important implication in practice, as it ensures that our model can correctly select the edges that are affected by the subject covariates.

\section{Simulations}
\label{sec::sim}

\noindent
We carry out experiments to investigate the finite-sample performance of our proposed method, and to compare with some competing solutions. We focus on the symmetric matrices throughout the simulations. We first consider our proposed model \eqref{eqn::model} and then the CISE model of \citet{wang2019common}, where our model structure is not satisfied. We further consider a stochastic blockmodel \citep{holland1983stochastic} and a latent factor model \citep{minhas2016inferential} (shown in the supplement).  We have found our method performs competitively in all settings, even under potential model misspecification. In all simulations, we tune the rank $r$ and sparsity $s$ using the eBIC criterion.

\subsection{Generalized Matrix Response Model}
\label{sec::model1}

\noindent
We first simulate the connectivity matrix with binary edges from our proposed model \eqref{eqn::model}, $g\{ \bmu^{(i)} \} = \bTheta + \bcB\times_3 \x_i$, where $g(\cdot)$ is the logit link function. We generate the covariates from $\mathcal{N}(0,1)$ and standardize the columns of the design matrix to have zero mean and unit standard deviation. For $\bTheta=\U\bLambda\U^\top$, we set $\bLambda$ as an $r\times r$ identity matrix, and generate the entries of $\U\in\mathbb{R}^{n\times r}$ from $\mathcal{N}(0,1)$. For $\bcB$, we randomly set a proportion of its entries to be 2, and the rest to zero; let $s_0=s/(n^2p)$ denote this proportion of the nonzero entries. We set the number of nodes $n = 50$, the number of covariates $p=10$, and vary the number of subjects $N=200,400$, the rank $r=2,5$, and the sparsity proportion $s_0=0.1,0.3$, respectively. 

We compare with three alternative methods. The first is the element-wise penalized GLM method of \citet{firth1993bias}, which fits a penalized GLM to each entry of $\A_{jj'}$, for all $j, j'$.
This approach has been shown to be effective in reducing the small sample bias \citep{firth1993bias}. The second method is similar to the first one, except that it uses an elastic-net penalty \citep{zou2005regularization}. 
The third is the common and individual structure explained method proposed by \citet{wang2019common} coupled with a GLM, and the tuning is done using the elbow method as described in \citet{wang2019common}.

\begin{table}[t!]
\caption{Simulation results under the low-rank and sparse model, with the varying sample size $N$, rank $r$ and sparsity proportion $s_0$.  The F1 score is calculated as $2TP/(2TP+FP+FN)$, where TP is the true positive count, FP is the false positive count, and FN is the false negative count.
The four methods under comparison are: the element-wise penalized GLM with the Jeffreys invariant prior penalty (denoted as GLM$_\text{JP}$), the element-wise penalized GLM with the elastic-net penalty (GLM$_\text{EN}$), the common and individual structure explained method (CISE), and the proposed generalized connectivity matrix response model (GLSNet). }
\label{tab1}
\setlength{\tabcolsep}{5pt}
\centering
\begin{tabular}{lllcllll|}
\hline
\multicolumn{1}{|l|}{$N$}                     & \multicolumn{1}{l|}{$r$}                  & \multicolumn{1}{l|}{$s_0$}                   & Method  & Error of $\bmu^{(i)}$ & Error of $\bTheta$ & Error of $\bcB$  & F1 score \\ \hline
\multicolumn{1}{|l|}{\multirow{12}{*}{200}} & \multicolumn{1}{l|}{\multirow{6}{*}{2}} & \multicolumn{1}{l|}{\multirow{3}{*}{0.1}}  & GLM$_\text{JP}$ & 1.106 (0.009)  & 47.09 (1.531)         & 35.09 (0.389) & - \\
\multicolumn{1}{|l|}{}                      & \multicolumn{1}{l|}{}                   & \multicolumn{1}{l|}{}                    & GLM$_\text{EN}$  & 1.063 (0.011)           & 47.50 (1.865)               & 28.38 (0.201) & 0.709 (0.002) \\
\multicolumn{1}{|l|}{}                      & \multicolumn{1}{l|}{}                   & \multicolumn{1}{l|}{}                    & CISE &  0.638 (0.001)           & -                & - & - \\ 
\multicolumn{1}{|l|}{}                      & \multicolumn{1}{l|}{}                   & \multicolumn{1}{l|}{}                    & GLSNet &  0.152 (0.002)    & 3.49 (0.129) & 25.79 (0.426) & 0.964 (0.002) \\ \cline{2-8} 

\multicolumn{1}{|l|}{}                      & \multicolumn{1}{l|}{}                   & \multicolumn{1}{l|}{\multirow{3}{*}{0.3}} & GLM$_\text{JP}$ & 1.101 (0.008)           & 45.06 (1.454)        & 52.53 (0.696) & - \\
\multicolumn{1}{|l|}{}                      & \multicolumn{1}{l|}{}                   & \multicolumn{1}{l|}{}                    & GLM$_\text{EN}$ & 1.062 (0.008)            & 45.61 (1.561)            & 46.73 (0.345) & 0.905 (0.001) \\
\multicolumn{1}{|l|}{}                      & \multicolumn{1}{l|}{}                   & \multicolumn{1}{l|}{}                    & CISE & 0.818 (0.001)           & -                & - & - \\ 
\multicolumn{1}{|l|}{}                      & \multicolumn{1}{l|}{}                   & \multicolumn{1}{l|}{}                    & GLSNet &  0.207 (0.002)           & 4.15 (0.175)        & 35.35 (0.415) & 0.994 (0.001) \\ \cline{2-8} 

\multicolumn{1}{|l|}{}                      & \multicolumn{1}{l|}{\multirow{6}{*}{5}} & \multicolumn{1}{l|}{\multirow{3}{*}{0.1}}  & GLM$_\text{JP}$ & 1.353 (0.005)           & 93.38 (1.604)                & 35.55 (0..486) & -\\
\multicolumn{1}{|l|}{}                      & \multicolumn{1}{l|}{}                   & \multicolumn{1}{l|}{}                    & GLM$_\text{EN}$ & 1.328 (0.006)           & 94.64 (1.599)     & 29.40 (0.215) & 0.736 (0.002) \\
\multicolumn{1}{|l|}{}                      & \multicolumn{1}{l|}{}                   & \multicolumn{1}{l|}{}                    & CISE & 0.631 (0.001)           & -                & - & - \\ 
\multicolumn{1}{|l|}{}                      & \multicolumn{1}{l|}{}                   & \multicolumn{1}{l|}{}                    & GLSNet & 0.154 (0.001)           & 6.51 (0.232)       & 26.86 (0.346) & 0.960 (0.002) \\ \cline{3-8} 

\multicolumn{1}{|l|}{}                      & \multicolumn{1}{l|}{}                   & \multicolumn{1}{l|}{\multirow{3}{*}{0.3}} & GLM$_\text{JP}$ & 1.311 (0.005)           & 88.39 (1.648)       & 52.30 (0.613) & -\\
\multicolumn{1}{|l|}{}                      & \multicolumn{1}{l|}{}                   & \multicolumn{1}{l|}{}                    & GLM$_\text{EN}$ & 1.287 (0.005)           & 87.92 (1.625)          & 47.63 (0.295) & 0.916 (0.001) \\
\multicolumn{1}{|l|}{}                      & \multicolumn{1}{l|}{}                   & \multicolumn{1}{l|}{}                    & CISE & 0.838 (0.001)           & -        & - & - \\ 
\multicolumn{1}{|l|}{}                      & \multicolumn{1}{l|}{}                   & \multicolumn{1}{l|}{}                    & GLSNet & 0.211 (0.001)   & 9.79 (0.488)     & 37.27 (0.337) & 0.981 (0.001)\\ \hline

\multicolumn{1}{|l|}{\multirow{12}{*}{400}} & \multicolumn{1}{l|}{\multirow{6}{*}{2}} & \multicolumn{1}{l|}{\multirow{3}{*}{0.1}} & GLM$_\text{JP}$ & 0.774 (0.007)  & 39.04 (1.892)  & 25.05 (0.132) & - \\
\multicolumn{1}{|l|}{}                      & \multicolumn{1}{l|}{}                   & \multicolumn{1}{l|}{}                    & GLM$_\text{EN}$ & 0.756 (0.007)           & 40.58 (1.815)          & 18.35 (0.101) & 0.700 (0.002) \\
\multicolumn{1}{|l|}{}                      & \multicolumn{1}{l|}{}                   & \multicolumn{1}{l|}{}                    & CISE & 0.457 (0.001)           & -                & - & - \\ 
\multicolumn{1}{|l|}{}                      & \multicolumn{1}{l|}{}                   & \multicolumn{1}{l|}{}                    & GLSNet & 0.055 (0.000)    & 2.44 (0.110)  & 13.61 (0.185) & 0.997 (0.000) \\ \cline{3-8} 

\multicolumn{1}{|l|}{}                      & \multicolumn{1}{l|}{}                   & \multicolumn{1}{l|}{\multirow{3}{*}{0.3}} & GLM$_\text{JP}$ & 0.769 (0.006)           & 36.47 (1.388)   & 33.27 (0.130) & - \\
\multicolumn{1}{|l|}{}                      & \multicolumn{1}{l|}{}                   & \multicolumn{1}{l|}{}                    & GLM$_\text{EN}$ & 0.752 (0.006)           & 38.48 (1.135)         &  30.47 (0.117) & 0.901 (0.001) \\
\multicolumn{1}{|l|}{}                      & \multicolumn{1}{l|}{}                   & \multicolumn{1}{l|}{}                    & CISE & 0.577 (0.001)           & -                & - &- \\ 
\multicolumn{1}{|l|}{}                      & \multicolumn{1}{l|}{}                   & \multicolumn{1}{l|}{}                    & GLSNet &  0.088 (0.000)  & 3.04 (0.131)    & 22.51 (0.151) & 0.998 (0.000) \\ \cline{2-8} 

\multicolumn{1}{|l|}{}                      & \multicolumn{1}{l|}{{\multirow{6}{*}{5}}}                   & \multicolumn{1}{l|}{\multirow{3}{*}{0.1}} & GLM$_\text{JP}$ & 0.974 (0.004)           & 81.80 (1.923)   & 26.82 (0.110) & - \\
\multicolumn{1}{|l|}{}                      & \multicolumn{1}{l|}{}                   & \multicolumn{1}{l|}{}                    & GLM$_\text{EN}$ & 0.964 (0.004)           & 82.52 (1.893)         &  18.86 (0.087) & 0.716 (0.002) \\
\multicolumn{1}{|l|}{}                      & \multicolumn{1}{l|}{}                   & \multicolumn{1}{l|}{}                    & CISE & 0.452 (0.001)           & -                & - &- \\ 
\multicolumn{1}{|l|}{}                      & \multicolumn{1}{l|}{}                   & \multicolumn{1}{l|}{}                    & GLSNet & 0.061 (0.000)  & 4.57 (0.161)    & 14.97 (0.234) & 0.993 (0.001) \\ \cline{3-8}

\multicolumn{1}{|l|}{}                      & \multicolumn{1}{l|}{}                   & \multicolumn{1}{l|}{\multirow{3}{*}{0.3}} & GLM$_\text{JP}$ & 0.939 (0.004)           & 77.18 (1.466)   & 34.65 (0.156) & - \\
\multicolumn{1}{|l|}{}                      & \multicolumn{1}{l|}{}                   & \multicolumn{1}{l|}{}                    & GLM$_\text{EN}$ & 0.927 (0.004)           & 77.94 (1.435)         &  31.13 (0.136) & 0.908 (0.001) \\
\multicolumn{1}{|l|}{}                      & \multicolumn{1}{l|}{}                   & \multicolumn{1}{l|}{}                    & CISE & 0.513 (0.001)           & -                & - &- \\ 
\multicolumn{1}{|l|}{}                      & \multicolumn{1}{l|}{}                   & \multicolumn{1}{l|}{}                    & GLSNet &  0.094 (0.001)  & 7.87 (0.274)    & 25.47 (0.273) & 0.995 (0.000) \\ \hline
\end{tabular}
\end{table}

To evaluate the estimation accuracy, we report the estimation errors, $N^{-1} \sum_{i=1}^N\| \bmu^{(i)} - \hat\bmu^{(i)} \|_F$, $\| \bTheta-\hat\bTheta \|_F$, and $\| \bcB-\hat\bcB \|_F$, where $\hat\bmu^{(i)}=g^{-1}( \hat\bTheta+\hat\bcB\times_3\x_i )$. To evaluate the edge selection accuracy, we report the F1 score. 
Since the method of \citet{firth1993bias} does not consider entry-wise sparsity, its F1 score is not reported. 
Since the method of \citet{wang2019common} could only estimate $\bmu^{(i)}$, the estimation errors for $\bTheta$ and $\bcB$ are not reported. 
Table~\ref{tab1} reports the average criteria, with the standard errors in the parentheses, over 50 data replications. 
Our proposed method is seen to achieve the best performance among all competing methods, in terms of both estimation accuracy and selection accuracy, and this holds true for different sample sizes $N$, ranks $r$ and sparsity levels $s_0$. Moreover, we see the estimation error of our method decreases as $N$ increases, or as $r$ and $s_0$ decrease. Such observations agree with our theoretical results in Theorem~\ref{thm3}. 
We further report the heat map of the eBIC over varying $r$ and $s_0$ values in Section \ref{sec::bic} of the supplement.

\subsection{Common and Individual Structure Explained Model}
\label{sec::model2}

\noindent
Next we consider the performance of our method under a potentially misspecified model, and compare with the individual structure explained method of \citet{wang2019common}. The CISE model assumes the entries in $\A^{(i)}$ are independent Bernoulli random variables with 
\vspace{-0.05in}
\begin{equation}\label{eqn:cise}
\text{logit}\{ \bmu^{(i)} \} = \bTheta + \D_i, \quad i=1,\ldots,N,
\end{equation}
where $\bmu^{(i)}$ is as defined in \eqref{eqn::model}, $\bm\Theta$ characterizes the common connectivity pattern, and $\D_i$ represents the subject-specific deviation; the subject-specific deviation $\D_i$ is assumed to be low-rank while no structure assumption is placed on $\bm\Theta$. 
We simulate binary networks from the CISE model in \eqref{eqn:cise} with details given in Table \ref{tab2}. The CISE model cannot incorporate subject covariates, and hence $\bcB$ is not included. Moreover, $\bTheta + \D_i$ is not necessarily low-rank. As such, our model assumption may not be satisfied. We set $n = 50$, $N=200,400$, and $r=5,20$. Table \ref{tab2} reports the estimation errors based on 50 data replications for the CISE method and our proposed method. It is seen that, under this potentially misspecified model, our method still achieves a comparable performance as \citet{wang2019common}. 

\begin{table}[t!]
\caption{Simulation results under the common and individual structure explained model, with the varying sample size $N$ and rank $r$. For $\bTheta=\U\bLambda\U^\top$, we set $\bLambda=\I_{r\times r}$, and generate the entries of $\U\in\mathbb{R}^{n\times r}$ from $\mathcal{N}(0,1)$. We set $\D_i=\d_i\otimes\d_i$, where $\otimes$ is outer product, and generate the entries of $\d_i\in\mathbb{R}^n$ from $\mathcal{N}(0,1)$. The two methods under comparison are: the common and individual structure explained method (CISE) and the proposed generalized connectivity matrix response model (GLSNet). }
\label{tab2}
\centering
\begin{tabular}{|c|c|l|l|l|l|}
\hline
\multirow{2}{*}{$r$}    & \multicolumn{1}{c|}{\multirow{2}{*}{Method}} & \multicolumn{2}{c|}{$N=200$}                  & \multicolumn{2}{c|}{$N=400$}                  \\ \cline{3-6} 
                        & \multicolumn{1}{c|}{} & Error of $\bmu^{(i)}$ & Error of $\bm\Theta$ & Error of $\bmu^{(i)}$ & Error of $\bm\Theta$ \\ \hline
\multirow{2}{*}{$5$}  & CISE  & 0.435 (0.000) & 46.08 (0.553) &   0.301 (0.000)  & 44.74 (0.651)  \\
                        & GLSNet & 0.506 (0.002) & 16.50 (0.256) & 0.359 (0.002)  & 16.27 (0.289) \\ \hline
\multirow{2}{*}{$20$} & CISE & 0.306 (0.000) & 157.2 (1.130) & 0.440 (0.000) & 155.8 (1.195)\\
                        & GLSNet    & 0.294 (0.001) & 104.8 (1.367) & 0.423 (0.002) & 100.6 (1.463) \\ \hline
\end{tabular}
\end{table}

\section{Applications to Brain Connectivity Analysis}
\label{sec::brain}

\noindent
We applied the proposed method to two brain connectivity studies. The first is a study of brain functional connectivity based on resting-state fMRI, where the edge is \emph{binary} resulting from a thresholded partial correlation matrix. The second is a study of brain structural connectivity based on DTI, where the edge is the \emph{count} of white matter fibers between pairs of brain regions.

\subsection{Functional Connectivity Analysis}
\label{sec::functional}

\noindent
We first analyzed an fMRI dataset from ADHD-200 (\url{http://fcon_1000.projects.nitrc.org/indi/adhd200/}). We focused on $N = 319$ healthy control subjects, aging between 7.09 to 21.8 years old, with $46.4\%$ females and $53.6\%$ males. Each subject received a resting-state fMRI scan, and the image was preprocessed, including slice timing correction, motion correction, spatial smoothing, denoising by regressing out motion parameters, white matter, and cerebrospinal fluid time, and band-pass filtering. Each fMRI image was then summarized in the form of a binary network, with the nodes corresponding to 264 seed regions of interest in the brain \citet{Power2011}, and the edges recording the binary indicator of the thresholded partial correlations. We applied our proposed model to this data with a logit link function. We standardized the covariates, age and sex, to have mean zero and variance one. The rank was selected as $r = 9$ and the sparsity proportion as $s_0 = 0.02$ based on eBIC. 

\begin{figure}[t!]
\centering
\includegraphics[trim=1cm 5mm 0 0, scale=0.4]{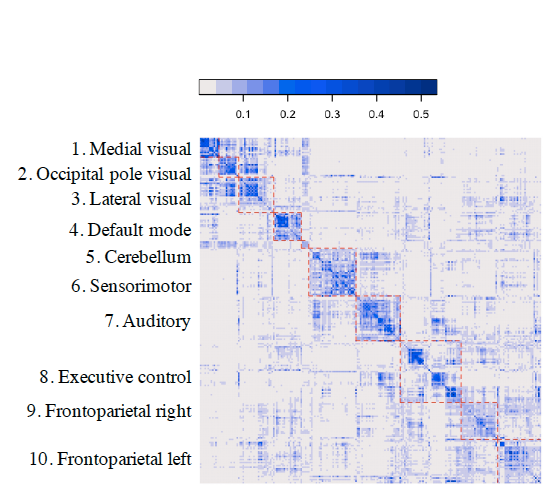}
\caption{The functional connectivity study. Heatmap of the $264\times264$ matrix $g^{-1}(\bm{\widehat\Theta})$ with rows and columns ordered according to the pre-specified functional module membership. The red dashed lines mark the boundaries of the ten functional modules. }
\label{heatmap1}
\end{figure}

We first examine the estimate $\hat\bTheta$. In the neuroscience literature, those 264 nodes have been partitioned into 10 functional modules \citep{Smith2009}. Each module possesses a relatively autonomous functionality, and complex brain tasks are carried out through coordinated collaborations among those modules. Figure~\ref{heatmap1} shows the heatmap of $g^{-1}(\hat\bTheta)$, with the nodes ordered according to the functional modules. Here the function $g^{-1}(\cdot)$ maps a value from the real line to $[0,1]$ so to facilitate data visualization. From this figure, we see that our estimate agrees reasonably well with the pre-specified functional modules by \cite{Smith2009}. We observe larger values of $\hat\bTheta$ located within the diagonal blocks, which indicates higher functional connectivities within those functional modules. Furthermore, there are high connectivities among modules 1-3, namely, the medial visual, occipital pole visual and lateral visual modules. 
These visual modules appear to have high connectivities with the cerebellum, but generally low connectivities with the rest of functional modules. We also observe a high connectivity between modules 9-10, namely, the frontoparietal right and frontoparietal left modules. These two modules are important in attention control and can generate a diverse range of control signals depending on task demands \citep{scolari2015functions}. 

We next examine the estimate $\hat\bcB$. In $\hat\bcB_{\cdot\cdot 1}$, i.e., the coefficient matrix for the sex covariate. The non-sparse entries are located within the lateral visual module, and those values are negative, ranging from $-0.777$ to $-0.506$. This indicates that male subjects have lower connectivities in those regions within the lateral visual module. This result agrees with the existing finding that developing females outperform developing males on tasks related to emotion identification and reasoning \citep{satterthwaite2014linked}. The non-sparse entries of $\hat\bcB$ mostly concentrate in $\hat\bcB_{\cdot\cdot 2}$, i.e., the coefficient matrix for the age covariate. In $\hat\bcB_{\cdot\cdot 2}$, the positive entries are located within the occipital pole visual, default mode, executive control and frontoparietal left modules, with values ranging from $0.454$ to $0.902$, indicating the connectivities within those modules increase with age. We also observe positive entries located in the default model to executive control and the default mode to frontoparietal right, which agrees with the literature that the default mode module has increasingly synchronized connections to other modules with increasing age \citep{grayson2017development}. 
We also find negative entries located in the medial visual to lateral visual, the executive control to frontoparietal right, and the default mode to auditory modules, with values ranging from $-1.350$ to $-1.095$, suggesting the connectivities between those modules also decrease with age. These findings suggest some interesting patterns that warrant further investigation and validation.

\subsection{Structural Connectivity Analysis}
\label{sec::structural}
\noindent
We next analyzed a structural DTI dataset from KKI-42 (\url{http://openconnecto.me/data/public/MR/archive/}). We focused on 21 subjects with no history of neurological conditions, aging from 22 to 61 years old, with $47.6\%$ females and $52.4\%$ males. Each subject received a resting-state DTI scan, which was a magnetic resonance imaging technique that enables measurement of the diffusion of water. Estimates of white matter connectivity patterns can be obtained using the diffusion anisotropy and the principal diffusion directions. In the KKI-42 study, a scan-rescan imaging session was conducted on each subject, leading to two images for each subject, and a total of $N=42$ for the study. For simplicity, we treated those images as if they formed independent samples. Each DTI image was preprocessed, and summarized in the form of a count network, with $n=68$ nodes defined following the Desikan Atlas, and the edges recording the total number of white matter fibers between the pair of nodes. See \citet{landman2011multi} for more information about data collection and brain networks construction using DTI scans. We applied our proposed method to this data, with a log link function. The covariates, age and sex, were standardized to have mean zero and variance one. The rank was selected as $r=5$ and the sparsity proportion $s_0=0.31$ using eBIC.

\begin{figure}[t!]
\begin{minipage}[t]{.45\linewidth}
\vspace{0pt}
\centering
\includegraphics[trim=0 0 0 1cm, scale=0.5]{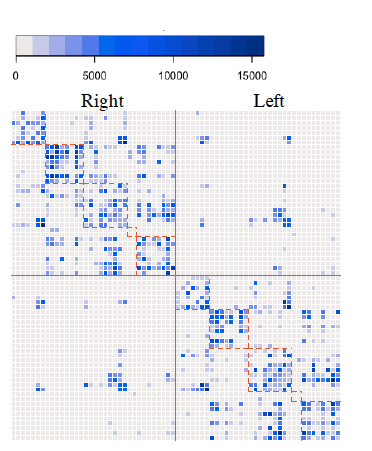}
\end{minipage}
\begin{minipage}[t]{.375\linewidth}
\vspace{0pt}
\centering
{\small\renewcommand{\arraystretch}{0.55}
\setlength{\tabcolsep}{3pt}
\begin{tabular}{|l|l|}
\hline
1 & \begin{tabular}[c]{@{}l@{}}frontalpole, parsopercularis, lateralorbitalfrontal, \\parstriangularis, medialorbitalfrontal, \\rostralanteriorcingulate, rostralmiddlefrontal\end{tabular}                   \\ \hline
2 & \begin{tabular}[c]{@{}l@{}}bankssts, fusiform, inferiortemporal, lingual, \\middletemporal, parsorbitalis, superiortemporal,\\ transversetemporal\end{tabular}                                             \\ \hline
3 & \begin{tabular}[c]{@{}l@{}}caudalanteriorcingulate, caudalmiddlefrontal, \\parahippocampal, posteriorcingulate, precentral, \\isthmuscingulate,  precuneus, corpuscallosum, \\superiorfrontal\end{tabular} \\ \hline
4 & entorhinal, temporalpole                                                                                                                                                                                  \\ \hline
5 & \begin{tabular}[c]{@{}l@{}}inferiorparietal, lateraloccipital, paracentral, \\cuneus, pericalcarine, postcentral, \\superiorparietal, supramarginal\end{tabular}                                           \\ \hline
\end{tabular}}
\end{minipage}
\caption{The structural connectivity study. Left panel: heatmap of the $68\times68$ matrix $g^{-1}(\bm{\widehat\Theta})$ with rows and columns ordered according to the $K$-means clustering result. Right and left hemispheres are marked in the plot. The red dashed lines mark the boundaries of the identified groups. Right panel: the anatomic regions of interest in the identified groups.}
\label{heatmap2}
\end{figure}

We first examine the estimate $\hat\bTheta$. To the best of our knowledge, communities in structural connectivity networks have not been studied before. We applied the $K$-means clustering algorithm to the estimate $\U^{(t)}$ from $\hat\bTheta$, and identified five clusters among the 68 anatomic regions of interest (ROIs). We selected the number of clusters based on the elbow plot. Figure~\ref{heatmap2}, right panel, reports the members of each cluster in the table. From an anatomical perspective, the first group of nodes are entirely contained in the frontal lobe, the second group are mostly contained in the temporal lobe, the fourth group are entirely contained in the temporal lobe, and the third and fifth groups contain nodes from the frontal, parietal, occipital and temporal lobes. Many of the 68 anatomic ROIs in the Desikan Atlas overlap with the resting-state functional modules. By exploring this overlap, we gained further insights of potential functions of those five groups. We found that group 1 is related to the dorsal attention and default mode modules, group 2 is related to the visual and auditory, group 3 is related to the default mode, and group 5 is related to the visual module. The resting-state functions of the nodes in Groups 4 are unidentified. Figure~\ref{heatmap2}, left panel, shows the heatmap of the estimated $\hat\bTheta$, with the nodes reordered according to the cluster membership. 

\begin{figure}[t!]
\centering
\includegraphics[trim=0 3cm 2cm 0, scale=0.4]{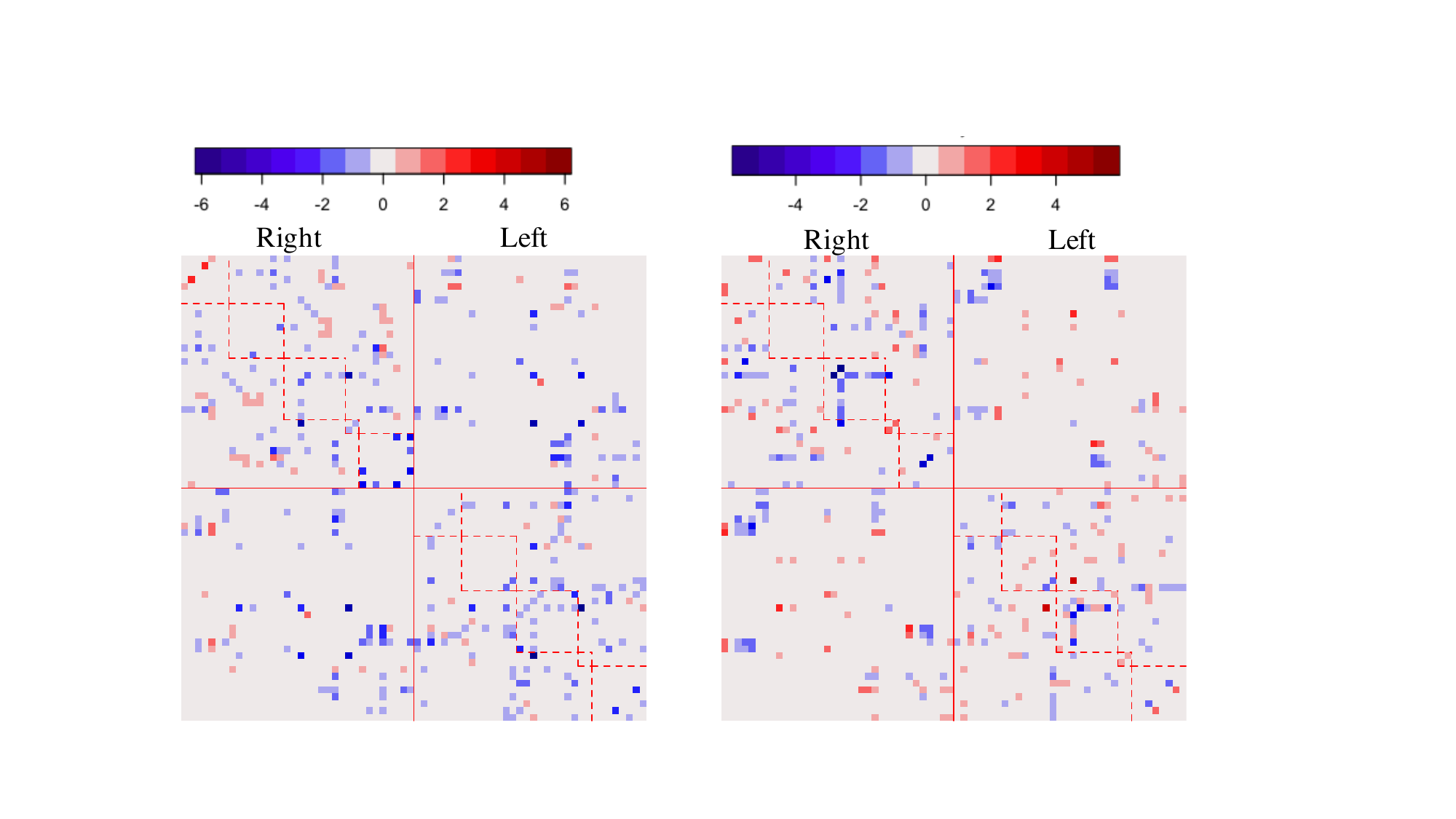}
\caption{The structural connectivity study.  Left panel: the age coefficient matrix. Right panel: the sex coefficient matrix. }
\label{heatmap3}
\end{figure}

We next examine the estimate $\hat\bcB$. Figure~\ref{heatmap3} shows the estimated subject covariates effect coefficients. From the left panel of Figure~\ref{heatmap3}, we see that, as age increases, the structural connectivity generally decreases both within and between the two hemispheres. This result agrees with existing neurological finding \citep{betzel2014changes}. From the right panel of Figure~\ref{heatmap3}, we see that male and female subjects have different structural connectivity patterns. Such differences are observed in the between-group connections within and between hemispheres, and in the within-group connections within each hemisphere. For instance, we see males have  lower between-hemisphere connectivities for the ROIs in Group 1. This observation agrees with the literature that males have lower connectivities between the left and right frontal regions \citep{ingalhalikar2014sex}.

\section{Discussion}
\label{sec::discuss}
{In this paper, we propose a generalized connectivity matrix response model that relates subject-specific connectivity matrices to external covariates. We investigate the statistical properties of our proposal including the statistical and computational error trade-off, community detection and edge selection consistency.
We briefly comment on potential future research. In our current community detection setup, we assume that the communities are fully determined by the population level connectivity matrix $\bTheta$, which is closely related to the stochastic blockmodel. It is possible to consider other specification of the community structure. For instance, one possibility is that the slope tensor $\bcB$ may have a community structure, in that the covariate effect on the connectivity between nodes $j$ and $j'$ is determined by the community labels of those two nodes. In this case, the coefficient matrix for the $l$th covariate, i.e., $\bcB_{..l}$, becomes a block matrix. This new structure requires a new estimation algorithm and theory, and we leave the full investigation as future research.}

\bibliographystyle{asa}
\begingroup
\bibliography{ref}
\endgroup

\newpage
\renewcommand{\thesection}{S\arabic{section}}
\renewcommand{\theequation}{S\arabic{equation}}
\setcounter{section}{0}  
\setcounter{equation}{0}
\setcounter{page}{1}
\def\eop
{\hfill $\Box$
}

\begin{center}
{\Large\bf Supplementary Materials} \\
\medskip
{\large Jingfei Zhang, Will Wei Sun, and Lexin Li}
\end{center}

\section{Supporting Lemmas} 
\renewcommand{\thelemma}{S\arabic{lemma}}

\noindent
We first state a number of supporting lemmas that are useful for our proofs. The proof of Lemma \ref{au1} is given in Section \ref{sec:pau1}.

\begin{lemma}[Matrix Bernstein inequality in \cite{tropp2012user}]
\label{bern}
Consider an independent sequence $(\Y_k)_{k\ge1}$ of random matrices in $\mathbb{R}^{n\times n}$ that satisfy $E(\Y_k)=\textbf{0}$ and $\|\Y_k\|_2\le z_0$ almost surely for each $k$. Let $\sigma_y^2=\|\sum_kE(\Y_k^2)\|_2$. For any $t\ge0$, we have 
$$
P\left(\|\sum_k\Y_k\|_2\ge t\right)\le n\exp\left(-\frac{t^2/2}{\sigma_y^2+z_0t/3}\right).
$$
\end{lemma}

\begin{lemma}[Proposition 5.16 in \cite{vershynin2010introduction}]
\label{sub}
Let $X_1,\ldots,X_N$ be independent centered sub-exponential random variables, and $z_1=\max_i\|X_i\|_{\psi_1}$, where $\|X_i\|_{\psi_1}$ denotes the sub-exponential norm, $\|X_i\|_{\psi_1}=\sup_{p\ge1}p^{-1}(E|X_i|^p)^{1/p}$.  For any $t>0$, there is a constant $c$ such that 
$$
P\left(\left|\sum_{i=1}^NX_i\right|\ge t\right)\le 2\exp\left\{ -c\min\left(\frac{t^2}{z_1^2N},\frac{t}{z_1}\right) \right\}. 
$$
\end{lemma}

\begin{lemma}[Theorem 2.1.5 in \cite{nesterov2013introductory}]
\label{B1}
For a function $f$, and $x,y\in\mathbb{R}^n$ 
$$
0\le f(y)-f(x)-\langle f'(x),y-x\rangle\le\frac{L}{2}\|x-y\|_2^2,
$$
if and only if
$$
f(x)+\langle f'(x),y-x\rangle+\frac{1}{2L}\|f'(x)-f'(y)\|_2^2\le f(y).
$$
\end{lemma}

\begin{lemma}[Lemma 3.3 in \cite{li2016nonconvex}]
\label{B2}
Let $\bm\theta^*\in\mathbb{R}^d$ be an unknown sparse vector with $\|\bm\theta^*\|_0\le k^*$, and Truncate$(\cdot)$ be a hard-thresholding operator that only keeps the largest $k$ entries in absolute values and sets the rest to zero. For any $k\ge k^*$ and any $\bm\theta\in\mathbb{R}^d$, we have
$$
\|\text{Truncate}(\bm\theta)-\bm\theta^*\|_2^2\le \frac{2\sqrt{k-k^*}+\sqrt{k^*}}{2\sqrt{k-k^*}-\sqrt{k^*}}\|\bm\theta-\bm\theta^*\|_2^2
$$
\end{lemma}

\begin{lemma} [Lemma 5.4 in \cite{tu2016low}]
\label{B3}
For any $\U, \V\in\mathbb{R}^{n\times r}$, we have
$$
d^2(\U,\V)\le \frac{1}{2(\sqrt{2}-1)\sigma_r^2(\V)}\|\U\U^\top-\V\V^\top\|_F^2
$$
\end{lemma}

\begin{lemma}
\label{au1}
Let $\bTheta,\bTheta^*\in\mathbb{R}^{n\times n}$ be two rank-$r$ matrices. Write $\bTheta = \U \bLambda \U^\top$ and $\bTheta^* = \U^* \bLambda {\U^*}^\top$, where $\U,\U^* \in \mathbb{R}^{n\times r}$ and $\bLambda$ is a $r\times r$ diagonal matrix with diagonal entries in $\{-1,1\}$. Define $d(\U,\U^*)=\min_{\bm\Gamma\in\mathbb{Q}_r}\|\U-\U^*\bm\Gamma\|_F$. We have
$$
\|\bTheta-\bTheta^*\|_F^2\ge 2(\sqrt{2}-1)\sigma^*_{r}d^2(\U,\U^*),
$$
where $\sigma^*_{r}$ is the $r$th singular value of $\bTheta^*$.
\end{lemma}

\section{Key Technical Results Under a General Loss Function} 
\renewcommand{\thelemma}{\arabic{lemma}}
\setcounter{lemma}{0}
In this section, we first provide the non-asymptotic error bound of the proposed estimator for a general loss function. 
Let $\ell_g(\bTheta,\bcB)$ denote a general loss function defined with respect to a low-rank matrix $\bTheta_{n\times n}$ and a sparse tensor $\bcB_{n\times n\times p}$, and is evaluated based on $N$ sample observations. Let $\bTheta^*$ denote the true coefficient matrix with rank $r^*$, and $\bcB^*$ the true coefficient tensor with $s^*$ nonzero entries. We introduce a set of regularity conditions. 

\begin{enumerate}
\item[(A1)] Assume that the loss function $\ell_g$ satisfies, with respect to $\bTheta$, the restricted strong convexity with parameter $\mu_1>0$, and the restricted strong smoothness with parameter $\alpha_1>0$,  in that, for any $\bcB\in\mathbb{B}_{\bcB^*}(\kappa_2)$ with at most $s$ nonzero entries and for any matrices $\bTheta_1,\bTheta_2\in\mathbb{B}_{\bTheta^*}(\kappa_1)$ with rank at most $r$,
\begin{eqnarray*}
\frac{\mu_1}{2}\|\bTheta_2-\bTheta_1\|_F^2\le \ell_g(\bTheta_2,\bcB)-\ell_g(\bTheta_1,\bcB)-\langle\nabla_{\bTheta}\ell_g(\bTheta_1,\bcB),\bTheta_2-\bTheta_1\rangle\le\frac{\alpha_1}{2}\|\bTheta_2-\bTheta_1\|_F^2
\end{eqnarray*} 
holds with probability at least $1-p_0$ for any $p_0\in(0,1)$.

\item[(A2)] Assume that the loss function $\ell_g$ satisfies, with respect to $\bcB$, the restricted strong convexity with parameter $\mu_2>0$, and the restricted strong smoothness with parameter $\alpha_2>0$, in that, for any $\bTheta\in\mathbb{B}_{\bTheta^*}(\kappa_1)$ with rank at most $r$ and for any $\bcB_1,\bcB_2\in\mathbb{B}_{\bcB^*}(\kappa_2)$ with at most $s$ nonzero entries,
\begin{eqnarray*}
\frac{\mu_2}{2}\|\bcB_2-\bcB_1\|_F^2\le \ell_g(\bTheta,\bcB_2)- \ell_g(\bTheta,\bcB_1)-\langle\nabla_{\bcB}\ell_g(\bTheta,\bcB_1),\bcB_2-\bcB_1\rangle\le\frac{\alpha_2}{2}\|\bcB_2-\bcB_1\|_F^2
\end{eqnarray*}
holds with probability at least $1-p_0$ for any $p_0\in(0,1)$.

\item[(A3)] For $\bTheta$ with rank at most $r$ and $\bcB$ with at most $s$ nonzero entries, assume that 
\begin{eqnarray*}
|\langle\nabla_{\bTheta}\ell_g(\bTheta^*,\bcB)-\nabla_{\bTheta}\ell_g(\bTheta^*,\bcB^*),\bTheta\rangle| & \le & \kappa\|\bTheta\|_F\cdot\|\bcB-\bcB^*\|_F, \\
|\langle\nabla_{\bcB}\ell_g(\bTheta,\bcB^*)-\nabla_{\bcB}\ell_g(\bTheta^*,\bcB^*),\bcB\rangle| & \le & \kappa\|\bTheta-\bTheta^*\|_F\cdot\|\bcB\|_F
\end{eqnarray*}
hold with probability at least $1-p_0$ for any $p_0\in(0,1)$. Here $\kappa$ is the Lipschitz gradient parameter that depends on $r$, $s$, $n$ and $N$.

\item[(A4)] For a tolerance parameter $\delta\in(0,1)$, assume there exist $\epsilon_N$ and $\xi_N$ such that 
\begin{eqnarray*}
\|\nabla_{\bTheta}\ell_g(\bTheta^*,\bcB^*)\|_2\le\epsilon_N\quad\text{and}\quad\|\nabla_{\bcB}\ell_g(\bTheta^*,\bcB^*)\|_{\infty}\le\xi_N,
\end{eqnarray*}
hold with probability at least $1-p_0$ for any $p_0\in(0,1)$. Here $\epsilon_N$ and $\xi_N$ depend on $N$ and $\delta$.
\end{enumerate}

Denote the nonzero singular values of $\bTheta^*$ as $\sigma^*_1\ge\ldots\ge\sigma^*_{r^*}>0$. In the asymmetric case, we may write $\bTheta^*=\U^*{\V^*}^\top$ such that the singular values of $\U^*\in\mathbb{R}^{n\times r^*}$ are the same as the singular values of $\V^*\in\mathbb{R}^{n\times r^*}$.
Write $\M^*=[\U^*;\V^*]$. For any $\M=[\U;\V]$ and $\bcB$, define the distance metric, 
\begin{eqnarray*}
D\left\{\M,\bcB\right\}=d^2(\M,\M^*) + \frac{1}{\sigma^*_1} \|\bcB-\bcB^*\|_F^2, \;\text{ where }d(\M, \M^*)=\min_{\bm\Gamma\in\mathbb{Q}_{r^*}}\|\M-\M^*\bm\Gamma\|_F, 
\end{eqnarray*}
and $\mathbb{Q}_{r^*}$ denotes the set of $r^*\times r^*$ orthonormal matrices. The next proposition states the non-asymptotic rate of the estimator from the algorithm; its proof is given in Section \ref{sec:pthm1}.

\begin{proposition}
Suppose the general loss function $\ell_g$ satisfies Conditions (A1)-(A4), with $\kappa_1=\kappa_2=\sqrt{\sigma^*_{r^*}}/3$. Let $c_1$ and $c_2$ be constants such that $c_1\le\min\{1/32,\mu_1/(192\alpha_1^2)\}$, and $6c_1\alpha_2\le c_2\le\min\left\{1/3,\sqrt{\mu'_1/(4+6\alpha_1+2\kappa^2/\mu_2)}\right\}$, where $\mu_1'=\min\{\mu_1,2\}$. Let the step sizes $\delta=c_1/\sigma^*_1$, $\tau=c_2/\alpha_2$, and $s=\gamma s^*$, where $\gamma \ge1+\{(3\alpha_2+\mu_2c_2)/(\mu_2c_2)\}^2$. Let the sample size $N$ be large enough such that $r^*\phi_1\epsilon^2_N+s^*\phi_2\xi^2_N\le(1-\rho)c^2_2\sigma^*_1\sigma^*_{r^*}$ and $\kappa\le\sqrt{(\mu_1\mu_2/2)}/12$, where $\rho=\max\{1-\delta\mu_1\sigma^*_{r^*}/16,1-\tau\mu_2/18\}\in(0,1)$ is a contraction parameter, and $\phi_1$ and $\phi_2$ are constants that depend on $c_1$, $c_2$, $\gamma$, $\mu_1$, $\mu_2$, $\alpha_1$ and $\alpha_2$. Then for a tolerance parameter $\delta\in(0,1)$, for any initial estimator $\{ \M^{(0)},\bcB^{(0)} \}$ satisfying $D\{ \M^{(0)},\bcB^{(0)} \} \le c_2^2\sigma^*_{r^*}$, we have, with probability at least $1-\delta$, 
\begin{equation} \label{linear1}
D\left\{ \M^{(t)},\bcB^{(t)} \right\} \le \rho^t D\left\{ \M^{(0)},\bcB^{(0)} \right\} + \frac{r^*\phi_1\epsilon^2_N+s^*\phi_2\xi^2_N}{(1-\rho)\sigma^*_1}. 
\end{equation} 
\label{thm1}
\end{proposition}

Proposition \ref{thm1} is a key result and is shown using the two following important lemmas on the one-step convergence of the low-rank component and the sparse component, respectively. Their proofs are given in Sections \ref{sec:plm1} and \ref{sec:plm2}.

\begin{lemma}[Convergence for the asymmetric low-rank component $\bTheta$] 
\label{lm1}
Suppose the general loss function $\ell_g$ satisfies Conditions (A1) and (A3). Let $c_1$ and $c_2$ be constants such that $c_1\le\min\{1/32,\mu_1/(192\alpha_1^2)\}$ and $c_2\le\sqrt{\mu'_1/(6\alpha_1+4+2\kappa^2/\mu_2)}$, where $\mu'_1=\min\{\mu_1,2\}$. Set the step size $\delta=c_1/\sigma^*_1$. If $d( \M^{(t)},\M^* ) < c_2\sqrt{\sigma^*_{r^*}}$, then the output $\M^{(t+1)}=[\U^{(t+1)};\V^{(t+1)}]$ from Algorithm \ref{algo_asym} satisfies that 
$$
d^2( \M^{(t+1)},\M^*\} \le \rho_1d^2\{ \M^{(t)},\M^* ) - \frac{\delta\mu_1}{4}\|\bTheta^{(t)}-\bTheta^*\|_F^2+C_1\|\bcB^{(t)}-\bcB^*\|_F^2+C_2\|\nabla_{\bTheta}\ell_g(\bTheta^*,\bcB^*)\|_2^2,
$$
where the contraction parameter $\rho_1=1-\delta\mu_1\sigma^*_{r^*}/16$, $C_1 = 48\kappa^2\delta^2\sigma^*_1 + 2\delta(\mu_2/8+4\kappa^2/\mu_1)$, and $C_2=48r^*\delta^2\sigma^*_1+2\delta(8r^*/\mu_1+r^*/\alpha_1)$.
\end{lemma}

\begin{lemma}[Convergence for the sparse component $\bcB$] 
\label{lm2}
Suppose the general loss function $\ell_g$ satisfies Conditions (A2) and (A3). For $\gamma \ge 1+\{(9\alpha_2-4\mu_2c_2)/(8\mu_2c_2)\}^2$ and the step size $\tau \le 1/(3\alpha_2)$, the output $\bcB^{(t)}$ from Algorithm \ref{algo_asym} satisfies that
$$
\|\bcB^{(t+1)}-\bcB^*\|_F^2\le\rho_2\|\bcB^{(t)}-\bcB^*\|_F^2+C_3\|\bTheta^{(t)}-\bTheta^*\|_F^2+C_4\|\nabla_{\bcB}\ell_g(\bTheta^*,\bcB^*)\|_{\infty}^2,
$$
where the contraction parameter $\rho_2=\frac{2\sqrt{\gamma-1}+1}{2\sqrt{\gamma-1}-1}(1-8\mu_2\tau/9)\le1$ and
$$
C_3=\frac{2\sqrt{\gamma-1}+1}{2\sqrt{\gamma-1}-1} \left(\frac{9\tau \kappa^2}{\mu_2}+3\tau^2\kappa^2\right),\quad C_4=\frac{2\sqrt{\gamma-1}+1}{2\sqrt{\gamma-1}-1} \left\{ \frac{9\tau(\gamma+1)s^*}{\mu_2}+3\tau^2s^* \right\}.
$$
\end{lemma}

\bigskip
Next, we consider the symmetric case, Write $\bTheta^*=\U^*\bLambda{\U^*}^\top$, where $\U^*\in\mathbb{R}^{n\times r^*}$ and $\bLambda$ is a $r^*\times r^*$ diagonal matrix with diagonal entries in $\{-1,1\}$, collecting signs of the singular values of $\bTheta^*$.
For any $\U$ and $\bcB$, define distance
\begin{eqnarray*}
D\left\{\U,\bcB\right\}=d^2(\U,\U^*)+\|\bcB-\bcB^*\|_F^2/\sigma^*_1, \;\; \text{ where } \; d(\U,\U^*)=\min_{\bm\Gamma\in\mathbb{Q}_r}\|\U-\U^*\bm\Gamma\|_F, 
\end{eqnarray*}
and $\mathbb{Q}_{r^*}$ is as defined before. The next proposition states the non-asymptotic rate of the estimator from the algorithm; its proof is given in Section \ref{sec:pcoro1}.

\begin{proposition}
\label{thm2}
Suppose the general loss function $\ell_g$ satisfies Conditions (A1)-(A4), with $\kappa_1=\kappa_2=\sqrt{\sigma^*_{r^*}}/3$. Let $c_1$ and $c_2$ be constants such that $c_1\le\mu_1/(96\alpha_1^2)$, and $3c_1\alpha_2\le c_2\le\min\left\{1/3,\sqrt{\mu_1/(9\alpha_1/2+8\kappa^2/\mu_2)}\right\}$. Let the step sizes $\delta=c_1/\sigma^*_1$, $\tau=c_2/\alpha_2$, and $s=\gamma s^*$, where $\gamma \ge 1+\{(3\alpha_2+\mu_2c_2)/(\mu_2c_2)\}^2$. Let the sample size $N$ be large enough such that $r^*\phi_1\epsilon^2_N+s^*\phi_2\xi^2_N\le(1-\rho)c_2\sigma^*_1\sigma^*_{r^*}$ and $\kappa\le\sqrt{\mu_1\mu_2}/9$, where $\rho=\max\{1-\delta\mu_1\sigma^*_{r^*}/16,1-\tau\mu_2/18\}\in(0,1)$ is a  contraction parameter, and $\phi_1$ and $\phi_2$ are constants that depend on $c_1$, $c_2$, $\gamma$, $\mu_1$, $\mu_2$, $\alpha_1$ and $\alpha_2$.  Then for a tolerance parameter $\delta\in(0,1)$, for any initial estimator $\left\{ \U^{(0)},\bcB^{(0)} \right\}$ satisfying $D\{ \U^{(0)},\bcB^{(0)} \} \le c_2^2\sigma^*_{r^*}$, we have, with probability at least $1-\delta$, 
\begin{equation*}
D\left\{ \U^{(t)},\bcB^{(t)} \right\} \le \rho^t D\left\{ \U^{(0)},\bcB^{(0)} \right\}+\frac{r^*\phi_1\epsilon^2_N+s^*\phi_2\xi^2_N}{(1-\rho)\sigma^*_1},
\end{equation*}
\end{proposition}

Proposition \ref{thm2} is shown using Lemma \ref{lm2} the following important lemma on the one-step convergence. Its proof is given in Sections \ref{sec:plm3}.

\begin{lemma}[Convergence for the symmetric low-rank component $\bTheta$]
\label{lm3}
Suppose the general loss function $\ell_g$ satisfies Conditions (A1) and (A3). Let $\bTheta^*=\U^*\bLambda{\U^*}^\top$ be the unknown rank-$r^*$ symmetric matrix. Let $c_1$ and $c_2$ be constants such that $c_1\le\mu_1/(96\alpha_1^2)$, and $c_2\le\sqrt{\mu_1/(9\alpha_1/2+8\kappa^2/\mu_2)}$. Set the step size $\delta=c_1/\sigma^*_1$. Assuming $\bLambda$ is known, and $d(\U^{(t)},\U^*) < c_2\sqrt{\sigma^*_{r^*}}$, then the output $\U^{(t+1)}$ from Algorithm \ref{algo_symm} satisfies that 
$$
d^2( \U^{(t+1)},\U^*) \le \rho_1d^2(\U^{(t)},\U^*)-\frac{\delta\mu_1}{16}\|\bTheta^{(t)}-\bTheta^*\|_F^2+C_1\|\bcB^{(t)}-\bcB^*\|_F^2+C_2\|\nabla_{\bTheta}\ell_g(\bTheta^*,\bcB^*)\|_2^2,
$$
where the contraction parameter $\rho_1=1-\delta\mu_1\sigma^*_{r^*}/16$, $C_1=3\kappa^2\delta^2\sigma^*_1+\delta(\mu_2/8+2\kappa^2/\mu_1)$, and $C_2=3r^*\delta^2\sigma^*_1+\delta(4r^*/\mu_1+2r^*/\alpha_1)$.
\end{lemma}

\section{Proofs}
\subsection{\bf Proof of Proposition~\ref{thm1}}
\label{sec:pthm1}
\noindent
At iteration $t$ of Algorithm \ref{algo_asym}, based on the results from Lemma~\ref{lm1} and Lemma~\ref{lm2}, we have 
\begin{eqnarray*}
d^2(\M^{(t+1)},\M^*)+\frac{1}{\sigma^*_1}\|\bcB^{(t+1)}-\bcB^*\|_F^2\le\rho_1d^2(\M^{(t)},\M^*)+\frac{1}{\sigma^*_1}(\rho_2+C_1\sigma^*_1)\|\bcB^{(t)}-\bcB^*\|_F^2\\
+\left(\frac{C_3}{\sigma^*_1}-\frac{\delta\mu_1}{4}\right)\|\bTheta-\bTheta^*\|_F^2+C_2\|\nabla_{\bTheta}\ell_g(\bTheta^*,\bcB^*)\|_2^2+\frac{C_4}{\sigma^*_1}\|\nabla_{\bcB}\ell_g(\bTheta^*,\bcB^*)\|_{\infty}^2.
\end{eqnarray*}
When $\gamma\ge1+\{(3\alpha_2+\mu_2c_2)/(\mu_2c_2)\}^2$, it is straightforward to verify that the condition $\gamma\ge1+\{(9\alpha_2-4\mu_2c_2)/(8\mu_2c_2)\}^2$ in Lemma~\ref{lm2} is satisfied. Additionally, we have
$$
\rho_2=\frac{2\sqrt{\gamma-1}+1}{2\sqrt{\gamma-1}-1}\left(1-\frac{8\mu_2\tau}{9}\right)\le\left(1+\frac{7\mu_2\tau}{9}\right)\left(1-\frac{8\mu_2\tau}{9}\right)\le1-\frac{\mu_2\tau}{9}.
$$
Under the condition on the Lipschitz gradient parameter $\kappa$ that $\kappa\le\sqrt{(\mu_1\mu_2/2)}/12$, we can verify the following inequalities, 
$$
C_1\sigma^*_1=c_1\left(\frac{\mu_2}{4}+\frac{8\kappa^2}{\mu_1}\right)+48c_1^2\kappa^2\le\frac{\mu_2c_1}{3},
$$
$$
C_3=\frac{2\sqrt{\gamma-1}+1}{2\sqrt{\gamma-1}-1}\left(\frac{9\tau \kappa^2}{\mu_2}+3\tau^2\kappa^2\right)\le\frac{c_1\mu_1}{4}.
$$
Therefore, if we set $c_1\le\tau/6$, we have 
\begin{eqnarray}
\label{error1}
d^2(\M^{(t+1)},\M^*)+\frac{1}{\sigma^*_1}\|\bcB^{(t+1)}-\bcB^*\|_F^2\le\rho_1d^2(\M^{(t)},\M^*)+\frac{1}{\sigma^*_1}\left(1-\frac{\mu_2\tau}{18}\right)\|\bcB^{(t)}-\bcB^*\|_F^2\\\nonumber
+C_2\|\nabla_{\bTheta}\ell_g(\bTheta^*,\bcB^*)\|_2^2+\frac{C_4}{\sigma^*_1}\|\nabla_{\bcB}\ell_g(\bTheta^*,\bcB^*)\|_{\infty}^2.
\end{eqnarray}
Define $D\{\M^{(t)},\bcB^{(t)}\}=d^2(\M^{(t)},\M^*)+\frac{1}{\sigma^*_1}\|\bcB^{(t)}-\bcB^*\|_F^2$ and $\rho=\max\{1-\delta\mu_1\sigma^*_{r^*}/16,1-\tau\mu_2/18\}$, we rewrite (\ref{error1}) as
$$
D\left\{\M^{(t+1)},\bcB^{(t+1)}\right\}\le\rho D\left\{\M^{(t)},\bcB^{(t)}\right\}+C_2\|\nabla_{\bTheta}\ell_g(\bTheta^*,\bcB^*)\|_2^2+\frac{C_4}{\sigma^*_1}\|\nabla_{\bcB}\ell_g(\bTheta^*,\bcB^*)\|_{\infty}^2.
$$
Under Condition (A4), if the sample size $N$ is large enough, we have  
$$
C_2\|\nabla_{\bTheta}\ell_g(\bTheta^*,\bcB^*)\|_2^2+\frac{C_4}{\sigma^*_1}\|\nabla_{\bcB}\ell_g(\bTheta^*,\bcB^*)\|_{\infty}^2
\le C_1\epsilon_N^2+\frac{C_4}{\sigma^*_1}\xi_N^2\le(1-\rho)c_2^2\sigma^*_{r^*}.
$$
From this inequality, we see that, as long as $D\{\M^{(0)},\bcB^{(0)}\}\le c_2^2\sigma^*_{r^*}$, we have $D\{\M^{(t)},\bcB^{(t)}\}\le c_2^2\sigma^*_{r^*}$ for any $t\ge 0$. This implies that $d(\M^{(t)},\M^*)<c_2\sqrt{\sigma^*_{r^*}}$. Therefore, we have 
$$
D\left\{\M^{(t+1)},\bcB^{(t+1)}\right\}\le\rho^t D\left\{\M^{(0)},\bcB^{(0)}\right\}+\frac{C_2}{1-\rho}\|\nabla_{\bTheta}\ell_g(\bTheta^*,\bcB^*)\|_2^2+\frac{C_4}{\sigma^*_1(1-\rho)}\|\nabla_{\bcB}\ell_g(\bTheta^*,\bcB^*)\|_{\infty}^2.
$$
This completes the proof of Proposition~\ref{thm1}. 
\eop

\subsection{\bf Proof of Proposition~\ref{thm2}} 
\label{sec:pcoro1}
\noindent
The proof of Proposition~\ref{thm2} is similar to that of Proposition \ref{thm1}. However, when $\bTheta^*$ is symmetric, the convergence for the low-rank component cannot be trivially extended from the asymmetric case, and Lemma \ref{lm3} is required to replace Lemma \ref{lm1}. Specifically, at iteration $t$ of Algorithm \ref{algo_symm}, based on the results from Lemma~\ref{lm2} and Lemma~\ref{lm3}, we have 
\begin{eqnarray*}
d^2(\U^{(t+1)},\U^*)+\frac{1}{\sigma^*_1}\|\bcB^{(t+1)}-\bcB^*\|_F^2\le\rho_1d^2(\U^{(t)},\U^*)+\frac{1}{\sigma^*_1}(\rho_2+C_1\sigma^*_1)\|\bcB^{(t)}-\bcB^*\|_F^2\\
+\left(\frac{C_3}{\sigma^*_1}-\frac{\delta\mu_1}{4}\right)\|\bTheta-\bTheta^*\|_F^2+C_2\|\nabla_{\bTheta}\ell_g(\bTheta^*,\bcB^*)\|_2^2+\frac{C_4}{\sigma^*_1}\|\nabla_{\bcB}\ell_g(\bTheta^*,\bcB^*)\|_{\infty}^2.
\end{eqnarray*}
When $\gamma\ge1+\{(3\alpha_2+\mu_2c_2)/(\mu_2c_2)\}^2$, it is straightforward to verify that the condition $\gamma\ge1+\{(9\alpha_2-4\mu_2c_2)/(8\mu_2c_2)\}^2$ in Lemma~\ref{lm2} is satisfied. Additionally, we have $\rho_2\le1-\mu_2\tau/9$. Under the condition on the Lipschitz gradient parameter $\kappa$ that $\kappa\le\sqrt{\mu_1\mu_2}/9$,, we can verify the following inequalities, 
$$
C_1\sigma^*_1=c_1\left(\frac{\mu_2}{8}+\frac{2\kappa^2}{\mu_1}\right)+3c_1^2\kappa^2\le\frac{\mu_2c_1}{6},
$$
$$
C_3=\frac{2\sqrt{\gamma-1}+1}{2\sqrt{\gamma-1}-1}\left(\frac{4\tau \kappa^2}{\mu_2}+3\tau^2\kappa^2\right)\le\frac{c_1\mu_1}{4}.
$$
Therefore, if we set $c_1\le\tau/3$, we have 
\begin{eqnarray}
\label{error2}
d^2(\U^{(t+1)},\U^*)+\frac{1}{\sigma^*_1}\|\bcB^{(t+1)}-\bcB^*\|_F^2\le\rho_1d^2(\U^{(t)},\U^*)+\frac{1}{\sigma^*_1}\left(1-\frac{\mu_2\tau}{18}\right)\|\bcB^{(t)}-\bcB^*\|_F^2\\\nonumber
+C_2\|\nabla_{\bTheta}\ell_g(\bTheta^*,\bcB^*)\|_2^2+\frac{C_4}{\sigma^*_1}\|\nabla_{\bcB}\ell_g(\bTheta^*,\bcB^*)\|_{\infty}^2.
\end{eqnarray}
Define $D\{\U^{(t)},\bcB^{(t)}\}=d^2(\U^{(t)},\U^*)+\frac{1}{\sigma^*_1}\|\bcB^{(t)}-\bcB^*\|_F^2$, and $\rho=\max\{1-\delta\mu_1\sigma^*_{r^*}/16,1-\tau\mu_2/18\}$, we have write (\ref{error2}) as
$$
D\left\{\U^{(t+1)},\bcB^{(t+1)}\right\}\le\rho D\left\{\U^{(t)},\bcB^{(t)}\right\}+C_2\|\nabla_{\bTheta}\ell_g(\bTheta^*,\bcB^*)\|_2^2+\frac{C_4}{\sigma^*_1}\|\nabla_{\bcB}\ell_g(\bTheta^*,\bcB^*)\|_{\infty}^2.
$$
Under Condition 4, if the sample size $N$ is large enough, we have 
$$
C_2\|\nabla_{\bTheta}\ell_g(\bTheta^*,\bcB^*)\|_2^2+\frac{C_4}{\sigma^*_1}\|\nabla_{\bcB}\ell_g(\bTheta^*,\bcB^*)\|_{\infty}^2
\le C_1\epsilon_N^2+\frac{C_4}{\sigma^*_1}\xi_N^2\le(1-\rho)c_2^2\sigma^*_{r^*}.
$$
From this inequality, we see that, as long as $D\{\U^{(0)},\bcB^{(0)}\}\le c_2^2\sigma^*_{r^*}$, we have $D\{\U^{(t)},\bcB^{((t))}\}\le c_2^2\sigma^*_{r^*}$ for any $t\ge 0$. This implies that $d(\U^{(t)},\U^*)<c_2\sqrt{\sigma^*_{r^*}}$. Therefore, we have 
$$
D\left\{\U^{(t+1)},\bcB^{(t+1)}\right\}\le\rho^t D\left\{\U^{(t)},\bcB^{(t)}\right\}+\frac{C_2}{1-\rho}\|\nabla_{\bTheta}\ell_g(\bTheta^*,\bcB^*)\|_2^2+\frac{C_4}{\sigma^*_1(1-\rho)}\|\nabla_{\bcB}\ell_g(\bTheta^*,\bcB^*)\|_{\infty}^2.
$$
This completes the proof of Proposition~\ref{thm2}. 
\eop

\subsection{\bf Proof of Theorem~\ref{thm3}} 
\label{sec:pthm3}
\noindent
Under the proposed GLM model, the sample loss function is calculated as the negative loglikelihood \eqref{loglikelihood}, up to a constant. 
Next, we verify that, under the GLM setup and Conditions (B1)-(B3), Conditions (A1)-(A4) are satisfied.

\medskip
\noindent\textbf{Condition (A1):} We verify that the GLM loss function $\ell(\bTheta,\bcB)$ has strong convexity and strong smoothness with respect to the low-rank component $\bTheta$. Applying Taylor's theorem, 
\begin{eqnarray*}
&&\ell(\bTheta_2,\bcB)-\ell(\bTheta_1,\bcB)-\langle\nabla_{\bTheta}\ell(\bTheta_1,\bcB),\bTheta_2-\bTheta_1\rangle\\
&=&\frac{1}{2}\{\text{vec}(\bTheta_2)-\text{vec}(\bTheta_1)\}^\top\nabla^2_{\text{vec}(\bTheta)}\ell(\tilde\bTheta,\bcB)\{ \text{vec}(\bTheta_2)-\text{vec}(\bTheta_1) \}\\
&=&\frac{1}{2N}\sum_{i=1}^N\sum_{j,j'=1}^n\psi''(\tilde\delta^{(i)}_{jj'})(\bTheta_{2,jj'}-\bTheta_{1,jj'})^2,
\end{eqnarray*}
where $\tilde{\bm\delta}_i=\tilde\bTheta+\bcB\times_3\x_i$ and $\tilde\bTheta_{jj'}=\zeta_{jj'}\bTheta_{1,jj'}+(1-\zeta_{jj'})\bTheta_{2,jj'}$, $\zeta_{jj'}\in(0,1)$. For $\bTheta_1,\bTheta_2\in\mathbb{B}_{\bTheta^*}(\kappa_1)$, we have $\tilde\bTheta\in\mathbb{B}_{\bTheta^*}(\kappa_1)$. 
Therefore, the second order derivative of the cumulant function is bounded as $1/ \nu_0 \le \psi''(\delta^{(i)}_{jj'}) \le \nu_0$.
Consequently, we have
$$
\frac{\mu_1}{2}\|\bTheta_2-\bTheta_1\|^2\le \ell(\bTheta_2,\bcB)-\ell(\bTheta_1,\bcB)-\langle\nabla_{\bTheta}\ell(\bTheta_1,\bcB),\bTheta_2-\bTheta_1\rangle\le\frac{\alpha_1}{2}\|\bTheta_2-\bTheta_1\|^2,
$$
where $\mu_1= 1 / \nu_0$ and $\alpha_1=\nu_0$.

\medskip
\noindent\textbf{Condition (A2):} We verify that the GLM loss function $\ell(\bTheta,\bcB)$ has strong convexity and strong smoothness with respect to the sparse component $\bcB$. Write $\bDelta=\bcB_1-\bcB_2$ and define $\delta_{\bcB} \ell(\bTheta,\bDelta)=\ell(\bTheta,\bcB_2)-\ell(\bTheta,\bcB_1)-\langle\nabla_{\bcB}\ell(\bTheta,\bcB_1),\bDelta\rangle$. Applying Taylor's theorem, we have
\begin{equation*} \label{cond21}
\delta_{\bcB} \ell(\bTheta,\bDelta)=\frac{1}{2}\{ \text{vec}(\bcB_2)-\text{vec}(\bcB_1) \}^\top\nabla^2_{vec(\bcB)}\ell(\bTheta,\mathcal{\tilde B})\{\text{vec}(\bcB_2)-\text{vec}(\bcB_1)\}
\end{equation*}
where $\widetilde\bcB_{jj'l}=\zeta'_{jj'l}\bcB_{1,jj'l}+(1-\zeta'_{jj'l})\bcB_{2,jj'l}$, $\zeta'_{jj'l}\in(0,1)$. For $\bcB_1,\bcB_2\in\mathbb{B}_{\bcB^*}(\kappa_2)$, we have $\widetilde\bcB \in \mathbb{B}_{\bcB^*}(\kappa_2)$. Therefore, the second order derivative of the cumulant function is bounded as $1/ \nu_0 \le \psi''(\delta^{(i)}_{jj'}) \le \nu_0$. Let $\X\in\mathbb{R}^{N \times p}$ denote the design matrix. With some straightforward calculation, we obtain that 
$$
\frac{1}{\nu_0N}\sum_{j,j'=1}^n\|\X\bDelta_{jj'\cdot}\|_2^2\le\delta_{\bcB} \ell(\bTheta,\bDelta)\le\frac{\nu_0}{N}\sum_{j,j'=1}^n\|\X\bDelta_{jj'\cdot}\|_2^2.
$$
Applying Proposition 1 of \cite{loh2013regularized}, the following inequality holds with probability at least $1-\nu_1\exp(-\nu_2N)$, 
$$
\lambda_{\min}(\Sigma)\left(\frac{1}{2}\|\bDelta_{jj'\cdot}\|_2^2-\frac{\log n}{N}\|\bDelta_{jj'\cdot}\|_1^2\right)\le\frac{\|\X\bDelta_{jj'\cdot}\|_2^2}{N}\le\lambda_{\max}(\Sigma)\left(\frac{3}{2}\|\bDelta_{jj'\cdot}\|_2^2+\frac{\log n}{N}\|\bDelta_{jj'\cdot}\|_1^2\right).
$$
Using the inequality that $P(\cap_{i=1}^m A_i)\ge \sum_{i=1}^mP(A_i)-(m-1)$, we have the following upper and lower bounds, with probability at least $1-\nu_1n^2\exp(-\nu_2N)$, 
\begin{eqnarray}
\label{cond22}
&&\frac{\sum_{j,j'=1}^n\|\X\bDelta_{jj'\cdot}\|_2^2}{N}\ge\sum_{j,j'=1}^n\lambda_{\min}(\Sigma)\left(\frac{1}{2}\|\bDelta_{jj'\cdot}\|_2^2-\frac{\log n}{N}\|\bDelta_{jj'\cdot}\|_1^2\right),\\
&&\frac{\sum_{j,j'=1}^n\|\X\bDelta_{jj'\cdot}\|_2^2}{N}\le\sum_{j,j'=1}^n\lambda_{\max}(\Sigma)\left(\frac{3}{2}\|\bDelta_{jj'\cdot}\|_2^2+\frac{\log n}{N}\|\bDelta_{jj'\cdot}\|_1^2\right).
\label{cond23}
\end{eqnarray}
By definition, $\bDelta$ has most $2s$ nonzero entries. Therefore, we have $\sum_{j,j'=1}^n\|\bDelta_{jj'\cdot}\|_1^2\le2s\sum_{j,j'=1}^n\|\bDelta_{jj'\cdot}\|_2^2$. Plugging this into (\ref{cond22}) and (\ref{cond23}), we have that, when $N\ge c_7s\log n$, the following inequality is true with probability at least $1-c_8/n$,
 \begin{equation}
 \label{cond24}
\frac{\lambda_{\min}(\bSigma)}{4}\sum_{j,j'=1}^n\|\bDelta_{jj'\cdot}\|^2_2\le\frac{\sum_{j,j'=1}^n\|\X\bDelta_{jj'\cdot}\|_2^2}{N}\le\frac{7\lambda_{\max}(\bSigma)}{4}\sum_{j,j'=1}^n\|\bDelta_{jj'\cdot}\|^2_2.
\end{equation}
By setting $\mu_2=\lambda_{\min}(\bSigma)/(4\nu_0)$ and $\alpha_2=7\nu_0\lambda_{\max}(\bSigma)/4$, we verify Condition 2.

\medskip
\noindent\textbf{Condition (A3):} We verify that the GLM loss function $\ell$ satisfies the Lipschitz gradient condition with respect to the low-rank component and the sparse-component, respectively. Given the smoothness of the function $\psi''$, Condition (B2), and the compactness of the parameter space, we have that $1 / \nu_0 \le\psi''(\bTheta_{jj'}+\x_i^\top\bcB_{jj'})\le \nu_0$ for some constant $\nu_0>0$ and any $(\bTheta,\bcB)$ in the parameter space. 

Considering the first Lipschitz gradient condition, we have that 
\begin{eqnarray*}
& & \langle\nabla_{\bTheta}\ell(\bTheta^*,\bcB)-\nabla_{\bTheta}\ell(\bTheta^*,\bcB^*),\bTheta\rangle \\
& = & \frac{1}{N}\sum_{i=1}^N\sum_{j,j'=1}^n\left[\psi'(\bTheta^*_{jj'}+\x_i^\top\bcB_{jj'\cdot})-\psi'(\bTheta^*_{jj'}+\x_i^\top\bcB^*_{jj'\cdot})\right]\bTheta_{jj'}\\
& = &\frac{1}{N}\sum_{i=1}^N\sum_{j,j'=1}^n\sum_{l=1}^px_{il}\psi''(\bTheta^*_{jj'}+\x_i^\top\tilde\bcB_{jj'\cdot})(\bcB_{jj'l}-\bcB^*_{jj'l})\bTheta_{jj'}\overset{\Delta}{=}\frac{1}{N}\sum_{i=1}^N\W_i,
\end{eqnarray*}
where $\W_i=\langle\psi''(\tilde \bfeta^{(i)})\circ\bTheta,(\bcB-\bcB^*)\times_3\x_i\rangle$, and $\tilde\bfeta^{(i)}=\bTheta^*+\tilde\bcB\times_3\x_i$.
Note that $\x_i$ and $\psi''(\tilde \bfeta^{(i)})$ are sub-Gaussian, and hence $\W_i$ is a sub-exponential. Next, we have
\begin{eqnarray*}
\|\W_i-\mathbb{E}(\W_i)\|_{\psi_1}&\le& 2\|\W_i\|_{\psi_1}\\
&=&2\|\sum_{j,j'=1}^n\sum_{l=1}^px_{il}\psi''(\tilde\bfeta^{(i)}_{jj'})(\bcB_{jj'l}-\bcB^*_{jj'l})\bTheta_{jj'}\|_{\psi_1}\\
&\le& 2\sum_{j,j'=1}^n\sum_{l=1}^p\|x_{il}(\bcB_{jj'l}-\bcB^*_{jj'l})\|_{\psi_2}\cdot\|\psi''(\tilde\bfeta^{(i)}_{jj'})\bTheta_{jj'}\|_{\psi_2}\\
&\le& 2C^2_5\|\bcB-\bcB^*\|_F\cdot\|\bTheta\|_F=z_1,
\end{eqnarray*}
where the second inequality is due to the relationship between sub-exponential norm and sub-Gaussian norm, and the last inequality is due to that $\x_i$ and $\psi''(\bfeta^{(i)}_{jj'})$ are sub-Gaussian. 
Thus by Lemma \ref{sub}, we have that 
\[
\mathbb{P}\left\{\left|\frac{1}{N}\sum_{i=1}^N(\W_i-\mathbb{E}(\W_i))\right|\ge t\right\}\le 2\exp\left\{-c\min\left(\frac{Nt^2}{z^2_1},\frac{Nt}{z_1}\right)\right\}.
\]
Using the covering number argument for low-rank matrices in \cite{candes2011tight} and for sparse matrices in \cite{vershynin2009role}, we have 
\begin{eqnarray*}
&&\mathbb{P}\left\{\sup_{\bTheta\in\mathcal{N}_{\epsilon}^{3r^*},(\bcB-\bcB^*)\in\mathcal{N}_{\epsilon}^{cs^*}}\left|\frac{1}{N}\sum_{i=1}^N(\W_i-\mathbb{E}(\W_i))\right|\ge t\right\}\\
&\le&2|\mathcal{N}_{\epsilon}^{3r^*}||\mathcal{N}_{\epsilon}^{cs^*}|\exp\left\{-c\min\left(\frac{Nt^2}{z^2_1},\frac{Nt}{z_1}\right)\right\}\\
&\le& 2\left(\frac{c_2n^2p}{cs^*\epsilon}\right)^{cs^*} \left(\frac{9}{\epsilon}\right)^{(2n+1)3r^*} \exp\left\{-c\min\left(\frac{Nt^2}{z^2_1},\frac{Nt}{z_1}\right)\right\}\\
&\le& \exp\left\{c_3(s^*\max\{\log n,\log(1/\epsilon)\}+r^*n\log(1/\epsilon))-c\min\left(\frac{Nt^2}{z^2_1},\frac{Nt}{z_1}\right)\right\}\\
&\le& \exp(-c_4n),
\end{eqnarray*}
where $c_2$, $c_3$, $c_4$ are constants, and the last inequality holds when $t=c_5\sqrt{(r^*n+s^*)\log n}/\sqrt{N}$.
Let $\mathcal{M}_{3r^*}$ denote the set of matrices with at most rank $3r^*$ and $\mathcal{M}_{cs^*}$ denote the set of tensors with at most $cs^*$ nonzero entries, respectively.
For any $\bTheta\in\mathcal{M}_{3r^*}$ and $\bcB-\bcB^*\in\mathcal{M}_{cs^*}$, there exists $\bTheta_1\in\mathcal{N}_{\epsilon}^{3r^*}$ and $\bcB_1\in\mathcal{N}_{\epsilon}^{cs^*}$ such that $\|\bTheta-\bTheta_1\|_F\le\epsilon$ and $\|\bcB-\bcB_1\|_F\le\epsilon$.
Therefore, we have
\begin{eqnarray*}
&&|\langle\nabla_{\bTheta}\ell(\bTheta^*,\bcB)-\nabla_{\bTheta}\ell(\bTheta^*,\bcB^*),\bTheta\rangle-\langle\nabla_{\bTheta}\ell(\bTheta^*,\bcB_1)-\nabla_{\bTheta}\ell(\bTheta^*,\bcB^*),\bTheta_1\rangle|\\
&\le&\left|\frac{1}{N}\sum_{i=1}^N\sum_{j,j'=1}^n\sum_{l=1}^p\left\{\psi'(\bTheta^*_{jj'}+\x_i^\top\bcB_{1,jj'\cdot})-\psi'(\bTheta^*_{jj'}+\x_i^\top\bcB_{jj'\cdot})\right\}\bTheta_{1,jj'}\right|\\
&&+\left|\frac{1}{N}\sum_{i=1}^N\sum_{j,j'=1}^n\sum_{l=1}^p\left\{\psi'(\bTheta^*_{jj'}+\x_i^\top\bcB_{jj'\cdot})-\psi'(\bTheta^*_{jj'}+\x_i^\top\bcB^*_{jj'\cdot})\right\}(\bTheta_{jj'}-\bTheta_{1,jj'})\right|\\
&\le&M_x\nu_0\|\bcB-\bcB_1\|_F^2\cdot\|\bTheta_1\|_F+M_x\nu_0\|\bcB-\bcB^*\|_F^2\cdot\|\bTheta-\bTheta_1\|_F\le2c_6M_x\nu_0\epsilon.
\end{eqnarray*}
Similarly, we have
\begin{eqnarray*}
&&|\mathbb{E}\langle\nabla_{\bTheta}\ell(\bTheta^*,\bcB)-\nabla_{\bTheta}\ell(\bTheta^*,\bcB^*),\bTheta\rangle-\mathbb{E}\langle\nabla_{\bTheta}\ell(\bTheta^*,\bcB_1)-\nabla_{\bTheta}\ell(\bTheta^*,\bcB^*),\bTheta_1\rangle|\\
&\le&\left|\frac{1}{N}\sum_{i=1}^N\sum_{j,j'=1}^n\sum_{l=1}^p\mathbb{E}\left\{\psi'(\bTheta^*_{jj'}+\x_i^\top\bcB_{1,jj'\cdot})-\psi'(\bTheta^*_{jj'}+\x_i^\top\bcB_{jj'\cdot})\right\}\bTheta_{1,jj'}\right|\\
&&+\left|\frac{1}{N}\sum_{i=1}^N\sum_{j,j'=1}^n\sum_{l=1}^p\mathbb{E}\left\{\psi'(\bTheta^*_{jj'}+\x_i^\top\bcB_{jj'\cdot})-\psi'(\bTheta^*_{jj'}+\x_i^\top\bcB^*_{jj'\cdot})\right\}(\bTheta_{jj'}-\bTheta_{1,jj'})\right|\\
&\le&M_x\nu_0\|\bcB-\bcB_1\|_F^2\cdot\|\bTheta_1\|_F+M_x\nu_0\|\bcB-\bcB^*\|_F^2\cdot\|\bTheta-\bTheta_1\|_F\le2c_6M_x\nu_0\epsilon.
\end{eqnarray*}
Combining the above results and using the triangle inequality, we have
\begin{eqnarray*}
&&\sup_{\bTheta\in\mathcal{M}_{3r^*},(\bcB-\bcB^*)\in\mathcal{M}_{cs^*}}\left|\langle\nabla_{\bTheta}\ell(\bTheta^*,\bcB)-\nabla_{\bTheta}\ell(\bTheta^*,\bcB^*),\bTheta\rangle-\mathbb{E}\langle\nabla_{\bTheta}\ell(\bTheta^*,\bcB)-\nabla_{\bTheta}\ell(\bTheta^*,\bcB^*),\bTheta\rangle\right|\\
&&\le t+4c_6M_x\nu_0\epsilon,
\end{eqnarray*}
with probability at least $1-\exp(-c_4n)$. Thus, together with condition (B3), we establish the first Lipschitz condition with $\kappa\le\sqrt{\mu_1\mu_2}/9$ and setting $\epsilon=t/(8c_6M_x\nu_0)$. Following similar arguments, we can also show the second Lipschitz condition in (A3).

\medskip
\noindent\textbf{Condition (A4):} We seek the sample error of our model, and derive the upper bound on the spectral norm of $\nabla_{\bTheta}\ell(\bTheta^*,\bcB^*)$ and the infinity norm of $\nabla_{\bcB}\ell(\bTheta^*,\bcB^*)$, respectively.

We first obtain the upper bound on the spectral norm of $\nabla_{\bTheta}\ell(\bTheta^*,\bcB^*)$. We decompose $\nabla_{\bTheta}\ell(\bTheta^*,\bcB^*)$ as 
$$
\nabla_{\bTheta}\ell(\bTheta^*,\bcB^*)=\frac{1}{N}\sum_{i=1}^N\left\{E(\A_i)-\A_i\right\} = \frac{1}{N}\sum_{i=1}^N\sum_{j=1}^n\left[E\left\{\Y^{(ij)}\right\} - \Y^{(ij)}\right],
$$
where $\Y^{(ij)}_{kl}= E(\A_{jli})-\A_{jli}$ when $k=j$ and $0$ otherwise. By definition, we have $E\{ \Y^{(ij)} \} = 0$, and $\|\Y^{(ij)}\|^2_2=\sum_{l=1}^n\{ E(A_{jli})-A_{jli} \}^2\le\nu_3n$ almost surely for some constant $\nu_3>0$. Let $\bm\delta_i^*=\bTheta^*+\bcB^*\times_3\x_i$, we have
$$
\sum_{i=1}^N\sum_{j=1}^nE\{\Y^{(ij)}\}^2 =
\begin{pmatrix}
\sum_{i=1}^N\sum_{j=1}^n\psi''(\delta^*_{1ji}) & 0 & \cdots & 0 \\
0 & \sum_{i=1}^N\sum_{j=1}^n\psi''(\delta^*_{2ji}) & \cdots & 0 \\
\vdots  & \vdots  & \ddots & \vdots  \\
0 & 0 & \cdots & \sum_{i=1}^N\sum_{j=1}^n\psi''(\delta^*_{nji})
\end{pmatrix}.
$$
Therefore, we have $\sigma_y^2=\|\sum_{ij}E(\Y^{(ij)})^2\|_2\le \nu^2_0nN$. Applying Lemma~\ref{bern}, we have that 
\begin{eqnarray*}
P\left(\|\nabla_{\bTheta}\ell(\bTheta^*,\bcB^*)\|_2\ge t\right)&=&P\left(\left\|\frac{1}{N}\sum_{i=1}^N\sum_{j=1}^nE(\Y^{(ij)})^2\right\|_2\ge t\right)\\
&=&n\exp\left(-\frac{t^2N}{2\nu_0n+2t\sqrt{n}/3}\right).
\end{eqnarray*}
Letting $t=2\nu_0\sqrt{n\log n/N}$, we have $\|\nabla_{\bTheta}\ell(\bTheta^*,\bcB^*)\|_2\le2\nu_0\sqrt{n\log n/N}$ with probability at least $1-1/n$.

Next, we consider the upper bound on the infinity norm of $\nabla_{\bTheta}\ell(\bTheta^*,\bcB^*)$. We note that 
\begin{eqnarray*}
\|\nabla_{\bcB}\ell(\bTheta^*,\bcB^*)\|_{\infty}&=&\max_{s,j,j'}\left|\frac{1}{N}\sum_{i=1}^Nx_{is}\left[E(A^{(i)}_{jj'})-A^{(i)}_{jj'}\right]\right|\\
&\le&M_x\cdot\max_{s,j,j'}\left|\frac{1}{N}\sum_{i=1}^N\left[E(A^{(i)}_{jj'})-A^{(i)}_{jj'}\right]\right|.
\end{eqnarray*}

In a usual GLM, $\A^{(i)}_{jj'}$ follows an exponential distribution. By Lemma \ref{sub} and the union bound, we have 
$$
P\left(\max_{s,j,j'}\left|\frac{1}{N}\sum_{i=1}^N\left[E(A^{(i)}_{jj'})-A^{(i)}_{jj'}\right]\right|\ge t\right)\le 2pn^2\exp\left[-c\min\left(\frac{t^2}{z_1^2},\frac{t}{z_1}\right)N\right].
$$
Letting $t=c_9\sqrt{\log n/N}$, we have the following inequality holds with probability at least $1-c_{10}/n$, 
$$
\|\nabla_{\bcB}\ell(\bTheta^*,\bcB^*)\|_{\infty}\le c_9\sqrt{\frac{\log n}{N}}.
$$
This completes the proof of Theorem~\ref{thm3}. 
\eop

\subsection{\bf Proof of Theorem~\ref{thm:cluster}}
\noindent
Based on Theorem \ref{thm3}, after $t \ge \log_{\rho}\left( e_0 / D\{\U^{(0)},\bcB^{(0)}\} \right)$ iterations, the computational error $\rho^tD\{\U^{(0)},\bcB^{(0)}\}$ is to be dominated by the statistical error $e_0$. Henceforth, $D\{\U^{(t)},\bcB^{(t)}\}\le2e_0$, and $d(\U^{(t)}, \U^*) \le 2e_0$. 
Let $\hat{\U}=\U^{(t)}$ and denote its column representation as $\hat{\U} = (\hat{\U}_{\cdot 1}, \ldots, \hat{\U}_{\cdot r^*}) \in \mathbb R^{n\times r^*}$. The normalized $\hat{\U}$ can be written as
$$
\hat{\overline \U} = \underbrace{\left(\frac{\hat{\U}_{\cdot 1}}{\|\hat{\U}_{\cdot 1}\|_2}, \ldots, \frac{\hat{\U}_{\cdot r^*}}{\|\hat{\U}_{\cdot r^*}\|_2}\right)}_{\textrm{column representation}} = \underbrace{(\hat{\overline\u}_1, \ldots, \hat{\overline\u}_n)^{\top}}_{\textrm{row representation}}.
$$
Similarly, the column representation of $\widetilde\U^*$ can be written as
$$
\widetilde\U^* = \U^* diag(\sigma^{-1/2}_1,\ldots,\sigma^{-1/2}_{r^*}) = \left(\frac{{\U}^*_{\cdot 1}}{\|\U^*_{\cdot 1}\|_2}, \ldots, \frac{{\U}^*_{\cdot r^*}}{\|\U^*_{\cdot r^*}\|_2}\right)
$$
by noting that $\|{\U}^*_{\cdot j}\|_2 = \sqrt{\sigma_j}$ for $j=1,\ldots, r^*$. 

Note that the orthonormal transformation of data points does not affect their pairwise distances, and hence any orthonormal transformation preserves the cluster structure. Without loss of generality, we consider the orthonormal transformation $\Gamma$ to be an identity matrix such that $d(\hat\U,\U^*) = \|\hat\U-\U^*\bm\|_F$. We next derive the upper bound of $\max_{i=1\ldots, n}\|\hat{\overline\u}_i - \overline\u^*_i\|_2^2$ through the upper bound of the column vectors $\|\hat{\overline\U}_{\cdot j} - \widetilde\U^*_{\cdot j}\|_2^2$, $j=1,\ldots,r^*$.

By definition, we have 
\begin{eqnarray*}
\|\hat{\overline \U} -  \widetilde\U^*\|_F^2 &=& \sum_{j=1}^{r^*} \left \|  \frac{\hat{\U}_{\cdot j}}{\|\hat{\U}_{\cdot j}\|_2} - \frac{{\U}^*_{\cdot j}}{\|\U^*_{\cdot j}\|_2} \right \|_2^2\\
&=& \sum_{j=1}^{r^*} \left \|  \frac{\hat{\U}_{\cdot j}}{\|\hat{\U}_{\cdot j}\|_2} -   \frac{\U^*_{\cdot j}}{\|\hat{\U}_{\cdot j}\|_2} + \frac{\U^*_{\cdot j}}{\|\hat{\U}_{\cdot j}\|_2}  -  \frac{{\U}^*_{\cdot j}}{\|\U^*_{\cdot j}\|_2} \right \|_2^2\\
&\le& \sum_{j=1}^{r^*} \left( \frac{ \|  \hat{\U}_{\cdot j} - {\U}^*_{\cdot j}\|_2}{\|\hat{\U}_{\cdot j}\|_2} + \frac{ |\|\hat{\U}_{\cdot j}\|_2 - \|{\U}^*_{\cdot j}\|_2|}{\|\hat{\U}_{\cdot j}\|_2} \right)^2\\
&\le& \sum_{j=1}^{r^*} \frac{ 4\|  \hat{\U}_{\cdot j} - {\U}^*_{\cdot j}\|_2^2}{\|\hat{\U}_{\cdot j}\|_2^2},
\end{eqnarray*}
where the inequalities are due to the triangle inequality. We next find the lower bound of the denominator $\|\hat{\U}_{\cdot j}\|_2^2$. Applying the triangle inequality again, we have $\|\hat{\U}_{\cdot j}\|_2 \ge |\| {\U}^*_{\cdot j}\|_2 -    \|  \hat{\U}_{\cdot j} - {\U}^*_{\cdot j}\|_2 | \ge | \| \min_{j} \| {\U}^*_{\cdot j}\|_2 - \max_{j} \|  \hat{\U}_{\cdot j} - {\U}^*_{\cdot j}\|_2 | \ge | \sqrt{\sigma^*_{r^*}} - \|\U-\U^*\bm\|_F | \ge  \sqrt{\sigma^*_{r^*}} - e_0 > \sqrt{c_5}/2$ for a sufficiently large $N$, by noting that $e_0 \rightarrow 0$ as $N\rightarrow \infty$ and the minimal singular value $\sigma^*_{r^*}$ is lower bounded by some constant $c_5>0$. Therefore, we have
\begin{eqnarray*}
\max_{i=1\ldots, n}\|\hat{\overline\u}_i - \overline\u^*_i\|_2^2 &\le& \sum_{i=1}^n \|\hat{\overline\u}_i - \overline\u^*_i\|_2^2 \le \|\hat{\overline \U} -  \widetilde\U^*\|_F^2 \le 16/c_6 \sum_{j=1}^{r^*} \|  \hat{\U}_{\cdot j} - {\U}^*_{\cdot j}\|_2^2\le 16 e_0 /c_6.
\end{eqnarray*}
Therefore, as long as $\min_{k\ne k^{'}} |\u^*_{k} - \u^*_{k^{'}}| > c_6 e_0$ with $c_6 > 64/c_5$, for a node $i \in \cA_k^*$, and any node $j, j^{'}$ such that $j \in \cA_k^*$ and $j^{'} \notin \cA_k^*$, we have $\|\hat{\overline\u}_i - \hat{\overline\u}_{j^{'}}\|_2 \ge \|{\overline\u^*}_i - \overline\u^*_{j^{'}}\|_2 -  \|\hat{\overline\u}_i - {\overline\u^*}_i \|_2 -  \|\hat{\overline\u}_{j^{'}} - \overline\u^*_{j^{'}}\|_2 > 32c_5e_0$, and $\|\hat{\overline\u}_i - \hat{\overline\u}_{j}\|_2 \le \|{\overline\u^*}_i - \overline\u^*_{j}\|_2 +  \|\hat{\overline\u}_i - {\overline\u^*}_i \|_2 +  \|\hat{\overline\u}_{j} - \overline\u^*_{j}\|_2 \le 32c_5e_0$. Therefore, $P(\widehat{\cA}_k^{(t)} = \cA_k^*) \ge 1-c_4/n$ for any $k=1,\ldots,K$, and henceforth, 
$$
P\left(\widehat{\cA}_k^{(t)} = \cA_k^*, \textrm{for all~} k\right) \ge 1-P\left(\cup_{k=1}^K \{\widehat{\cA}_k^{(t)} \ne \cA_k^*\}\right)
\ge 1- \sum_{k=1}^K P(\widehat{\cA}_k^{(t)} \ne \cA_k^*) \ge 1 - \frac{c_4K}{n}. 
$$
This completes the proof of Theorem~\ref{thm:cluster}. \eop

\subsection{\bf Proof of Lemma \ref{lm1}}
\label{sec:plm1}
\noindent Denote the optimal rotation of $\M^*$ that minimizes the Frobenius norm distance of $\M$ and the rotated $\M^*$ as $\bm\Gamma=\min_{\bm\Gamma\in\mathbb{Q}_{r^*}}\|\M-\M^*\bm\Gamma\|_F$, and $\H=\M-\M^*\bm\Gamma$. Write the regularized objective function as
$$
\tilde \ell_g(\M,\bcB)=\ell_g(\U\V^\top,\bcB)+\frac{1}{8}\|\U^\top\U-\V^\top\V\|_F^2.
$$
We divide the proof into three steps. In the first step, we investigate the local curvature of $\tilde \ell_g$ with respect to the low-rank component. In the second step, we investigate the local smoothness of $\tilde \ell_g$ with respect to the low-rank component. In the third step, we  show the convergence for the low-rank component $\bTheta$.

\medskip
\noindent\textbf{Step 1:} We show that
\begin{eqnarray*}
\langle\nabla_{\M}\tilde \ell_g(\M,\bcB),\H\rangle&\ge&\frac{\mu_1}{4}\|\bTheta-\bTheta^*\|_F^2+\frac{1}{16}\|\widetilde\M^\top\M\|_F^2+\frac{\mu'_1\sigma^*_{r^*}}{16}\|\H\|_F^2-\left(\frac{3\alpha_1+2}{16}+\frac{\kappa^2}{8\mu_2}\right)\|\H\|_F^4\\
&&-\left(\frac{\mu_2}{8}+\frac{4\kappa^2}{\mu_1}\right)\|\bcB-\bcB^*\|_F^2-\left(\frac{8r^*}{\mu_1}+\frac{r^*}{\alpha_1}\right)\|\nabla_{\bTheta}\ell_g(\bTheta^*,\bcB^*)\|_2^2.
\end{eqnarray*}
where $\widetilde\M=[\U;-\V]$ and $\mu'_1=\min\{\mu_1,2\}$. First, we see that $\|\U^\top\U-\V^\top\V\|^2_F$=$\|\widetilde\M^\top\M\|_F^2$. Henceforth, $\nabla_{\M}\|\U^\top\U-\V^\top\V\|^2_F=4\widetilde\M\widetilde\M^\top\M$. Note that  
\begin{equation}
\nabla_{\M}\tilde \ell_g(\M,\bcB)=
 \begin{bmatrix}
       \nabla_{\U}\ell_g(\U\V^\top,\bcB)+\frac{1}{2}\U(\U^\top\U-\V^\top\V)  \\
       \nabla_{\V}\ell_g(\U\V^\top,\bcB)+\frac{1}{2}\V(\U^\top\U-\V^\top\V)           
     \end{bmatrix}.
     \label{derivative}
\end{equation}
We decompose the term $\langle\nabla_{\M}\tilde \ell_g(\M,\bcB),\H\rangle$ as
$$
\langle\nabla_{\M}\tilde \ell_g(\M,\bcB),\H\rangle=\underbrace{\langle\nabla_{\U} \ell_g(\bTheta,\bcB),\H_{\U}\rangle+\langle\nabla_{\V} \ell_g(\bTheta,\bcB),\H_{\V}\rangle}_{I_1}+\underbrace{\frac{1}{2}\langle\widetilde\M\widetilde\M^\top\M,\H\rangle}_{I_2},
$$
where $\H_{\U}$ and $\H_{\V}$ denote the top $n\times r^*$ and bottom $n\times r^*$ sub-matrices of $\H$, respectively.  Moreover, it holds true that $\nabla_{\U} \ell_g(\bTheta,\bcB)=\nabla_{\bTheta} \ell_g(\bTheta,\bcB)\V$, and $\nabla_{\V} \ell_g(\bTheta,\bcB)=\nabla_{\bTheta} \ell_g(\bTheta,\bcB)^\top\U$. We further decompose the term $I_1$ as
\begin{eqnarray*}
I_1&=&\langle\nabla_{\bTheta} \ell_g(\bTheta,\bcB),\U\V^\top-\U^*{\V^*}^\top+\H_{\U}\H_{\V}^\top\rangle\\
&=&\underbrace{\langle\nabla_{\bTheta} \ell_g(\bTheta^*,\bcB^*),\bTheta-\bTheta^*+\H_{\U}\H_{\V}^\top\rangle}_{I_{11}}\\
&&+\underbrace{\langle\nabla_{\bTheta} \ell_g(\bTheta^*,\bcB)-\nabla_{\bTheta} \ell_g(\bTheta^*,\bcB^*),\bTheta-\bTheta^*+\H_{\U}\H_{\V}^\top\rangle}_{I_{12}}\\
&&+\underbrace{\langle\nabla_{\bTheta} \ell_g(\bTheta,\bcB)-\nabla_{\bTheta} \ell_g(\bTheta^*,\bcB),\bTheta-\bTheta^*+\H_{\U}\H_{\V}^\top\rangle}_{I_{13}}.
\end{eqnarray*}
Next we bound the three terms $I_{11}$, $I_{12}$ and $I_{13}$ separately. Considering the term $I_{11}$, we have
\begin{eqnarray*}
|I_{11}|&\le&\|\langle\nabla_{\bTheta} \ell_g(\bTheta^*,\bcB^*)\|_2\cdot(\|\bTheta-\bTheta^*\|_*+\|\H_{\U}\H_{\V}^\top\|_*)\\
&\le&\|\langle\nabla_{\bTheta} \ell_g(\bTheta^*,\bcB^*)\|_2\cdot(\sqrt{2r^*}\|\bTheta-\bTheta^*\|_F+\sqrt{r^*}\|\H_{\U}\H_{\V}^\top\|_F)\\
&\le&\|\langle\nabla_{\bTheta} \ell_g(\bTheta^*,\bcB^*)\|_2\cdot(\sqrt{2r^*}\|\bTheta-\bTheta^*\|_F+\frac{\sqrt{r^*}}{2}\|\H\|^2_F)\\
&\le&\left(\frac{8r^*}{\mu_1}+\frac{r^*}{\alpha_1}\right)\|\nabla_{\bTheta}\ell_g(\bTheta^*,\bcB^*)\|_2^2+\frac{\mu_1}{16}\|\bTheta-\bTheta^*\|_F^2+\frac{\alpha_1}{16}\|\H\|^4_F,
\end{eqnarray*}
where the first inequality is due to the Von Neumann's trace inequality; the second inequality is due to that $\bTheta-\bTheta^*$ has at most rank $2r^*$, $\H_{\U}\H_{\V}^\top$ has at most rank $r^*$, and $\|\A\|_*\le\sqrt{r}\|\A\|_F$ in which $\A$ is a matrix with rank $r$; the last inequality is due to the fact that $2ab\le ta^2+b^2/t$. Considering the term $I_{12}$, we have
\begin{eqnarray*}
|I_{12}|&=&|\langle\nabla_{\bTheta} \ell_g(\bTheta^*,\bcB)-\nabla_{\bTheta} \ell_g(\bTheta^*,\bcB^*),\bTheta-\bTheta^*+\H_{\U}\H_{\V}^\top\rangle|\\
&\le&\kappa\|\bTheta-\bTheta^*+\H_{\U}\H_{\V}^\top\|_F\cdot\|\bcB-\bcB^*\|_F\\
&\le&\frac{\kappa}{2}\|\bcB-\bcB^*\|_F\cdot\|\H\|^2_F+\kappa\|\bcB-\bcB^*\|_F\cdot\|\bTheta-\bTheta^*\|_F\\
&\le&\left(\frac{\mu_2}{8}+\frac{4\kappa^2}{\mu_1}\right)\|\bcB-\bcB^*\|_F^2+\frac{\mu_1}{16}\|\bTheta-\bTheta^*\|_F^2+\frac{\kappa^2}{8\mu_2}\|\H\|_F^4,
\end{eqnarray*}
where the first inequality is due to the Lipschitz Condition (A3), and that $\bTheta-\bTheta^*+\H_{\U}\H_{\V}^\top$ has most rank $3r^*$, the second inequality is due to that $\|\H_{\U}\H_{\V}^\top\|_F\le\|\H_{\U}\|_F\cdot\|\H_{\V}\|_F\le\|\H\|_F^2/2$ and the last inequality is true due to the fact that $2ab\le ta^2+b^2/t$. Considering the term $I_{13}$, we first have 
$$
\ell_g(\bTheta^*,\bcB)\ge \ell_g(\bTheta,\bcB)+\langle\nabla_{\bTheta} \ell_g(\bTheta,\bcB),\bTheta^*-\bTheta\rangle+\frac{1}{2\alpha_1}\|\nabla_{\bTheta} \ell_g(\bTheta^*,\bcB)-\nabla_{\bTheta} \ell_g(\bTheta,\bcB)\|_F^2,
$$
due to Condition (A1) and Lemma \ref{B1}. Moreover, from Condition (A1), we have
$$
\ell_g(\bTheta,\bcB)\ge \ell_g(\bTheta^*,\bcB)+\langle\nabla_{\bTheta}\ell_g(\bTheta^*,\bcB),\bTheta-\bTheta^*\rangle+\frac{\mu_1}{2}\|\bTheta-\bTheta^*\|^2.
$$
Combining the above two inequalities, we have
\begin{equation}
\langle\nabla_{\bTheta}\ell_g(\bTheta,\bcB)-\ell_g(\bTheta^*,\bcB),\bTheta-\bTheta^*\rangle\ge\frac{\mu_1}{2}\|\bTheta-\bTheta^*\|^2+\frac{1}{2\alpha_1}\|\nabla_{\bTheta} \ell_g(\bTheta^*,\bcB)-\nabla_{\bTheta} \ell_g(\bTheta,\bcB)\|_F^2.
\label{I131}
\end{equation}
For the remainder term in $I_{13}$, we have
\begin{eqnarray}
\label{I132}
|\langle\nabla_{\bTheta} \ell_g(\bTheta,\bcB)-\nabla_{\bTheta} \ell_g(\bTheta^*,\bcB),\H_{\U}\H_{\V}^\top\rangle|&\le&\frac{1}{2}\|\nabla_{\bTheta} \ell_g(\bTheta^*,\bcB)-\nabla_{\bTheta} \ell_g(\bTheta,\bcB)\|_F\cdot\|\H\|_F^2\\
&\le&\frac{1}{2\alpha_1}\|\nabla_{\bTheta} \ell_g(\bTheta^*,\bcB)-\nabla_{\bTheta} \ell_g(\bTheta,\bcB)\|_F^2+\frac{\alpha_1}{8}\|\H\|_F^4\nonumber
\end{eqnarray}
Combining (\ref{I131}) and (\ref{I132}), we have
$$
I_{13}\ge\frac{\mu_1}{2}\|\bTheta-\bTheta^*\|^2-\frac{\alpha_1}{8}\|\H\|_F^4.
$$
Combining the bounds for the terms $I_{11}$, $I_{12}$ and $I_{13}$, we have
\begin{eqnarray*}
I_1&\ge&\frac{3\mu_1}{8}\|\bTheta-\bTheta^*\|_F^2-\left(\frac{3\alpha_1}{16}+\frac{\kappa^2}{8\mu_2}\right)\|\H\|_F^4\\
&&-\left(\frac{\mu_2}{8}+\frac{4\kappa^2}{\mu_1}\right)\|\bcB-\bcB^*\|_F^2-\left(\frac{8r^*}{\mu_1}+\frac{r^*}{\alpha_1}\right)\|\nabla_{\bTheta}\ell_g(\bTheta^*,\bcB^*)\|_2^2.
\end{eqnarray*}
Next we consider the term $I_2$. Since $\H$ can be written as $\H=\M-\M^*\bm\Gamma$, we have
\begin{eqnarray}
\label{I21}
\langle\widetilde\M\widetilde\M^\top\M,\H\rangle&=&\frac{1}{2}\langle\widetilde\M^\top\M,\widetilde\M^\top\M\rangle+\frac{1}{2}\langle\widetilde\M^\top\M,\widetilde\M^\top\M-2\widetilde\M^\top\M^*\bm\Gamma\rangle\\\nonumber
&=&\frac{1}{2}\|\widetilde\M^\top\M\|_F^2+\frac{1}{2}\langle\widetilde\M^\top\M,\widetilde\M^\top\H-\widetilde\M^\top\M^*\bm\Gamma\rangle.
\end{eqnarray}
Furthermore, we have that 
\begin{eqnarray}
\label{I22}
\langle\widetilde\M^\top\M,\widetilde\M^\top\M^*\bm\Gamma\rangle&=&\langle\widetilde\M^\top\M,\bm\Gamma^\top{\M^*}^\top\widetilde\M\bm\rangle\\\nonumber
&=&\langle\widetilde\M^\top\M,\bm\Gamma^\top\widetilde{\M^*}^\top\M\rangle=\langle\widetilde\M^\top\M,\bm\Gamma^\top\widetilde{\M^*}^\top\H\rangle\nonumber,
\end{eqnarray}
where the first equality is due to that $\widetilde\M^\top\M$ is symmetric, the second equality is due to that ${\M^*}^\top\widetilde\M={\widetilde{\M^*}}^\top\M$ and the last equality is due to that $\widetilde{\M^*}^\top\widetilde{\M^*}=0$. Combining (\ref{I21}) and (\ref{I22}), we have 
\begin{equation}
\label{I2}
\langle\widetilde\M^\top\M,\widetilde\M^\top\H-\widetilde\M^\top\M^*\bm\Gamma\rangle=\langle\widetilde\M^\top\M,(\widetilde\M-\widetilde{\M^*}\bm\Gamma)^\top\H\rangle\le\|\widetilde\M^\top\M\|_F\cdot\|\H\|_F^2,
\end{equation}
where the inequality is due to that $\|\widetilde\M-\widetilde{\M^*}\bm\Gamma\|_F^2=\|\M-{\M^*}\bm\Gamma\|_F^2=\|\H\|_F^2$.
From (\ref{I2}), we have 
$$
I_2\ge\frac{1}{2}\|\widetilde\M^\top\M\|_F^2-\frac{1}{2}\|\widetilde\M^\top\M\|_F\cdot\|\H\|_F^2\ge\frac{1}{4}\|\widetilde\M^\top\M\|_F^2-\frac{1}{4}\|\H\|_F^4.
$$
Combining the two bounds for $I_1$ and $I_2$, we have that 
\begin{eqnarray}
\langle\nabla_{\M}\tilde \ell_g(\M,\bcB),\H\rangle&\ge&\frac{3\mu_1}{8}\|\bTheta-\bTheta^*\|_F^2+\frac{1}{8}\|\widetilde\M^\top\M\|_F^2-\left(\frac{3\alpha_1+2}{16}+\frac{\kappa^2}{8\mu_2}\right)\|\H\|_F^4\\
&&-\left(\frac{\mu_2}{8}+\frac{4\kappa^2}{\mu_1}\right)\|\bcB-\bcB^*\|_F^2-\left(\frac{8r^*}{\mu_1}+\frac{r^*}{\alpha_1}\right)\|\nabla_{\bTheta}\ell_g(\bTheta^*,\bcB^*)\|_2^2\nonumber.
\label{step1.1}
\end{eqnarray}
Furthermore, we have 
\begin{eqnarray*}
\|\widetilde\M^\top\M\|_F^2&=&\langle\M\M^\top,\widetilde\M\widetilde\M^\top\rangle\\
&=&\langle\M\M^\top-\M^*{\M^*}^\top,\widetilde\M\widetilde\M^\top-\widetilde{\M^*}\widetilde{\M^*}^\top\rangle+\langle\M^*{\M^*}^\top,\widetilde\M\widetilde\M^\top\rangle+\langle\M\M^\top,\widetilde{\M^*}\widetilde{\M^*}^\top\rangle\\
&\ge&\langle\M\M^\top-\M^*{\M^*}^\top,\widetilde\M\widetilde\M^\top-\widetilde{\M^*}\widetilde{\M^*}^\top\rangle\\
&=&\|\U\U^\top-\U^*{\U^*}^\top\|_F^2+\|\V\V^\top-\V^*{\V^*}^\top\|_F^2-2\|\bTheta-\bTheta^*\|_F^2.
\end{eqnarray*}
Therefore, applying Lemma~\ref{B3}, we have $2\|\bTheta-\bTheta^*\|_F^2+\|\widetilde\M^\top\M\|_F^2\ge 4(\sqrt{2}-1)\sigma^*_{r^*}\|\H\|_F^2$. Plugging this back into (\ref{step1.1}), we have
\begin{eqnarray*}
\langle\nabla_{\M}\tilde \ell_g(\M,\bcB),\H\rangle&\ge&\frac{\mu_1}{4}\|\bTheta-\bTheta^*\|_F^2+\frac{1}{16}\|\widetilde\M^\top\M\|_F^2+\frac{\mu'_1\sigma^*_{r^*}}{16}\|\H\|_F^2-\left(\frac{3\alpha_1+2}{16}+\frac{\kappa^2}{8\mu_2}\right)\|\H\|_F^4\\
&&-\left(\frac{\mu_2}{8}+\frac{4\kappa^2}{\mu_1}\right)\|\bcB-\bcB^*\|_F^2-\left(\frac{8r^*}{\mu_1}+\frac{r^*}{\alpha_1}\right)\|\nabla_{\bTheta}\ell_g(\bTheta^*,\bcB^*)\|_2^2.
\end{eqnarray*}
This completes Step 1.

\medskip
\noindent\textbf{Step 2:} We show that
\begin{eqnarray*}
\|\nabla_{\M}\tilde \ell_g(\M,\bcB)\|_F^2&\le& 12\alpha_1^2\|\bTheta-\bTheta^*\|_F^2\cdot\|\M\|_2^2+12\kappa^2\|\bcB-\bcB^*\|_F^2\cdot\|\M\|_2^2\\
&&+12r^*\|\nabla_{\bTheta}\ell_g(\bTheta^*,\bcB^*)\|_2^2\cdot\|\M\|_2^2+\|\U^\top\U-\V^\top\V\|_F^2\cdot\|\M\|_2^2.
\end{eqnarray*}
From (\ref{derivative}), we have that 
$$
\|\nabla_{\M}\tilde \ell_g(\M,\bcB)\|_F^2\le2\|\nabla_{\U} \ell_g(\bTheta,\bcB)\|_F^2+2\|\nabla_{\V} \ell_g(\bTheta,\bcB)\|_F^2+\|\U^\top\U-\V^\top\V\|_F^2\cdot\|\M\|_2^2,
$$
where inequality is due to that $\|\A\B\|_F\le\|\A\|_F\cdot\|\B\|_2$. Considering the first term $\|\nabla_{\U} \ell_g(\bTheta,\bcB)\|_F^2$, we have the following decomposition, 
\begin{eqnarray*}
\|\nabla_{\U} \ell_g(\bTheta,\bcB)\|_F^2&\le&3\underbrace{\|(\nabla_{\bTheta} \ell_g(\bTheta,\bcB)-\nabla_{\bTheta} \ell_g(\bTheta^*,\bcB))\V\|_F^2}_{I_1}\\
&&+3\underbrace{\|(\nabla_{\bTheta} \ell_g(\bTheta^*,\bcB)-\nabla_{\bTheta} \ell_g(\bTheta^*,\bcB^*))\V\|_F^2}_{I_2}+3\underbrace{\|\nabla_{\bTheta} \ell_g(\bTheta^*,\bcB^*)\V\|_F^2}_{I_3}.
\end{eqnarray*}
Next we bound the three terms $I_1$, $I_2$ and $I_3$ separately. We have
\begin{eqnarray*}
I_1&=&\|(\nabla_{\bTheta} \ell_g(\bTheta,\bcB)-\nabla_{\bTheta} \ell_g(\bTheta^*,\bcB))\V\|_F^2\\
&\le&\|(\nabla_{\bTheta} \ell_g(\bTheta,\bcB)-\nabla_{\bTheta} \ell_g(\bTheta^*,\bcB))\|_F^2\cdot\|\V\|_2^2.
\end{eqnarray*}
\begin{eqnarray*}
I_2&=&\|(\nabla_{\bTheta} \ell_g(\bTheta^*,\bcB)-\nabla_{\bTheta} \ell_g(\bTheta^*,\bcB^*))\V\|_F\\
&=&\sup_{\|\W\|_F=1}\langle(\nabla_{\bTheta} \ell_g(\bTheta^*,\bcB)-\nabla_{\bTheta} \ell_g(\bTheta^*,\bcB^*))\V,\W\rangle\\
&\le&\sup_{\|\W\|_F=1}\kappa\|\bcB-\bcB^*\|_F.\|\W\V^\top\|_F\le \kappa\|\bcB-\bcB^*\|_F.\|\V\|_2,
\end{eqnarray*}
where the first inequality is due to Condition (A3), and the last inequality is due to that $\|\A\B\|_F\le\|\A\|_F\cdot\|\B\|_2$. Lastly, we have
$$
I_3=\|\nabla_{\bTheta} \ell_g(\bTheta^*,\bcB^*)\V\|_F^2\le r^*\|\nabla_{\bTheta} \ell_g(\bTheta^*,\bcB^*)\|_2^2\cdot\|\V\|_2^2,
$$
where the inequality is due to that $\|\A\B\|_F\le\|\A\|_F\cdot\|\B\|_2$ and $\|\A\|_F\le\sqrt{r}\|\A\|_2$ in which $\A$ is a matrix with rank $r$.
Combining the above bounds for $I_1$, $I_2$ and $I_3$, we have
\begin{eqnarray*}
\|\nabla_{\U} \ell_g(\bTheta,\bcB)\|_F^2&\le&3\|(\nabla_{\bTheta} \ell_g(\bTheta,\bcB)-\nabla_{\bTheta} \ell_g(\bTheta^*,\bcB))\|_F^2\cdot\|\V\|_2^2\\
&&+3\kappa^2\|\bcB-\bcB^*\|_F^2.\|\V\|_2^2+3r^*\|\nabla_{\bTheta} \ell_g(\bTheta^*,\bcB^*)\|_2^2\cdot\|\V\|_2^2.
\end{eqnarray*}
Following similar arguments, we can derive that 
\begin{eqnarray*}
\|\nabla_{\V} \ell_g(\bTheta,\bcB)\|_F^2&\le&3\|(\nabla_{\bTheta} \ell_g(\bTheta,\bcB)-\nabla_{\bTheta} \ell_g(\bTheta^*,\bcB))\|_F^2\cdot\|\U\|_2^2\\
&&+3\kappa^2\|\bcB-\bcB^*\|_F^2.\|\U\|_2^2+3r^*\|\nabla_{\bTheta} \ell_g(\bTheta^*,\bcB^*)\|_2^2\cdot\|\U\|_2^2.
\end{eqnarray*}
Given Condition (A1) and Lemma \ref{B1}, we have
\begin{eqnarray*}
\|(\nabla_{\bTheta} \ell_g(\bTheta,\bcB)-\nabla_{\bTheta} \ell_g(\bTheta^*,\bcB))\|_F^2&\le&2\alpha_1\left(\ell_g(\bTheta^*,\bcB)-\ell_g(\bTheta,\bcB)-\langle \ell_g(\bTheta,\bcB),\bTheta^*-\bTheta\rangle\right)\\
&\le&\alpha_1^2\|\bTheta-\bTheta^*\|_F^2.
\end{eqnarray*}
Therefore, we have the following upper bound 
\begin{eqnarray*}
\|\nabla_{\M}\tilde \ell_g(\M,\bcB)\|_F^2&\le& 12\alpha_1^2\|\bTheta-\bTheta^*\|_F^2\cdot\|\M\|_2^2+12\kappa^2\|\bcB-\bcB^*\|_F^2\cdot\|\M\|_2^2\\
&&+12r^*\|\nabla_{\bTheta}\ell_g(\bTheta^*,\bcB^*)\|_2^2\cdot\|\M\|_2^2+\|\U^\top\U-\V^\top\V\|_F^2\cdot\|\M\|_2^2.
\end{eqnarray*}
This completes Step 2.

\medskip
\noindent\textbf{Step 3:} Consider the iteration step $t$ in Algorithm \ref{algo_asym}, and denote
\begin{eqnarray*}
&&\U^{(t+1)}=\U^{(t)}-\delta\nabla_{\U} \ell_g(\bTheta^{(t)},\bcB^{(t)})-\frac{1}{2}\delta\U^{(t)}({\U^{(t)}}^\top\U^{(t)}-{\V^{(t)}}^\top\V^{(t)}),\\
&&\V^{(t+1)}=\V^{(t)}-\delta\nabla_{\V} \ell_g(\bTheta^{(t)},\bcB^{(t)})-\frac{1}{2}\delta\V^{(t)}({\U^{(t)}}^\top\U^{(t)}-{\V^{(t)}}^\top\V^{(t)}).
\end{eqnarray*}
Denote $\M^{(t)}=[\U^{(t)};\V^{(t)}]$ and $\H^{(t)}=\M^{(t)}-\M^*\bm\Gamma^{(t)}$. We have
\begin{eqnarray}
\label{dm}
d^2(\M^{(t+1)},\M^*)&\le&\|\M^{(t+1)}-\M^*\bm\Gamma^{(t)}\|_F^2\\\nonumber
&=&\|\M^{(t)}-\delta\nabla_{\M}\tilde \ell_g(\M^{(t)},\bcB^{(t)})-\M^*\bm\Gamma^{(t)}\|_F^2\\\nonumber
&=&d^2(\M^{t},\M^*)-2\delta\langle\nabla_{\M}\tilde \ell_g(\M^{(t)},\bcB^{(t)}),\H^{(t)}\rangle+\delta^2\|\nabla_{\M}\tilde \ell_g(\M^{(t)},\bcB^{(t)})\|_F^2.
\end{eqnarray}
According to the result from Step 1, we have the following lower bound on the second term,
\begin{eqnarray*}
\langle\nabla_{\M}\tilde \ell_g(\M,\bcB),\H\rangle&\ge&\frac{\mu_1}{4}\|\bTheta-\bTheta^*\|_F^2+\frac{1}{16}\|\widetilde\M^\top\M\|_F^2+\frac{\mu'_1\sigma^*_{r^*}}{16}\|\H\|_F^2-\left(\frac{3\alpha_1+2}{16}+\frac{\kappa^2}{8\mu_2}\right)\|\H\|_F^4\\
&&-\left(\frac{\mu_2}{8}+\frac{4\kappa^2}{\mu_1}\right)\|\bcB-\bcB^*\|_F^2-\left(\frac{8r^*}{\mu_1}+\frac{r^*}{\alpha_1}\right)\|\nabla_{\bTheta}\ell_g(\bTheta^*,\bcB^*)\|_2^2.
\end{eqnarray*}
According to the result from Step 2, we have the following upper bound on the third term, 
\begin{eqnarray*}
\|\nabla_{\M}\tilde \ell_g(\M^{(t)},\bcB)\|_F^2&\le& 12\alpha_1^2\|\bTheta^{(t)}-\bTheta^*\|_F^2\cdot\|\M^{(t)}\|_2^2+12\kappa^2\|\bcB^{(t)}-\bcB^*\|_F^2\cdot\|\M^{(t)}\|_2^2\\
&&+12r^*\|\nabla_{\bTheta}\ell_g(\bTheta^*,\bcB^*)\|_2^2\cdot\|\M^{(t)}\|_2^2+\|{\U^{(t)}}^\top\U^{(t)}-{\V^{(t)}}^\top\V^{(t)}\|_F^2\cdot\|\M^{(t)}\|_2^2.
\end{eqnarray*}
Moreover, we have that 
$$
\|\M^{(t)}\|_2\le\|\M^*\|_2+\|\M^{(t)}-\M^*\bm\Gamma\|_2\le\sqrt{2\sigma^*_1}+c_2\sqrt{\sigma^*_{r^*}}\le 2\sqrt{\sigma^*_1}.
$$
Let $\delta=c_1/\sigma^*_1$. If we set $c_1\le \min\{1/32,\mu_1/(192\alpha_1^2)\}$, we have the following inequality, 
\begin{eqnarray*}
-2\delta\langle\nabla_{\M}\tilde \ell_g(\M^{(t)},\bcB^{(t)}),\H^{(t)}\rangle+\delta^2\|\nabla_{\M}\tilde \ell_g(\M^{(t)},\bcB^{(t)})\|_F^2\le-\frac{\delta\mu_1}{4}\|\bTheta^{(t)}-\bTheta^*\|_F^2-\frac{\delta\mu'_1\sigma^*_{r^*}}{8}\|\H\|_F^2\\
+\delta\left(\frac{3\alpha_1+2}{8}+\frac{\kappa^2}{8\mu_2}\right)\|\H^{(t)}\|_F^4+C_1\|\bcB-\bcB^*\|_F^2+C_2\|\nabla_{\bTheta}\ell_g(\bTheta^*,\bcB^*)\|_2^2,
\end{eqnarray*}
where $C_1= 2\delta(\mu_2/8+4\kappa^2/\mu_1)+48\delta^2\kappa^2\sigma^*_1$ and $C_2=2\delta(8r^*/\mu_1+r^*/\alpha_1)+48\delta^2r^*\sigma^*_1$.
Additionally, according to our assumption that $d(\M^{(t)},\M^*)<c_2\sqrt{\sigma^*_{r^*}}$, we have $\|\H^{(t)}\|_F^2\le c_2^2\sigma^*_{r^*}$ and $\kappa<1$. Therefore, if we set $c_2\le\sqrt{\mu'_1/(6\alpha_1+4+2\kappa^2/\mu_2)}$, we have that 
\begin{eqnarray}
\label{dm2}
-2\delta\langle\nabla_{\M}\tilde \ell_g(\M^{(t)},\bcB^{(t)}),\H^{(t)}\rangle+\delta^2\|\nabla_{\M}\tilde \ell_g(\M^{(t)},\bcB^{(t)})\|_F^2\le-\frac{\delta\mu_1}{4}\|\bTheta^{(t)}-\bTheta^*\|_F^2\\\nonumber
-\frac{\delta\mu_1\sigma^*_{r^*}}{16}\|\H\|_F^2+C_1\|\bcB-\bcB^*\|_F^2+C_2\|\nabla_{\bTheta}\ell_g(\bTheta^*,\bcB^*)\|_2^2.
\end{eqnarray}
Plugging (\ref{dm2}) into (\ref{dm}), we complete the proof of Lemma~\ref{lm1}.
\eop

\subsection{\bf Proof of Lemma \ref{lm2}}
\label{sec:plm2}
At iteration step $t$, for the update on the sparse component in Algorithm \ref{algo_asym}, we have,
$$
\bcB^{(t+1)}=Truncate\left[ \bcB^{(t)}-\tau\nabla_{\bcB}\ell_g(\bTheta^{(t)},\bcB^{(t)}),s \right],
$$
where $s=\gamma s^*$. Define $\widetilde\bcB^{(t+1)}=\bcB^{(t)}-\tau\nabla_{\bcB}\ell_g(\bTheta^{(t)},\bcB^{(t)})$, and we have that $\bcB^{(t+1)}=Truncate\{ \widetilde\bcB^{(t+1)},\gamma s^* \}$. Based on Lemma~\ref{B2}, we have
$$
\|\bcB^{(t+1)}-\bcB^*\|_F^2\le\frac{2\sqrt{\gamma-1}+1}{2\sqrt{\gamma-1}-1}\cdot\|\widetilde\bcB^{(t+1)}-\bcB^*\|_F^2.
$$
Therefore, in order to bound the term $\|\bcB^{(t+1)}-\bcB^*\|_F^2$, we can bound the term $\|\widetilde\bcB^{(t+1)}-\bcB^*\|_F^2$. We have the following decomposition
\begin{eqnarray*}
\|\widetilde\bcB^{(t+1)}-\bcB^*\|_F^2&=&\|\bcB^{(t)}-\tau\nabla_{\bcB}\ell_g(\bTheta^{(t)},\bcB^{(t)})-\bcB^*\|_F^2\\
&\le&\|\bcB^{(t)}-\bcB^*\|_F^2-2\tau\underbrace{\langle\nabla_{\bcB}\ell_g(\bTheta^{(t)},\bcB^{(t)}),\bcB^{(t)}-\bcB^*\rangle}_{I_1}+\tau^2\underbrace{\|\nabla_{\bcB}\ell_g(\bTheta^{(t)},\bcB^{(t)})\|_F^2}_{I_2}.
\end{eqnarray*}
Next, we bound the terms $I_1$ and $I_2$ separately. Considering the term $I_1$, we further decompose it as
\begin{eqnarray*}
I_1&=&\underbrace{\langle\nabla_{\bcB}\ell_g(\bTheta^{(t)},\bcB^{(t)})-\nabla_{\bcB}\ell_g(\bTheta^{(t)},\bcB^*),\bcB^{(t)}-\bcB^*\rangle}_{I_{11}}+\underbrace{\langle\nabla_{\bcB}\ell_g(\bTheta^{(t)},\bcB^*)-\nabla_{\bcB}\ell_g(\bTheta^*,\bcB^*),\bcB^{(t)}-\bcB^*\rangle}_{I_{12}}\\
&&+\underbrace{\langle\nabla_{\bcB}\ell_g(\bTheta^*,\bcB^*),\bcB^{(t)}-\bcB^*\rangle}_{I_{13}}.
\end{eqnarray*}
For the term $I_{11}$, given Condition (A2) and Lemma~\ref{B1}, we have 
\begin{eqnarray*}
\ell_g(\bTheta^{(t)},\bcB^*)&\ge& \ell_g(\bTheta^{(t)},\bcB^{(t)})+\langle\nabla_{\bcB} \ell_g(\bTheta^{(t)},\bcB^{(t)}),\bcB^*-\bcB^{(t)}\rangle\\
&+&\frac{1}{2\alpha_2}\|\nabla_{\bcB} \ell_g(\bTheta^{(t)},\bcB^*)-\nabla_{\bcB} \ell_g(\bTheta^{(t)},\bcB^{(t)})\|_F^2.
\end{eqnarray*}
Moreover, from the strong convexity property in Condition (A2), we have
$$
\ell_g(\bTheta^{(t)},\bcB^{(t)})\ge \ell_g(\bTheta^{(t)},\bcB^*)+\langle\nabla_{\bcB}\ell_g(\bTheta^{(t)},\bcB^*),\bcB^{(t)}-\bcB^*\rangle+\frac{\mu_2}{2}\|\bcB^{(t)}-\bcB^*\|^2.
$$
Combining the above two inequalities, we have
$$
I_{11}\ge\frac{\mu_2}{2}\|\bcB^{(t)}-\bcB^*\|^2+\frac{1}{2\alpha_2}\|\nabla_{\bcB} \ell_g(\bTheta^{(t)},\bcB^*)-\nabla_{\bcB} \ell_g(\bTheta^{(t)},\bcB^{(t)})\|_F^2.
$$
For the term $I_{12}$, we have the following result due to Condition (A3), 
$$
|I_{12}|=|\langle\nabla_{\bcB}\ell_g(\bTheta^{(t)},\bcB^*)-\nabla_{\bcB}\ell_g(\bTheta^*,\bcB^*),\bcB^{(t)}-\bcB^*\rangle|\le \kappa\|\bTheta^{(t)}-\bTheta^*\|_F\cdot\|\bcB^{(t)}-\bcB^*\|_F.
$$
For the term $I_{13}$, we have 
\begin{eqnarray*}
|I_{13}|=|\langle\nabla_{\bcB}\ell_g(\bTheta^*,\bcB^*),\bcB^{(t)}-\bcB^*\rangle|
&\le& \|\nabla_{\bcB}\ell_g(\bTheta^*,\bcB^*)\|_{\infty}\cdot\|\bcB^{(t)}-\bcB^*\|_1\\
&\le& \|\nabla_{\bcB}\ell_g(\bTheta^*,\bcB^*)\|_{\infty}\cdot\sqrt{(\gamma+1)s^*}\|\bcB^{(t)}-\bcB^*\|_F,
\end{eqnarray*}
where $\|\B\|_1$ denotes the $\ell_1$-norm of $\B$ and the last inequality is true because the term $\bcB^{(t)}-\bcB^*$ has at more $(\gamma+1)s^*$ nonzero entries. Putting together $I_{11}$, $I_{12}$ and $I_{13}$, we have 
\begin{eqnarray*}
I_1\ge\frac{4\mu_2}{9}\|\bcB^{(t)}-\bcB^*\|^2+\frac{1}{2\alpha_2}\|\nabla_{\bcB} \ell_g(\bTheta^{(t)},\bcB^*)-\nabla_{\bcB} \ell_g(\bTheta^{(t)},\bcB^{(t)})\|_F^2\\
-\frac{9\kappa^2}{2\mu_2}\|\bTheta-\bTheta^*\|^2_F-\frac{9(\gamma+1)s^*}{2\mu_2}\|\nabla_{\bcB}\ell_g(\bTheta^*,\bcB^*)\|_{\infty}^2,
\end{eqnarray*}
where the inequality is due to $2ab\le a^2t+b^2/t$. 

Next, considering the term $I_2$, we have the following decomposition
\begin{eqnarray*}
|I_2|&\le&3\|\nabla_{\bcB}\ell_g(\bTheta^{(t)},\bcB^{(t)})-\nabla_{\bcB}\ell_g(\bTheta^{(t)},\bcB^*)\|_F^2+3\underbrace{\|\nabla_{\bcB}\ell_g(\bTheta^{(t)},\bcB^*)-\nabla_{\bcB}\ell_g(\bTheta^*,\bcB^*)\|_F^2}_{I_{21}}\\
&&+3\underbrace{\|\nabla_{\bcB}\ell_g(\bTheta^*,\bcB^*)\|_F^2}_{I_{22}}.
\end{eqnarray*}
For the term $I_{21}$, we have 
\begin{eqnarray*}
I_{21}&=&\|\nabla_{\bcB}\ell_g(\bTheta^{(t)},\bcB^*)-\nabla_{\bcB}\ell_g(\bTheta^*,\bcB^*)\|_F\\
&=&\sup_{\substack{\mathcal{W}\in\mathbb{R}^{n\times n\times p} \\ \|\mathcal{W}\|_F\le1}}\langle\nabla_{\bcB}\ell_g(\bTheta^{(t)},\bcB^*)-\nabla_{\bcB}\ell_g(\bTheta^*,\bcB^*),\mathcal{W}\rangle\\
&=&\sup_{\substack{\mathcal{W}\in\mathbb{R}^{n\times n\times p} \\ \|\mathcal{W}\|_F\le1}}
\langle\nabla_{\bcB}\ell_g(\bTheta^{(t)},\bcB^*)-\nabla_{\bcB}\ell_g(\bTheta^*,\bcB^*),\mathcal{W}\rangle\\
&\le&\sup_{\substack{\mathcal{W}\in\mathbb{R}^{n\times n\times P} \\ \|\mathcal{W}\|_F\le1}}\kappa\|\bTheta^{(t)}-\bTheta^*\|_F\cdot\|\mathcal{W}\|_F\le \kappa\|\bTheta^{(t)}-\bTheta^*\|_F.
\end{eqnarray*}
Furthermore, we have that
$$
I_{22}=\|\nabla_{\bcB}\ell_g(\bTheta^*,\bcB^*)\|_F^2\le s^*\|\nabla_{\bcB}\ell_g(\bTheta^*,\bcB^*)\|_{\infty}^2.
$$
Putting together $I_{21}$ and $I_{22}$, we have 
$$
|I_2|\le3\|\nabla_{\bcB}\ell_g(\bTheta^{(t)},\bcB^{(t)})-\nabla_{\bcB}\ell_g(\bTheta^{(t)},\bcB^*)\|_F^2+3\kappa^2\|\bTheta^{(t)}-\bTheta^*\|_F^2+3s^*\|\nabla_{\bcB}\ell_g(\bTheta^*,\bcB^*)\|_{\infty}^2.
$$
If we set $\tau<1/(3\alpha_2)$, then we have 
\begin{eqnarray*}
&&-2\tau\langle\nabla_{\bcB}\ell_g(\bTheta^{(t)},\bcB^{(t)}),\bcB^{(t)}-\bcB^*\rangle+\tau^2\|\nabla_{\bcB}\ell_g(\bTheta^{(t)},\bcB^{(t)})\|_F^2\le-\frac{8\mu_2\tau}{9}\|\bcB^{(t)}-\bcB^*\|^2\\
&&+\left(\frac{9\tau \kappa^2}{\mu_2}+3\tau^2\kappa^2\right)\|\bTheta-\bTheta^*\|_F^2
+\left(\frac{9\tau(\gamma+1)s^*}{\mu_2}+3\tau^2s^*\right)\|\nabla_{\bcB}\ell_g(\bTheta^*,\bcB^*)\|_{\infty}^2,
\end{eqnarray*}
which leads to the desired result and completes the proof of Lemma~\ref{lm2}.
\eop

\subsection{\bf Proof of Lemma \ref{lm3}}
\label{sec:plm3}
Denote the optimal rotation of $\U^*$ with respect to $\U$ as $\bm\Gamma=\min_{\bm\Gamma\in\mathbb{Q}_{r^*}}\|\U-\U^*\bm\Gamma\|_F$, and $\H=\U-\U^*\bm\Gamma$. We follow a similar structure as the proof of Lemma \ref{lm1}, and divide the proof of this lemma into three steps. In the first step, we investigate the local curvature of $\ell_g$ with respect to the low-rank component. In the second step, we show the local smoothness of $\ell_g$ with respect to the low-rank component. In the third step, we put together results from the first two steps and show the convergence for the low-rank symmetric component $\bTheta$. Note that in the symmetric case, the decomposition $\bTheta^*=\widetilde\U^*\bLambda\widetilde\U^{*\top}$ is unique up to a rotation of $\U^*$. Therefore, we no longer need to add a regularizer to the objective function $\ell_g$ to guarantee the uniqueness.

\medskip
\noindent\textbf{Step 1:} We show that
\begin{eqnarray*}
\langle\nabla_{\U}\ell_g(\bTheta,\bcB),\H\rangle&\ge&\frac{3\mu_1}{32}\|\bTheta-\bTheta^*\|_F^2-\left(\frac{9\alpha_1}{64}+\frac{\kappa^2}{4\mu_2}\right)\|\H\|_F^4\\
&&-\left(\frac{\kappa^2}{\mu_1}+\frac{\mu_2}{16}\right)\|\bcB-\bcB^*\|_F^2-\left(\frac{2r^*}{\mu_1}+\frac{r^*}{\alpha_1}\right)\|\nabla_{\bTheta}\ell_g(\bTheta^*,\bcB^*)\|_2^2.
\end{eqnarray*}
First we have the decomposition, 
\begin{eqnarray*}
\langle\nabla_{\U}\ell_g(\bTheta,\bcB),\H\rangle&=&\frac{1}{2}\langle\nabla_{\bTheta}\ell_g(\bTheta,\bcB)\U\bLambda,\H\rangle
=\frac{1}{4}\langle\nabla_{\bTheta}\ell_g(\bTheta,\bcB),\bTheta-\bTheta^*+\H\bLambda\H^\top\rangle\\
&=&\frac{1}{4}\underbrace{\langle\nabla_{\bTheta}\ell_g(\bTheta^*,\bcB^*),\bTheta-\bTheta^*+\H\bLambda\H^\top\rangle}_{I_1}\\
&&+\frac{1}{4}\underbrace{\langle\nabla_{\bTheta}\ell_g(\bTheta^*,\bcB)-\langle\nabla_{\bTheta}\ell_g(\bTheta^*,\bcB^*),\bTheta-\bTheta^*+\H\bLambda\H^\top\rangle}_{I_2}\\
&&+\frac{1}{4}\underbrace{\langle\nabla_{\bTheta}\ell_g(\bTheta,\bcB)-\langle\nabla_{\bTheta}\ell_g(\bTheta^*,\bcB),\bTheta-\bTheta^*+\H\bLambda\H^\top\rangle}_{I_3}.
\end{eqnarray*}
Considering the term $I_1$, we have
\begin{eqnarray*}
|I_{1}|&\le&\|\nabla_{\bTheta}\ell_g(\bTheta^*,\bcB^*)\|_2\cdot(\|\bTheta-\bTheta^*\|_*+\|\H\bLambda\H^\top\|_*)\\
&\le&\|\nabla_{\bTheta}\ell_g(\bTheta^*,\bcB^*)\|_2\cdot(\sqrt{2r^*}\|\bTheta-\bTheta^*\|_F+\sqrt{r^*}\|\H\bLambda\H^\top\|_F)\\
&\le&\|\nabla_{\bTheta}\ell_g(\bTheta^*,\bcB^*)\|_2\cdot(\sqrt{2r^*}\|\bTheta-\bTheta^*\|_F+\sqrt{r^*}\|\H\|_F^2)\\
&\le&\left(\frac{8r^*}{\mu_1}+\frac{4r^*}{\alpha_1}\right)\|\nabla_{\bTheta}\ell_g(\bTheta^*,\bcB^*)\|_2^2+\frac{\mu_1}{16}\|\bTheta-\bTheta^*\|_F^2+\frac{\alpha_1}{16}\|\H\|_F^4,
\end{eqnarray*}
where the first equality is due to the Von Neumann's trace inequality, the second inequality is due to that $\bTheta-\bTheta^*$ has at most rank $2r^*$, $\H\bLambda\H^\top$ has at most rank $r^*$ and $\|\A\|_*\le\sqrt{r}\|\A\|_F$ in which $\A$ is a matrix with rank $r$. The last inequality is due to the fact that $2ab\le ta^2+b^2/t$. Considering the term $I_2$, we have
\begin{eqnarray*}
|I_2|&=&|\langle\nabla_{\bTheta}\ell_g(\bTheta^*,\bcB)-\langle\nabla_{\bTheta}\ell_g(\bTheta^*,\bcB^*),\bTheta-\bTheta^*+\H\bLambda\H^\top\rangle|\\
&\le& \kappa\|\bcB-\bcB^*\|_F\cdot\|\bTheta-\bTheta^*+\H\bLambda\H^\top\|_F\\
&\le& \kappa\|\bcB-\bcB^*\|_F\cdot(\|\bTheta-\bTheta^*\|_F+\|\H\bLambda\H^\top\|_F)\\
&\le& \left(\frac{4\kappa^2}{\mu_1}+\frac{\mu_2}{4}\right)\|\bcB-\bcB^*\|_F^2+\frac{\mu_1}{16}\|\bTheta-\bTheta^*\|_F^2+\frac{\kappa^2}{\mu_2}\|\H\|_F^4,
\end{eqnarray*}
where the first inequality is due to Condition (A3), the second inequality is due to the triangle inequality, and the last inequality is due to the fact that $2ab\le ta^2+b^2/t$. Considering the term $I_3$, following a similar argument as in (\ref{I131}) and (\ref{I132}), we have
$$
I_3\ge\frac{\mu_2}{2}\|\bTheta-\bTheta^*\|_F^2-\frac{\alpha_1}{2}\|\H\|_F^4.
$$
Combing the bounds for $I_1$, $I_2$ and $I_3$, we have
\begin{eqnarray*}
\langle\nabla_{\U}\ell_g(\bTheta,\bcB),\H\rangle&\ge&\frac{3\mu_1}{32}\|\bTheta-\bTheta^*\|_F^2-\left(\frac{9\alpha_1}{64}+\frac{\kappa^2}{4\mu_2}\right)\|\H\|_F^4\\
&&-\left(\frac{\kappa^2}{\mu_1}+\frac{\mu_2}{16}\right)\|\bcB-\bcB^*\|_F^2-\left(\frac{2r^*}{\mu_1}+\frac{r^*}{\alpha_1}\right)\|\nabla_{\bTheta}\ell_g(\bTheta^*,\bcB^*)\|_2^2.
\end{eqnarray*}
This completes Step 1.

\medskip
\noindent\textbf{Step 2:} We show that
\begin{eqnarray*}
\|\nabla_{\U}\ell_g(\bTheta,\bcB)\|_F^2&\le&\frac{3\alpha_1^2}{4}\|\bTheta-\bTheta^*\|_F^2\cdot\|\U\|_2^2+\frac{3\kappa^2}{16}\|\bcB-\bcB^*\|_F^2\cdot\|\U\|_2^2\\
&&+\frac{3r^*}{16}\|\nabla_{\bTheta}\ell_g(\bTheta^*,\bcB^*)\|_2^2\cdot\|\U\|_2^2
\end{eqnarray*}
We start with the decomposition, 
\begin{eqnarray*}
\|\nabla_{\U}\ell_g(\bTheta,\bcB)\|_F^2&\le&\frac{3}{4}\underbrace{\|(\nabla_{\bTheta}\ell_g(\bTheta,\bcB)-\nabla_{\bTheta}\ell_g(\bTheta^*,\bcB))\U\bLambda\|^2_F}_{I_1}\\
&&+\frac{3}{4}\underbrace{\|(\nabla_{\bTheta}\ell_g(\bTheta^*,\bcB)-\nabla_{\bTheta}\ell_g(\bTheta^*,\bcB^*))\U\bLambda\|_F^2}_{I_2}+\frac{3}{4}\underbrace{\|\nabla_{\bTheta}\ell_g(\bTheta^*,\bcB^*)\U\bLambda\|_F^2}_{I_3}.
\end{eqnarray*}
Following a similar argument as in Step 2 of Lemma \ref{lm1}, we obtain that 
$$
I_1\le\|\nabla_{\bTheta}\ell_g(\bTheta,\bcB)-\nabla_{\bTheta}\ell_g(\bTheta^*,\bcB)\|_F^2\cdot\|\U\|_2^2,
$$
$I_2\le \kappa\|\bcB-\bcB^*\|_F\cdot\|\U\|_2$ and $I_3\le r^*\|\nabla_{\bTheta}\ell_g(\bTheta^*,\bcB^*)\|_2^2\cdot\|\U\|_2^2$.
Given Condition (A1) and Lemma \ref{B1}, we have
\begin{eqnarray*}
\|(\nabla_{\bTheta} \ell_g(\bTheta,\bcB)-\nabla_{\bTheta} \ell_g(\bTheta^*,\bcB))\|_F^2&\le&2\alpha_1\left(\ell_g(\bTheta^*,\bcB)-\ell_g(\bTheta,\bcB)-\langle \ell_g(\bTheta,\bcB),\bTheta^*-\bTheta\rangle\right)\\
&\le&\alpha_1^2\|\bTheta-\bTheta^*\|_F^2.
\end{eqnarray*}
Therefore, we obtain the following upper bound 
\begin{eqnarray*}
\|\nabla_{\U} \ell_g(\U,\bcB)\|_F^2&\le& \frac{3\alpha_1^2}{4}\|\bTheta-\bTheta^*\|_F^2\cdot\|\U\|_2^2+\frac{3\kappa^2}{4}\|\bcB-\bcB^*\|_F^2\cdot\|\U\|_2^2\\
&&+\frac{3r^*}{4}\|\nabla_{\bTheta}\ell_g(\bTheta^*,\bcB^*)\|_2^2\cdot\|\U\|_2^2.
\end{eqnarray*}
This completes Step 2.

\medskip
\noindent\textbf{Step 3:} Consider the iteration step $t$ in Algorithm \ref{algo_symm}, and denote $\U^{(t+1)}=\U^{(t)}-\delta\nabla_{\U} \ell_g(\bTheta^{(t)},\bcB^{(t)})$, where $\bTheta^{(t)}=\U^{(t)}\bLambda{\U^{(t)}}^\top$. Denoting $\H^{(t)}=\U^{(t)}-\U^*\bm\Gamma^{(t)}$, we have
\begin{eqnarray*}
d^2(\U^{(t+1)},\U^*)&\le&\|\U^{(t+1)}-\U^*\bm\Gamma^{(t)}\|_F^2
\le\|\U^{(t)}-\delta\nabla_{\U}\ell_g(\bTheta^{(t)},\bcB^{(t)})-\U^*\bm\Gamma^{(t)}\|_F^2\\\nonumber
&=&d^2(\U^{t},\U^*)-2\delta\langle\nabla_{\U}\ell_g(\bTheta^{(t)},\bcB^{(t)}),\U^{(t)}\rangle+\delta^2\|\nabla_{\U}\ell_g(\bTheta^{(t)},\bcB^{(t)})\|_F^2.
\end{eqnarray*}
According to the result from Step 1 and Lemma~\ref{au1}, we have the following lower bound on the second term $\langle\nabla_{\U}\ell_g(\bTheta^{(t)},\bcB^{(t)}),\U^{(t)}\rangle$, 
\begin{eqnarray*}
\langle\nabla_{\U}\ell_g(\bTheta^{(t)},\bcB^{(t)}),\U^{(t)}\rangle&\ge&\frac{3\mu_1}{32}\|\bTheta-\bTheta^*\|_F^2-\left(\frac{9\alpha_1}{64}+\frac{\kappa^2}{4\mu_2}\right)\|\H\|_F^4\\
&&-\left(\frac{\kappa^2}{\mu_1}+\frac{\mu_2}{16}\right)\|\bcB-\bcB^*\|_F^2-\left(\frac{2r^*}{\mu_1}+\frac{r^*}{\alpha_1}\right)\|\nabla_{\bTheta}\ell_g(\bTheta^*,\bcB^*)\|_2^2\\
&\ge&\frac{\mu_1}{32}\|\bTheta-\bTheta^*\|_F^2+\frac{\mu_1\sigma^*_{r^*}}{32}\|\H\|_F^2-\left(\frac{9\alpha_1}{64}+\frac{\kappa^2}{4\mu_2}\right)\|\H\|_F^4\\
&&-\left(\frac{\kappa^2}{\mu_1}+\frac{\mu_2}{16}\right)\|\bcB-\bcB^*\|_F^2-\left(\frac{2r^*}{\mu_1}+\frac{r^*}{\alpha_1}\right)\|\nabla_{\bTheta}\ell_g(\bTheta^*,\bcB^*)\|_2^2,
\end{eqnarray*}
where the second inequality is due to Lemma~\ref{au1}. According to the result from Step 2, we have the following upper bound on the third term, 
\begin{eqnarray*}
\|\nabla_{\U} \ell_g(\U^{(t)},\bcB^{(t)})\|_F^2&\le& \frac{3\alpha_1^2}{4}\|\bTheta^{(t)}-\bTheta^*\|_F^2\cdot\|\U^{(t)}\|_2^2+\frac{3\kappa^2}{4}\|\bcB^{(t)}-\bcB^*\|_F^2\cdot\|\U^{(t)}\|_2^2\\
&&+\frac{3r^*}{4}\|\nabla_{\bTheta}\ell_g(\bTheta^*,\bcB^*)\|_2^2\cdot\|\U^{(t)}\|_2^2.
\end{eqnarray*}
Moreover, we have that 
$$
\|\U^{(t)}\|_2\le\|\U^*\|_2+\|\U^{(t)}-\U^*\bm\Gamma\|_2\le\sqrt{\sigma^*_1}+c_2\sqrt{\sigma^*_{r^*}}\le 2\sqrt{\sigma^*_1}.
$$
Let $\delta=c_1/\sigma^*_1$. If we set $c_1\le\mu_1/(96\alpha_1^2)$, we have the following inequality 
\begin{eqnarray*}
-2\delta\langle\nabla_{\M}\tilde \ell_g(\M^{(t)},\bcB^{(t)}),\M^{(t)}\rangle+\delta^2\|\nabla_{\M}\tilde \ell_g(\M^{(t)},\bcB^{(t)})\|_F^2\ge-\frac{\delta\mu_1}{16}\|\bTheta^{(t)}-\bTheta^*\|_F^2-\frac{\delta\mu_1\sigma^*_{r^*}}{16}\|\H\|_F^2\\
+\delta\left(\frac{9\alpha_1}{32}+\frac{\kappa^2}{2\mu_2}\right)\|\H^{(t)}\|_F^4+C_1\|\bcB-\bcB^*\|_F^2+C_2\|\nabla_{\bTheta}\ell_g(\bTheta^*,\bcB^*)\|_2^2,
\end{eqnarray*}
where $C_1= \delta(\mu_2/8+2\kappa^2/\mu_1)+3\delta^2\kappa^2\sigma^*_1$ and $C_2=\delta(4r^*/\mu_1+2r^*/\alpha_1)+3\delta^2r^*\sigma^*_1$.
Additionally, according to our assumption on $\U^{(t)}$, we have $\|\H^{(t)}\|_F^2\le c_2^2\sigma^*_{r^*}$. Therefore, if we set $c_2\le\sqrt{\mu_1/(9\alpha_1/2+8\kappa^2/\mu_2)}$, we have that 
\begin{eqnarray*}
-2\delta\langle\nabla_{\M}\tilde \ell_g(\M^{(t)},\bcB^{(t)}),\M^{(t)}\rangle+\delta^2\|\nabla_{\M}\tilde \ell_g(\M^{(t)},\bcB^{(t)})\|_F^2\ge-\frac{\delta\mu_1}{16}\|\bTheta^{(t)}-\bTheta^*\|_F^2\\\nonumber
-\frac{\delta\mu_1\sigma^*_{r^*}}{16}\|\H\|_F^2+C_1\|\bcB-\bcB^*\|_F^2+C_2\|\nabla_{\bTheta}\ell_g(\bTheta^*,\bcB^*)\|_2^2,
\end{eqnarray*}
which completes the proof of Lemma~\ref{lm3}.
\eop

\subsection{\bf Proof of Lemma \ref{au1}}
\label{sec:pau1}
First note that 
\begin{eqnarray*}
 \begin{bmatrix}\U^* & \U\\-\U^*\bLambda & \U\bLambda \end{bmatrix}\begin{bmatrix}\U^* & \U\\-\U^*\bLambda & \U\bLambda \end{bmatrix}^\top
 =\begin{bmatrix}\U\U^\top+\U^*{\U^*}^\top & \bTheta-\bTheta^* \\\bTheta^\top-{\bTheta^*}^\top & \U\U^\top+\U^*{\U^*}^\top \end{bmatrix}\\
 =\begin{bmatrix}\bm{0} & \bTheta-\bTheta^* \\ \bTheta^\top-{\bTheta^*}^\top & \bm{0} \end{bmatrix}+\begin{bmatrix}\U\U^\top+\U^*{\U^*}^\top & \bm{0} \\\bm{0} & \U\U^\top+\U^*{\U^*}^\top \end{bmatrix}.
\end{eqnarray*}
Applying the Weyl's inequality, we have that 
\begin{eqnarray*}
&&\sigma_{2r}\left(\begin{bmatrix}\U^* & \U\\-\U^*\bLambda & \U\bLambda \end{bmatrix}\begin{bmatrix}\U^* & \U\\-\U^*\bLambda & \U\bLambda \end{bmatrix}^\top\right)\\
&\ge&\sigma_{2r}\left(\begin{bmatrix}\U\U^\top+\U^*{\U^*}^\top & \bm{0} \\\bm{0} & \U\U^\top+\U^*{\U^*}^\top \end{bmatrix}\right)+\sigma_{2n}\left(\begin{bmatrix}\bm{0} & \bTheta-\bTheta^* \\ \bTheta^\top-{\bTheta^*}^\top & \bm{0} \end{bmatrix}\right)\\
&\ge&\sigma_{2r}\left(\begin{bmatrix}{\U^*}{\U^*}^\top & \bm{0} \\\bm{0} & {\U^*}{\U^*}^\top\end{bmatrix}\right)\ge\sigma^*_{r},
\end{eqnarray*}
where the last inequality is true due to $\sigma_{r}(\U^*{\U^*}^\top)\ge\sigma_{r}(\U^*\bLambda{\U^*}^\top).$
Applying Lemma~\ref{B3} to $\begin{bmatrix}\U^* & \U\\\U^*\bLambda & -\U\bLambda \end{bmatrix}$ and $\begin{bmatrix}\U^* & \U\\-\U^*\bLambda & \U\bLambda \end{bmatrix}$, we have that 
$
d^2\left(\begin{bmatrix}\U^* & \U\\\U^*\bLambda & -\U\bLambda \end{bmatrix},\begin{bmatrix}\U^* & \U\\-\U^*\bLambda & \U\bLambda \end{bmatrix}\right)\le\frac{2}{(\sqrt{2}-1)\sigma^*_{r}}\|\bTheta-\bTheta^*\|_F^2.
$
Let $\check\U\check\bSigma\check\V^\top$ be the singular value decomposition of ${\U^*}^\top\U$. It is easy to verify that the optimal rotation between $\begin{bmatrix}\U^* & \U\\\U^*\bLambda & -\U\bLambda \end{bmatrix}$ and $\begin{bmatrix}\U^* & \U\\-\U^*\bLambda & \U\bLambda \end{bmatrix}$ is $\check\Gamma=\begin{bmatrix}\bm{0} & \check\U\check\V^\top\\-\check\V\check\U^\top & \bm{0} \end{bmatrix}$. Therefore, we have
$$
d^2\left(\begin{bmatrix}\U^* & \U\\\U^*\bLambda & -\U\bLambda \end{bmatrix},\begin{bmatrix}\U^* & \U\\-\U^*\bLambda & \U\bLambda \end{bmatrix}\right)=2\|\U-\U^*\check\U\check\V^\top\|_F^2+2\|\U-\U^*\check\U\check\V^\top\|_F^2=4d^2(\U,\U^*).
$$
This completes the proof of Lemma~\ref{au1}.
\eop

\section{Computational Results}
\subsection{Gradients}
\label{sec::gradient}

\noindent
We given the explicit forms of the gradients in Algorithm \ref{algo_asym}. Specifically,
\begin{eqnarray*}
\nabla_{\U}\tilde\ell \left\{ \U{\V}^\top,\bcB \right\}&=&\frac{1}{N}\sum_{i=1}^N\left\{ -\A^{(i)}+\psi'(\bTheta+\bcB\times_3\x_i) \right\} \V+\frac{1}{2}\U(\U^\top\U-\V^\top\V),\\
\nabla_{\V}\tilde\ell \left\{ \U \V^\top,\bcB \right\}&=&\frac{1}{N}\sum_{i=1}^N\left\{ -\A^{(i)}+\psi'(\bTheta+\bcB\times_3\x_i) \right\}^\top \U+\frac{1}{2}\V(\U^\top\U-\V^\top\V),\\
\nabla_{\bcB}\tilde\ell \left\{ \U{\V}^\top,\bcB \right\}&=&\frac{1}{N}\sum_{i=1}^N\left\{ -\A^{(i)}\otimes \x_i+\psi'(\bTheta+\bcB\times_3\x_i)\otimes \x_i \right\},
\end{eqnarray*}
where $\otimes$ denotes the outer product, and $\psi'(\cdot)$ is applied element-wise.

\subsection{Parameter tuning}
\label{sec::bic}

\noindent
To investigate the robustness of the method in terms of the rank $r$ and sparsity proportion $s_0$, we plot in Figure \ref{fig:bic} the heat map of the average eBIC over 50 data replications for a range of values of $r$ and $s_0$. We employ the simulation model in Section \ref{sec::model1} with the sample size $N=200$. The left panel shows eBIC over the full range of $r \in \{1,2,\ldots,20\}$ and $s_0 \in 10^{\{-3,-2.9,\ldots,0.1,0\}}$, while the right panel zooms in and shows only for $r \in \{3,4,5,6,7\}$ and $s_0\in10^{\{-1.3,-1.2,-1.1,-1.0,-0.9,-0.8\}}$. It is seen that eBIC is the smallest at the true rank $r=5$ and sparsity proportion $s_0=0.1$. 

\begin{figure}[t!]
\centering
\includegraphics[scale=0.52, trim=4mm 0 0 0]{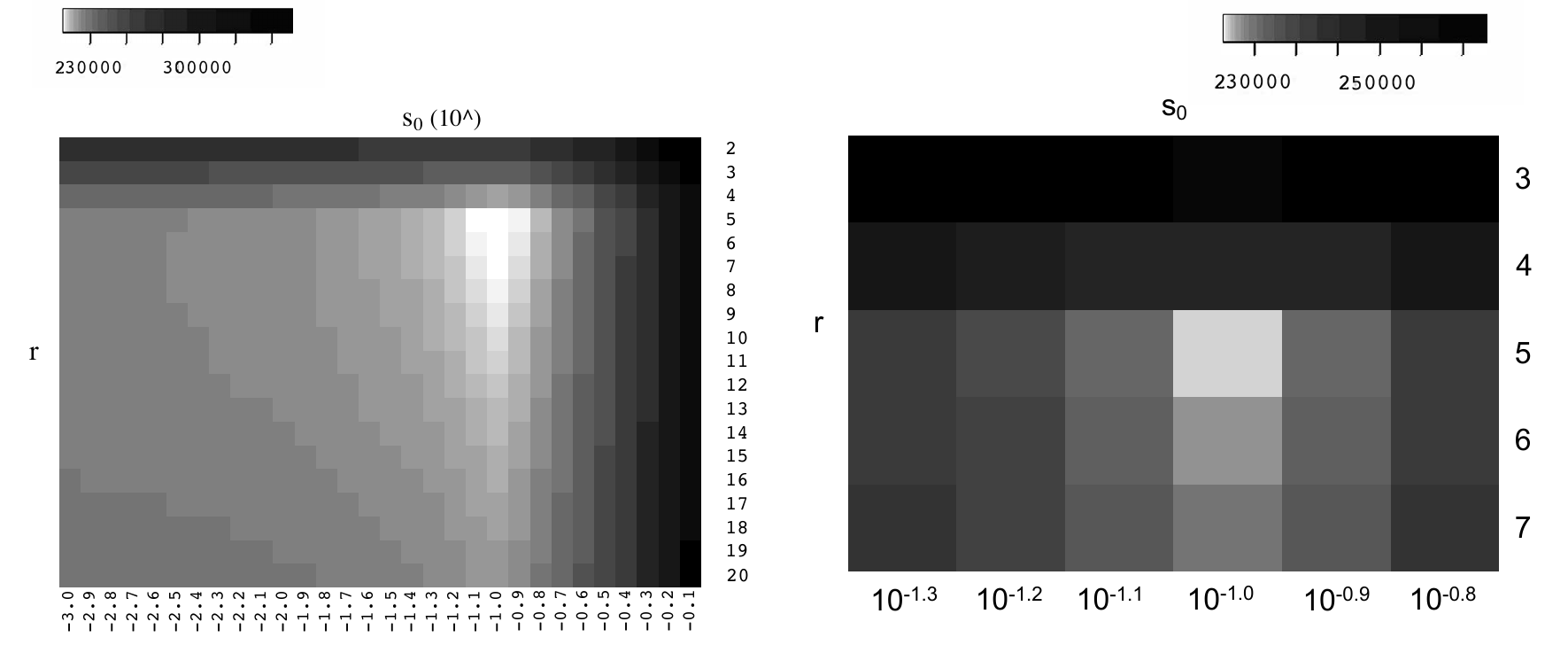}
\caption{\small Heat map of eBIC over a range of values of the rank $r$ and sparsity level $s_0$. The left panel shows for the full range of $r \in \{1,2,\ldots,20\}$ and $s_0 \in 10^{\{-3,-2.9,\ldots,0.1,0\}}$, and the right panel zooms in and shows only for $r \in \{3,4,5,6,7\}$ and $s_0\in10^{\{-1.3,-1.2,-1.1,-1.0,-0.9,-0.8\}}$.}
\label{fig:bic}
\end{figure}

\subsection{Additional simulation results}
\label{sec::model3}

\noindent
We simulate data from a stochastic blockmodel (SBM), and compare with the variational EM method specifically designed for SBM. Following the model setup in Section \ref{sec::con}, given the connecting probability matrix $\M$ and the community assignment matrix $\C$, we generate $\A_{jj'i}$'s as independent Bernoulli random variables, with $P(\A_{jj'i}=1)=\C_{\cdot j}\M\C_{\cdot j'}^\top$, $i=1,\ldots,N$. We set $n=100$ nodes, belonging to $K=3$ communities, and the number of nodes in each community is 50, 25, and 25, respectively. This determines the binary community membership $\C \in \mathbb{R}^{n \times K}$. We set $\M$ as a $3 \times 3$ matrix, with $w$ on the diagonal representing the within-community connecting probability, and $0.1$ on the off-diagonal representing the between-community connecting probability. We vary the value of $w$ from $0.15$ to $0.50$, with a larger value of $w$ implying a larger difference between the within-community and the between-community connecting probabilities, and thus a stronger signal. Since the classical SBM has been designed for a single network, it does not incorporate the subject covariates. Moreover, it usually assumes the number of communities $K$ is known. As such, we set $K$ at the true value for all estimation methods. We generate $N = 100$ samples. We fit the stochastic blockmodel to each of the $N$ samples, and report the average estimates, whereas the variational EM algorithm is initialized with spectral clustering.

\begin{figure}[t!]
\centering
\includegraphics[trim=1cm 1cm 1cm 0, scale=0.6]{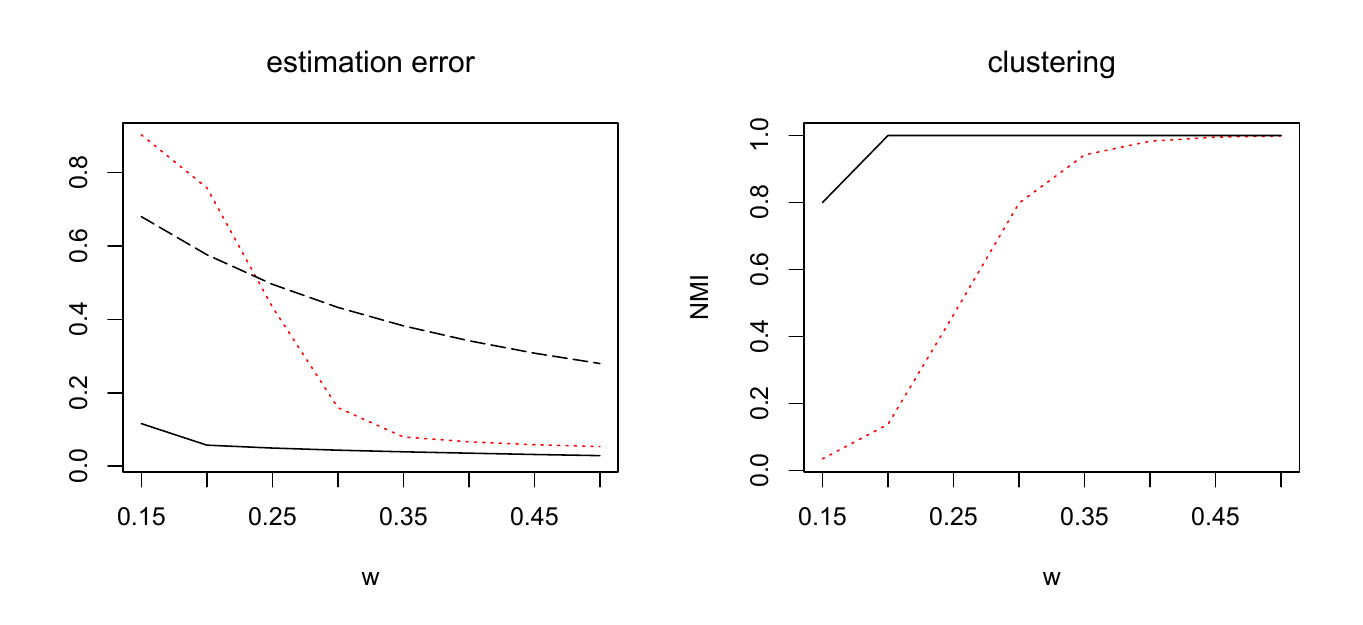}
\caption{\small Simulation results under the stochastic blockmodel, with varying signal strength $w$. The left panel reports the average estimation error, and the right panel the average clustering accuracy. The three methods under comparison are: the variational EM method (red dotted line), the CISE method of \citet{wang2019common} (black dashed line), and our proposed GLSNet method (black solid line).}
\label{SBMerrors}
\end{figure}

To evaluate the estimation accuracy, we report the normalized estimation error, $N^{-1} \sum_{i=1}^N$ $\| \bmu^{(i)}-\hat\bmu^{(i)} \|_F/ \|\bmu^{(i)}\|_F$. To evaluate the node clustering accuracy, we report the normalized mutual information between the estimated clustering membership and the true membership. Figure~\ref{SBMerrors} reports the average results based on 50 data replications. We also include the method of \citet{wang2019common} in the comparison, but only report its estimation error, as it cannot produce any community estimation. It is seen that our method outperforms the variational EM method in both estimation accuracy and clustering accuracy. This is because our method utilizes information from all $N$ subjects jointly, whereas the variational EM method utilizes each subject's information separately, and only averages the estimates in the final step. This difference is more pronounced when the signal strength is relatively weak, such as when $w$ is smaller than $0.3$. The superior empirical clustering accuracy also agrees with our theoretical findings in Theorem~\ref{thm:cluster}.

Next, we simulate data from the network latent factor model, and compare our method with the Bayesian MCMC method \citep{minhas2016inferential}. Again, following Section \ref{sec::con}, we simulate the node additive effect $\balpha_j$ and the node multiplicative effect $\c_j$, $1 \leq j \leq n$, from the standard normal distribution. We set the number of latent factors $K=5,10$, and vary the sample size $N$ from 5 to 100. Similar as in Section \ref{sec::model3}, the Bayesian MCMC method is applied to each sample separately, then the results are averaged. The method of \citet{wang2019common} is also included in the comparison. Figure~\ref{LFMerrors} reports the normalized estimation error averaged over 50 data replications. It is seen again that our method performs best, as it jointly models all $N$ samples and effectively borrows information from each other. As the sample size $N$ increases, the estimation error of our method decreases. As the number of latent factors $K$ increases, the rank of the low-rank representation increases and the estimation error increases. These observations agree with our theoretical results in Theorem~\ref{thm3}. We note that \cite{minhas2016inferential} has a constant error rate. This is because their method is designed for a single network sample. When applied to multiple samples, the method is implemented on each individual sample, then the results are averaged. 

\begin{figure}[t!]
\centering
\includegraphics[trim=1cm 1cm 1cm 0, scale=0.6]{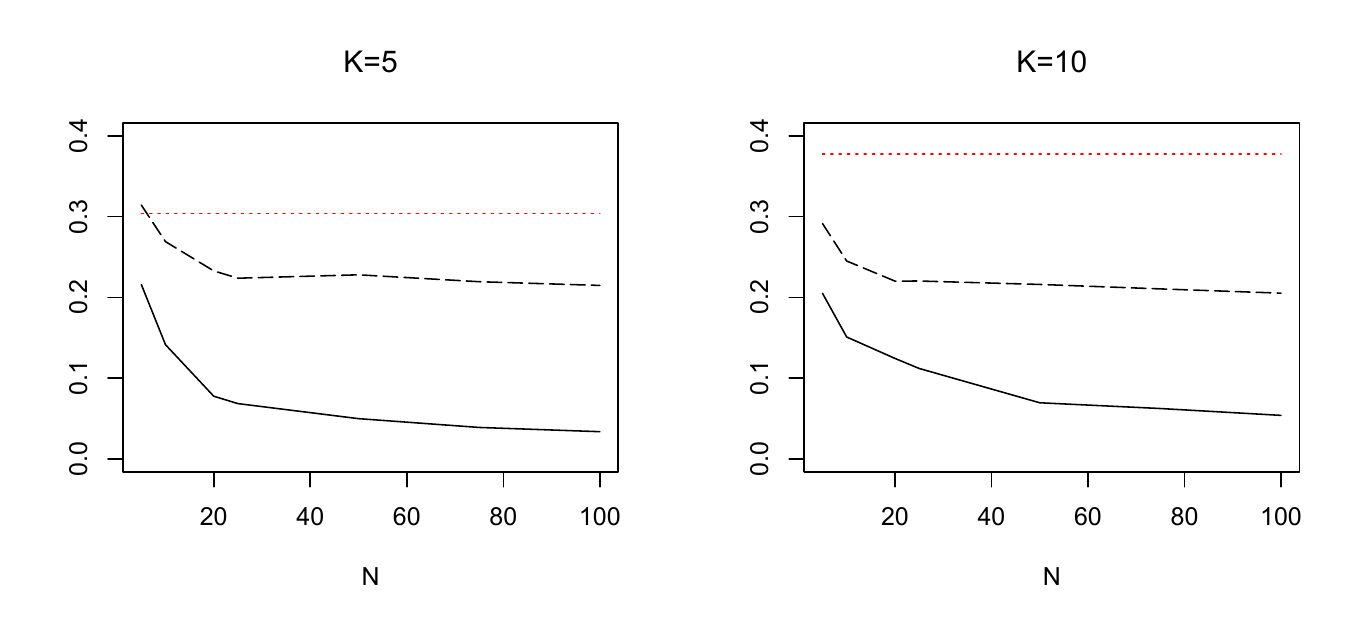}
\caption{\small Simulation results under the latent factor model, with the varying sample size $N$ and number of latent factors $K$. The left panel reports the average estimation error for $K = 5$, and the right panel for $K = 10$. The three methods under comparison are: the Bayesian MCMC method (red dotted line), the CISE method of \citet{wang2019common} (black dashed line), and our proposed GLSNet method (black solid line).}
\label{LFMerrors}
\end{figure}

\subsection*{Additional references}

\begin{description}\setlength{\itemsep}{-0.5ex}
\bibitem[Candes and Plan(2011)]{candes2011tight}
Candes, E.~J., and Plan, Y. (2011),
``Tight oracle inequalities for low-rank matrix recovery from a minimal number of noisy random measurements,"
\textit{IEEE Transactions on Information Theory}, 57, 2342--2359.

\bibitem[{Li et~al.(2016)Li, Arora, Liu, Haupt, and Zhao}]{li2016nonconvex}
Li, X., Arora, R., Liu, H., Haupt, J., and Zhao, T. (2016), ``Nonconvex
  sparse learning via stochastic optimization with progressive variance
  reduction," \textit{arXiv preprint arXiv:1605.02711}.
  
 \bibitem[{Loh and Wainwright(2013)}]{loh2013regularized}
Loh, P.-L., and Wainwright, M.~J. (2013), ``Regularized M-estimators with
  nonconvexity: Statistical and algorithmic theory for local optima," in
  \textit{Advances in Neural Information Processing Systems}, 476--484.

\bibitem[{Nesterov(2013)}]{nesterov2013introductory}
Nesterov, Y. (2013), \textit{Introductory lectures on convex optimization: A
  basic course}, vol.~87, Springer Science \& Business Media.
  
\bibitem[{Tropp(2012)}]{tropp2012user}
Tropp, J.~A. (2012),``User-friendly tail bounds for sums of random
  matrices," \textit{Foundations of computational mathematics}, 12, 389--434.

\bibitem[{Tu et~al.(2016)Tu, Boczar, Simchowitz, Soltanolkotabi, and Recht}]{tu2016low}
Tu, S., Boczar, R., Simchowitz, M., Soltanolkotabi, M., and Recht, B. (2016),
  ``Low-rank solutions of linear matrix equations via procrustes flow,"
  in \textit{Proceedings of the 33rd International Conference on International
  Conference on Machine Learning-Volume 48}, JMLR. org, pp. 964--973.
  
\bibitem[{Vershynin(2009)}]{vershynin2009role}
Vershynin, R. (2009), ``On the role of sparsity in compressed sensing and random matrix theory," 
\textit{2009 3rd IEEE International Workshop on Computational Advances in Multi-Sensor Adaptive Processing (CAMSAP)}, 189--192.
  
\bibitem[{Vershynin(2010)}]{vershynin2010introduction}
Vershynin, R. (2010), ``Introduction to the non-asymptotic analysis of
  random matrices," \textit{arXiv preprint arXiv:1011.3027}.
\end{description}

\end{document}